\documentclass[11pt,a4paper]{article} 
\pdfoutput=1
\usepackage{epsfig}
\usepackage{graphicx}
\usepackage{jheppub}
\usepackage{amsmath}
\usepackage{amsfonts}
\usepackage{amssymb}
\usepackage{graphicx}

\def\beq{\begin{equation}}
\def\eeq{\end{equation}}
\def\bsp#1\esp{\begin{split}#1\end{split}}
\def\beqa{\begin{eqnarray}}
\def\eeqa{\end{eqnarray}}

\def\M{{\cal M}}

\def \as {\relax\ifmmode\alpha_s\else{$\alpha_s${ }}\fi}
\def\e{\epsilon}

\def\eps{\epsilon}
\def\ord{{\cal O} }

\newcommand\sss{\scriptscriptstyle}

\def\CF{C_{\sss F}}
\def\CA{C_{\sss A}}

\newcommand{\lsim}{
\mathrel{\hbox{\rlap{\hbox{\lower4pt\hbox{$\sim$}}}\hbox{$<$}}}}
\newcommand{\gsim}{
\mathrel{\hbox{\rlap{\hbox{\lower4pt\hbox{$\sim$}}}\hbox{$>$}}}}

\def\sect#1{Sec.~{\ref{#1}}}
\def\fig#1{fig.~{\ref{#1}}}

\def\eqn#1{eq.~(\ref{#1})}
\def\Eqn#1{Equation~(\ref{#1})}
\def\eqns#1#2{eqs.~(\ref{#1}) and~(\ref{#2})}

\newcommand{\JHEP}[1]{JHEP {\bf #1}}

\title{The infrared structure of gauge theory amplitudes 
in the high-energy limit}

\author[a]{Vittorio Del Duca,}
\author[b]{Claude Duhr,}
\author[c]{Einan Gardi,}
\author[d]{Lorenzo Magnea,}
\author[e]{Chris D. White}

\affiliation[a]{INFN, Laboratori Nazionali Frascati, \\
00044 Frascati (Roma), Italy}
\affiliation[b]{Institute for Particle Physics Phenomenology,
University of Durham,\\ Durham, DH1 3LE, UK}
\affiliation[c]{The Tait Institute, School of Physics and Astronomy, 
The University of Edinburgh,\\
Edinburgh, EH9 3JZ, Scotland, UK}
\affiliation[d]{Dipartimento di Fisica Teorica, Universit{\`a} di Torino, and \\ INFN, Sezione di Torino, Via P. Giuria 1, I-10125 Torino, Italy}
\affiliation[e]{School of Physics and Astronomy, 
Scottish Universities Physics Alliance, University of Glasgow, \\
Glasgow, G12 8QQ, Scotland, UK}

\emailAdd{delduca@lnf.infn.it}
\emailAdd{claude.duhr@durham.ac.uk}
\emailAdd{Einan.Gardi@ed.ac.uk}
\emailAdd{magnea@to.infn.it}
\emailAdd{Christopher.White@glasgow.ac.uk}

\abstract{We develop an approach to the high-energy limit 
of gauge theories based on the universal properties
of their infrared singularities. Our main tool is the dipole formula, 
a compact ansatz for the all-order infrared singularity structure 
of scattering amplitudes of massless partons. By taking the high-energy 
limit, we show that the dipole formula implies Reggeization of 
infrared-singular contributions to the amplitude, at leading logarithmic 
accuracy, for the exchange of arbitrary color representations 
in the cross channel. We observe that the real part of the 
amplitude Reggeizes also at next-to-leading logarithmic 
order, and we compute the singular part  of the two-loop Regge 
trajectory, which is universally expressed in terms of the cusp 
anomalous dimension. Our approach provides tools to study the 
high-energy limit beyond the boundaries of Regge factorization:
thus we show that Reggeization generically breaks down 
at next-to-next-to-leading logarithmic accuracy, and provide a 
general expression for the leading Reggeization-breaking operator. 
Our approach applies to multiparticle amplitudes in multi-Regge 
kinematics, and it also implies new constraints on possible corrections to 
the dipole formula, based on the Regge limit.}


\begin{document}
\begin{flushright}
IPPP/11/48 \\ DCPT/11/96 \\ 
Edinburgh 2011/22 \\ DFTT-24/2011\\
\vspace*{-25pt}
\end{flushright}
\maketitle
\allowdisplaybreaks

\section{Introduction}
\label{sec:intro}

It is well known that the structure of gauge theory scattering 
amplitudes simplifies dramatically in the high-energy limit, in 
which the centre-of-mass energy $\sqrt{s}$ is much larger than 
the typical momentum transfer $\sqrt{- t}$, or, alternatively, $| s/t | 
\rightarrow \infty$, with $t$ held fixed. Studies of this limit predate 
QCD, and formed the basis of Gribov-Regge theory (see for 
example~\cite{Collins:1977jy,Landshoff,Gribov} and references 
therein), in which the analytic properties of amplitudes were 
considered in the complex angular momentum plane, independently
of any underlying field theory; the high-energy limit is thus often
referred to as the {\it Regge limit}. The asymptotic form of scattering 
amplitudes in this limit is determined by the structure of 
singularities in the complex angular momentum plane. If simple 
poles are present, for example, the amplitude takes the form
\beq
  {\cal A} (s, t) \, \, \xrightarrow{\left| \frac{s}{t} \right| 
  \rightarrow \infty} \, \,  f(t) \, s^{\epsilon(t)} \, ,
\label{ampregge}
\eeq
for some prefactor function $f(t)$, where $\epsilon(t)$ is the Regge
trajectory associated with the right-most pole in the complex angular 
momentum plane, which can be physically interpreted in terms of the 
exchange of a family of particles in the $t$-channel. Multiple poles or 
cuts give rise to a more complicated $s$-dependence, in addition to 
the power-like growth described by \eqn{ampregge}.

The high-energy limit has also been extensively studied within the 
context of perturbative quantum field theory. In a variety of theories,
amplitudes may display the phenomenon of {\it Reggeization}: 
specifically, amplitudes for $2 \rightarrow n$ scattering are 
dominated, in the Regge limit,  by $t$-channel exchanges of 
particles whose propagators become dressed according to the schematic form\footnote{For spin 1 gauge bosons, \eqn{propregg} 
may be taken to represent the Feynman gauge result.}
\beq
  \frac{1}{t} \, \longrightarrow \, \frac{1}{t} \, 
  \left( \frac{s}{- t} \right)^{\alpha(t)} \, .
\label{propregg}
\eeq
This perturbative result leads to amplitudes which are consistent 
with Regge theory expectations, {\it i.e.} having the form of 
\eqn{ampregge}, with the two functions $\alpha(t)$ and $\epsilon(t)$ 
related by an integer additive constant. The function $\alpha(t)$ is 
thus usually referred to as the {\it Regge trajectory} of the 
corresponding particle. 

The history of Reggeization studies in quantum field 
theory is by now a lengthy one, beginning with the work 
of~\cite{PhysRevLett.9.275,Mandelstam:1965zz,Abers:1967zz,Grisaru:1973vw,Grisaru:1974cf,Grisaru:1973ku}. 
In QED, at leading logarithmic (LL) accuracy, the electron is found 
to Reggeize~\cite{PhysRevLett.9.275,McCoy:1976ff}, but 
the photon does not~\cite{Mandelstam:1965zz,Frolov:1970ij,
Gribov:1970ik,Cheng:1969bf}. In QCD it has been shown that 
both the gluon~\cite{Balitsky:1979ap} and the 
quark~\cite{Bogdan:2006af} Reggeize at LL accuracy. 
Those proofs are based on the gluon~\cite{Tyburski:1975mr,
Lipatov:1976zz,Mason:1976fr,Cheng:1977gt,Fadin:1975cb,
Kuraev:1977fs,Kuraev:1976ge} and quark~\cite{Mason:1976ky,
Sen:1982xv,Fadin:1977jr} Regge trajectories to one-loop order, 
which are necessary to generate all leading logarithms of $s/t$ in 
the scattering amplitude\footnote{For a pedagogical review of 
Reggeization at one loop in QCD, see~\cite{DelDuca:1995hf,
Forshaw:1997dc}.}. The two-loop gluon~\cite{Fadin:1995xg,
Fadin:1996tb,Fadin:1995km,Blumlein:1998ib,DelDuca:2001gu}
and quark~\cite{Bogdan:2002sr} Regge trajectories have also 
been computed. Reggeization, however, has been proven to 
next-to-leading logarithmic order (NLL) only for the 
gluon~\cite{Fadin:2006bj}. Furthermore, contributions beyond 
NLL order have been considered in the simpler context of an 
${\cal N} = 4$ Super-Yang-Mills (SYM) theory (see section 5 in \cite{Drummond:2007aua} as well as Refs.~\cite{Bartels:2008ce,DelDuca:2008pj}). The proof of Reggeization to a given
logarithmic accuracy, but to all orders in perturbation theory 
(corresponding to a fixed loop order in the Regge trajectory) 
typically involves a careful iterative argument, which shows that 
kinematic information at any given fixed order is consistent with 
the form of \eqn{propregg}, and also that the color factor at each 
order is proportional to that of the appropriate single-particle 
exchange graph. 

An alternative approach~\cite{Korchemsky:1993hr,Sotiropoulos:1993rd,
Korchemskaya:1994qp,Korchemskaya:1996je,Kucs:2004gj,
Dokshitzer:2005ig} uses the fact that 
scattering in the Regge limit can be described by a pair of 
Wilson lines\footnote{A general discussion of the role of 
Wilson lines in the high-energy limit of QCD was given 
in~\cite{Balitsky:2001gj}.}, and exploits the renormalisation properties of 
the latter to derive the gluon Regge trajectory up to two 
loops, as well as an all-order expression~\cite{Korchemskaya:1996je}
for the singular part of the trajectory in terms of an integral over the cusp anomalous dimension~\cite{Korchemsky:1988si,Korchemsky:1988hd,Korchemsky:1987wg,Ivanov:1985np,Korchemsky:1985xj}. We shall reproduce these results in the present paper using a different theoretical framework.

Aside from being of conceptual interest in making contact 
between the high-energy limit of perturbative quantum field 
theories and the known constraints of Gribov-Regge theory, 
Reggeization is a highly important result both in view of 
phenomenological applications and as a tool for the theoretical
analysis of gauge theory amplitudes. For example, it is a crucial 
ingredient in the BFKL equation~\cite{Kuraev:1976ge,
Fadin:1975cb,Kuraev:1977fs,Balitsky:1978ic}, an integral 
equation for the gluon four-point function in the Regge limit, 
whose singlet solution describes the $t$-channel exchange of a Reggeized 
object having the quantum numbers of the vacuum (the Pomeron). Phenomenological applications of the BFKL equation (and related 
results concerning the factorized structure of scattering amplitudes 
in the Regge limit) are wide-ranging and constitute a field of
research too vast to be summarized here. 
On the theoretical side, the high-energy limit has been instrumental
in several recent studies concerning the all-order structure of 
gauge theory amplitudes: for example, corrections to the 
BDS conjecture~\cite{Bern:2005iz} for the iterative structure 
of scattering amplitudes in planar ${\cal N} = 4$ SYM were 
analyzed in~\cite{Bartels:2008ce,DelDuca:2008jg,Brower:2008nm,Brower:2008ia}, and properties of the Regge limit 
were used in~\cite{DelDuca:2009au,DelDuca:2010zg,
DelDuca:2010zp} for the calculation of light-like polygonal 
Wilson loops, which are conjectured to be dual to maximally helicity violating scattering amplitudes 
in this particular theory~\cite{Alday:2007hr,
Drummond:2007aua,Brandhuber:2007yx}.

In this paper we consider the high-energy limit of gauge theory 
amplitudes from a novel viewpoint, and we relate Regge 
factorization to the universal structure of infrared singularities 
of massless gauge theories~\cite{DelDuca:2011xm}. Our approach is motivated by the 
well-known observation that the Regge trajectory is infrared divergent 
in perturbation theory, since it arises formally as an integral over 
the loop transverse momentum, which always diverges in the presence 
of massless gauge bosons. As a consequence, it must be possible to 
employ our understanding of the universal properties of infrared 
radiation in order to study the high-energy limit in general, and Reggeization
in particular. Clearly, finite contributions 
to the Regge trajectory, which are known to arise starting at NLL, will 
be outside the domain of applicability of our method. As we will see, 
however, our approach will allow us to draw very broad conclusions
concerning both the generality and the limitations of the phenomenon 
of Reggeization, and will provide us with tools to analyze the
high-energy limit beyond the limits of Regge factorization.

The factorization and exponentiation of soft and collinear 
singularities have been actively studied for several decades 
(see, for example~\cite{Grammer:1973db,Mueller:1979ih,
Collins:1980ih,Sen:1981sd,Sen:1982bt,Gatheral:1983cz,
Frenkel:1984pz,Magnea:1990zb}); only recently, however, has
a general all-order understanding of the anomalous dimensions 
that govern infrared exponentiation for multiparticle amplitudes 
in non-abelian gauge theories begun to emerge. For massless 
gauge theories, current knowledge is summarized in the 
{\it dipole formula}~\cite{Becher:2009cu,Gardi:2009qi,
Becher:2009qa,Gardi:2009zv}, a recently proposed ansatz for 
the all-order infrared singularity structure of general fixed-angle 
scattering amplitudes of massless partons, which we will review in 
more detail in \sect{sec:sumodipoles}. Briefly, the essential idea 
of infrared exponentiation is that all IR singularities, which appear 
as poles in dimensional regularisation (in $d =  4 - 2 \epsilon$ 
dimensions, with $\epsilon < 0$) may be encapsulated in an 
exponential operator acting on a hard interaction which is finite 
as $\epsilon\rightarrow 0$. The exponent contains in principle 
terms which couple kinematic and color dependence of all hard 
partons. The dipole formula, however, posits that for massless 
amplitudes correlations exist only between pairs of hard particles 
({\it i.e.} there are no irreducible correlations between three 
or more partons, a fact which is not true for massive 
particles~\cite{Kidonakis:2009ev,Mitov:2009sv,Becher:2009kw,
Beneke:2009rj,Czakon:2009zw,Ferroglia:2009ep,Ferroglia:2009ii,
Chiu:2009mg,Mitov:2010xw}). Furthermore, the coefficients of 
these dipole correlations are governed purely by the cusp 
anomalous dimension~\cite{Korchemsky:1988si,Korchemsky:1988hd,Korchemsky:1987wg,Ivanov:1985np,Korchemsky:1985xj} and by the beta function. The dipole formula 
is known to be exact up to two loop order in the exponent, for 
any number of massless hard partons. Possible corrections at higher 
orders are strongly constrained~\cite{Gardi:2009qi,Becher:2009qa,
Dixon:2009ur}, as we discuss below in \sect{sec:sumodipoles}. 

By applying the dipole formula in the case of $2 \rightarrow 2$ scattering in the high-energy limit, we will show explicitly that, at 
leading-logarithmic accuracy\footnote{By LL accuracy we mean 
the ability to predict the coefficients of the largest powers of 
$\ln(s/t)$ arising to all orders, regardless of possible overall 
powers of $\alpha_s$ which might be present for a given 
observable. Thus if the first logarithm of $s/t$ arises at 
order $\alpha_s^{k +1}$, leading logarithms are of the form 
$\alpha_s^k \left(\alpha_s \ln(s/(-t)) \right)^p$.}, Reggeization 
of allowed $t$-channel exchanges is a completely general 
phenomenon, which takes place for arbitrary color representations 
exchanged in the crossed ($t$ or $u$) channel. The relevant one-loop 
Regge trajectory in each case involves the quadratic Casimir invariant 
of the appropriate representation of the gauge group, generalising 
what is already known about the quark and gluon Regge trajectories. 
Next, given the all-order nature of the dipole formula, we will also be 
able to examine Reggeization beyond leading logarithmic accuracy. 
We will show that, for the singular terms of the Regge trajectory, 
Reggeization holds also at NLL accuracy and with the same degree 
of generality, but only for the real part of the amplitude. We will 
also show that the singular terms of the Regge trajectory are 
given by a simple integral of the cusp anomalous dimension 
over the scale of the running coupling, as already derived
in~\cite{Korchemsky:1993hr,Sotiropoulos:1993rd,
Korchemskaya:1994qp,Korchemskaya:1996je} using Wilson lines. Beyond NLL 
order, however, we will provide evidence for the breakdown 
of Reggeization through the appearance in the perturbative 
exponent of color operators that survive in the high-energy 
limit but in general cannot be diagonalized simultaneously 
with the ones responsible for $t$ (or $u$) channel exchange. We 
further show that Reggeization breaking will take place at NNLL 
independently of the precise form of the three-loop soft anomalous 
dimension; conversely, we also use the Regge limit to derive new constraints on potential three-loop corrections to the soft anomalous
dimension going beyond the dipole formula, extending the results 
of Ref.~\cite{Dixon:2009ur}. Finally, we generalize our discussion 
to the case of $2 \rightarrow n$ scattering, in multi-Regge 
kinematics. Also in this general case, we show that the dipole 
formula implies LL Reggeization in the crossed channel, according 
to the standard ansatz employed in the multi-Regge 
limit~\cite{DelDuca:1995hf}, and we observe the same pattern 
of partial generalization to higher logarithmic accuracy that arises 
for the four-point amplitude. In general, our approach, while focused  
on divergent contributions to the high-energy limit, goes beyond
the limitations of Regge factorization, and gives results that
are valid to arbitrary logarithmic accuracy and for general color 
exchanges.

The structure of the paper is as follows. In the remainder of this 
introduction, we will review relevant information regarding both
the high-energy limit of amplitudes and the dipole formula. 
In \sect{sec:reggeproof} we will present our argument for 
Reggeization based on the dipole formula, by considering 
the Regge limit of the four-point amplitude and examining 
the action of the infrared-singular operator on hard interactions 
involving a definite $t$-channel exchange. Using this formalism we will 
reproduce in \sect{sec:hard} the known form of gluon-gluon scattering at leading logarithmic order, and show that this result generalises to exchanges involving particles in different representations of the gauge group. 
In \sect{sec:beyond} we consider the Regge trajectory at 
higher order in the perturbative expansion, and construct explicitly 
the color operator responsible for the expected breakdown of 
Reggeization at NNLL. In Sec.~\ref{sec:beyond_dipoles} we use 
the reverse logic and demonstrate that the Regge limit provides a 
useful constraint on potential corrections to the dipole formula at 
three loops and beyond. In \sect{sec:multiregge} we generalize our 
basic argument to the case of $n$-particle amplitudes in multi-Regge 
kinematics. Finally, we discuss our results in \sect{sec:discuss} 
before concluding.

\subsection{The high-energy limit: an outline}
\label{sec:Reggeization}

In this section, we give a slightly more detailed presentation of 
known results on the high-energy limit of scattering amplitudes, 
introducing concepts and notations that will be useful in the 
following sections. We will focus in particular on gluon and 
quark scattering, preparing for the more general discussion 
of \sect{sec:hard}. 

\begin{figure}
\begin{center}
\scalebox{1.5}{\includegraphics{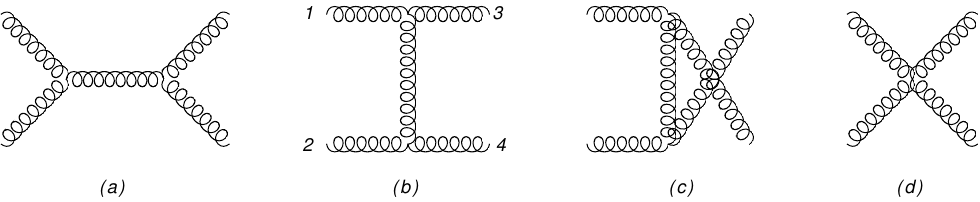}}
\caption{Leading order Feynman diagrams for $gg\rightarrow gg$ 
scattering: (a) $s$ channel; (b) $t$ channel; (c) $u$ channel; (d) four-gluon vertex. 
Only diagram (b) contributes in the Regge limit, $|t/s| \to 0$ (in an appropriate gauge).}
\label{LOgg}
\end{center}
\end{figure}

Let us consider first the case of $g g \rightarrow g g$ scattering, 
whose leading-order Feynman diagrams are shown in \fig{LOgg}. 
While the high-energy limit of the amplitude is, of course, gauge invariant, it is useful to refer to
a diagrammatic picture in which the dominant contribution at large $|s/t|$ (up to power-suppressed terms) is obtained from a single diagram. Indeed, in an appropriate gauge~(see e.g. Sec.~2.4 in  Ref.~\cite{DelDuca:1995hf}), of the four possible diagrams  -- involving $s$, $t$ and $u$-channel exchanges, and a four-gluon vertex, respectively -- only the $t$-channel exchange diagram of \fig{LOgg}(b) contributes in the Regge limit\footnote{Note that keeping only diagram (b) violates gauge invariance, but only 
by terms which are suppressed in the Regge limit.}, with the others suppressed by powers of $|t/s|$. The 
leading order (LO) contribution in the Regge limit thus has a single 
color structure, namely that associated with a $t$-channel color-octet exchange. 

Higher-order contributions in $g g \rightarrow g g$ scattering might 
in general involve additional color structures, besides the pure octet 
exchange observed at tree level, even in the Regge limit. There are 
two reasons for this: first, beyond LO there could be contributions 
not described by pure $t$-channel exchange; second, even if pure 
$t$-channel exchange dominates, there may be different possible 
color structures at higher orders. Indeed, one can enumerate the 
possible color quantum numbers exchanged in the $t$ channel by 
taking the product of the color representations of the particles 
labeled 1 and 3 in \fig{LOgg}(b), which in the present case are 
gluons, belonging to the adjoint representation of $SU(3)$, and 
decomposing it into irreducible representations as
\beq
  {\bf 8}_a \otimes {\bf 8}_a \, = \, {\bf 1} \oplus {\bf 8}_a 
  \oplus {\bf 8}_s
  \oplus {\bf 10} \oplus \overline{{\bf 10}} \oplus {\bf 27} \, ,
\label{8*8}
\eeq
where we introduced indices to distinguish the (antisymmetric) adjoint representation 
from the 8-dimensional symmetric representation. One sees explicitly 
that a color-octet exchange is only one of a number of possibilities, 
and predicting which ones will contribute in the Regge limit
requires further theoretical input. This input is provided by the 
observation that, at least for leading logarithms, the diagrams that 
contribute to the Regge limit correspond to the exchange of a gluon 
ladder in the $t$ channel~\cite{Gribov:1970ik,Lipatov:1976zz,
Kuraev:1976ge}. As we will see shortly, this also constrains the 
color structure of the amplitude. The non-trivial nature of the 
Reggeization property is thus twofold, involving both color and
kinematics: the leading kinematic behaviour at each order in the 
coupling constant involves logarithms of $s/t$ precisely so as 
to produce the power-like dependence given by \eqn{propregg}; 
also, the color factor at each order in perturbation theory is 
proportional to that of the tree level exchange. In other words, 
of the possible $t$-channel exchanges listed in \eqn{8*8}, only 
the octet exchange survives at leading logarithmic order in the 
Regge limit, with other possible color factors being kinematically 
suppressed, either by logarithms or powers of $t/s$.

Calculating the amplitude for exchange of a Reggeized gluon, to 
LL accuracy, gives a matrix element of the form~\cite{Lipatov:1976zz,Kuraev:1976ge}
\beq
  \! {\cal M}^{g g \rightarrow g g}_{a_1a_2 a_3 a_4} (s,t) 
  \, = \, 2 \, g_s^2 \, \frac{s}{t} \,
  \bigg[ (T^b)_{a_1a_3} C_{\lambda_1\lambda_3}(k_1, k_3) 
  \bigg] \, \left( \frac{s}{- t} \right)^{\alpha(t)} \,
  \bigg[ (T_b)_{a_2 a_4} C_{\lambda_2\lambda_4}(k_2, k_4) 
  \bigg] \,,
\label{Mgg}
\eeq
where $a_j$ and $p_j$ are the color index and momentum of gluon 
$j$ (with labelling as in figure~(\ref{LOgg}b)), and $T^b$ is a color 
generator in the adjoint representation, so that $(T^a)_{b c} = 
- {\rm i} f_{abc}$. The coefficient functions $C_{\lambda_i
\lambda_j}(k_i, k_j)$, usually referred to as {\it impact factors}, 
depend on the  helicities~\cite{DelDuca:1995zy,DelDuca:1996km} 
of the gluons (or on the spin polarizations~\cite{Kuraev:1976ge} 
in the case of quarks), and may contain collinear singularities 
associated with them, but, as the notation suggests, carry no 
$s$~dependence. In the high-energy limit helicity is conserved 
across the vertices, so only certain impact factors $C_{\lambda_i
\lambda_j}(k_i, k_j)$ are relevant (see~\cite{DelDuca:1996km} 
for more details). Equation~(\ref{Mgg}) is an example of {\it Regge
factorization}: the impact factors are universal (process-independent), 
reflecting the
properties of the scattered partons, while the states exchanged in
the $t$ channel appear only through their Reggeized propagator.

A further important ingredient in the computation of the Regge 
limit is the observation that the matrix element must have even
parity under $s \leftrightarrow u$ exchange, which follows from 
the assumption that only $t$-channel gluon ladders contribute. It 
is easy to see that, at leading logarithmic accuracy, the kinematic part of \eqn{Mgg} is 
odd under $s \leftrightarrow u$ exchange, due to the overall factor of $s/t$. 
Indeed, Mandelstam invariants satisfy, for massless particles, the 
momentum conservation relation
\beq
  s + t + u \, = \, 0 \, ,
\label{momcon}
\eeq
which in the Regge limit implies
\beq
  u \simeq - s \, ,
\eeq
leading to an overall sign change when $s$ is replaced by $u$. This,  
in turn, requires that the color structure of the amplitude should also 
be odd under the same exchange. Once again, this is true for 
Reggeized gluon exchange, since
\beq
  T^b_{i_1i_3} \, T^{b}_{i_2 i_4} \, = \, - \, T^b_{i_1 i_3} \, 
  T^{b}_{i_4 i_2} \, ,
\label{Tpar}
\eeq
if we take the generators in the (antisymmetric) color octet 
representation. The color factor on the right-hand side is that of 
the process $g(k_1) g(k_4) \rightarrow g(k_3) g(k_2)$, whose 
amplitude (by crossing symmetry) is equal to that of \eqn{Mgg} 
upon replacing $s$ with $u$. One may then rewrite~\eqn{Mgg} 
to display explicitly the symmetry under $s \leftrightarrow u$ 
exchange, as
\beqa
  {\cal M}^{g g \rightarrow g g}_{a_1a_2 a_3 a_4} (s,t)
  \, = \, g_s^2 \, \, \frac{s}{t} & \bigg[ (T^b)_{a_1a_3}
  C_{\lambda_1\lambda_3}(k_1, k_3) \bigg] & \left[ \left( 
  \frac{s}{- t} \right)^{\alpha(t)} \!\! + \left( \frac{- s}{- t} 
  \right)^{\alpha(t)} \right] \nonumber \\ \times &
  \bigg[ (T^b)_{a_2 a_4} C_{\lambda_2\lambda_4}
  (k_2, k_4) \bigg] & \, .
\label{Mgg2}
\eeqa
One observes that the symmetry requirement under $s 
\leftrightarrow u$ exchange, together with the negative parity 
(usually called `signature' in this context) of the kinematic part 
of the amplitude, force the color representation exchanged in the 
$t$ channel to be antisymmetric. Notice however that this 
requirement does not uniquely select the (antisymmetric)
octet  in \eqn{8*8}: either of the two decuplet representations 
would also be allowed. That only the octet actually contributes 
to the reggeized amplitude at LL  (and indeed at NLL as well) is 
a result of the detailed proof of Reggeization~\cite{Balitsky:1979ap,
Fadin:2006bj}.

Similar expressions are obtained for quark-quark or quark-gluon 
scattering, where the only modification in \eqn{Mgg} is that the 
color generators and the coefficient functions are replaced by 
those belonging to the appropriate representation. Crucially, 
however, the Regge trajectory $\alpha(t)$ is a universal object: 
it is a property of the particle exchanged in the $t$ channel, and 
does not depend on the identities of the external particles. 
More precisely, the gluon Regge trajectory $\alpha (t)$, appearing 
in \eqn{Mgg} and in \eqn{Mgg2} can be expressed as 
\beq
  \alpha(t) \, = \, \frac{\alpha_s (- t, \epsilon)}{4 \pi} 
  \, \, \alpha^{(1)} +
  \left( \frac{\alpha_s (- t, \epsilon)}{4 \pi} \right)^2 
  \, \alpha^{(2)} + \, \ord \left( \alpha_s^3 \right) \, ,
\label{alphb}
\eeq
where we expanded the trajectory in terms of the $d$-dimensional
running coupling
\beq
  \alpha_s (- t, \epsilon) = \left( \frac{\mu^2}{- t} 
  \right)^\epsilon \, \alpha_s (\mu^2) 
  + \, \ord \left( \alpha_s^2 \right) \, ,
\label{rescal}
\eeq 
with $d = 4 - 2 \eps$, and $\eps < 0$ for infrared regularization.
According to \eqn{Mgg2}, truncating \eqn{alphb} to first order, one 
finds that only the octet of negative signature gives contributions 
of the form $\alpha_s^n \, (\ln(s/|t|))^n$ in the $n$-loop amplitude. 
This is the statement of the Reggeization of the gluon to LL
accuracy~\cite{Balitsky:1979ap}. The result for the one-loop 
gluon trajectory is 
\beq
  \alpha^{(1)} \, = \, C_A \, 
  \frac{\widehat{\gamma}_K^{(1)}}{\epsilon} \, = \,
  C_A \, {2 \over \epsilon} \, ,
\label{alpha1}
\eeq
where $\CA = N_c$ is the quadratic Casimir invariant for 
the adjoint representation, as is appropriate for the exchange 
of a gluon, and we have introduced the one loop coefficient 
of the universal cusp anomalous dimension~\cite{Korchemsky:1988si,Korchemsky:1988hd,Korchemsky:1987wg,Ivanov:1985np,Korchemsky:1985xj}
(to be discussed in more detail in \sect{sec:sumodipoles}), 
$\widehat{\gamma}_K^{(1)} = 2$, following the notation of Ref.~\cite{Gardi:2009qi}. Together with the 
effective vertex for the emission of a gluon along the 
ladder~\cite{Lipatov:1976zz}, Reggeization  of the gluon is a key prerequisite
for the derivation of the BFKL equation 
at LL accuracy~\cite{Kuraev:1977fs,Balitsky:1978ic}.

\begin{figure}
\begin{center}
\scalebox{0.9}{\includegraphics{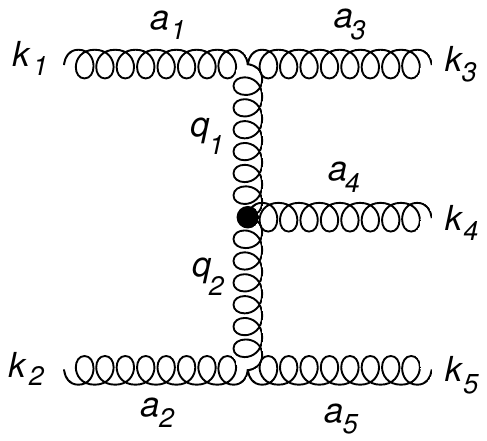}}
\caption{The $gg\rightarrow ggg$ scattering process in the Regge 
limit, where momenta $k_i$ and color indices $i$ are shown. The 
intermediate gluon on the right-hand side couples with an effective 
vertex $V_{\lambda_4}$, as written in \eqn{Mggg}.}
\label{3gfig}
\end{center}
\end{figure}

In order to generalize the idea of Reggeization to multiparticle
emission, one may begin by considering the amplitude for $gg 
\to ggg$ scattering. Each emitted gluon may be characterized by 
its rapidity in the center-of-mass frame of the collision, given by 
$y_i = \frac{1}{2}\ln\left(\frac{E_i+p_{l,i}}{E_i-p_{l,i}}\right)$, 
with $p_{l,i}$ the longitudinal momentum 
of the $i$-th gluon. High-energy logarithms arise then in the 
limit of strongly ordered rapidities of the outgoing gluons, with 
the transverse momenta of comparable size,
\beq
  y_3 \gg y_4 \gg y_5\,, \qquad 
  |k_3^\perp| \simeq |k_4^\perp| \simeq |k_5^\perp|  \, ,
\label{3rapord}
\eeq
where, without loss of generality, we have taken the rapidities as 
decreasing. Note that we have labelled momenta and color indices 
as in \fig{3gfig}.
With these conventions, the amplitude for $gg\to ggg$ scattering
can be written as~\cite{Lipatov:1976zz}
\beqa
  \lefteqn{ {\cal M}^{g g \rightarrow g g g}_{a_1a_2 a_3 a_4 a_5} 
  \, = \, 2 \, g_s^3 \, s \,
  \bigg[ (T^b)_{a_1a_3} \, C_{\lambda_1\lambda_3}
  (k_1, k_3) \bigg]
  \left[ \frac1{t_1} \left( \frac{s_{34}}{- t_1} \right)^{\alpha(t_1)}
  \right]  } \nonumber\\
  &\times&  \bigg[ (T^{a_4})_{bc} \, V_{\lambda_4}(q_1, q_2) 
  \bigg]
  \left[ \frac1{t_2} \left( \frac{s_{45}}{- t_2} \right)^{\alpha(t_2)} 
  \right]
  \bigg[ (T^c)_{a_2 a_5} \, C_{\lambda_2\lambda_5}(k_2, k_5) 
  \bigg] \,,
\label{Mggg}
\eeqa
with $k_4 = q_1 - q_2$, $k_3^\perp = - q_1^\perp$, 
$k_5^\perp = q_2^\perp$ and $t_i \simeq - |q_i^\perp|^2$ 
for $i = 1, 2$. The effective vertex for the emission of a positive 
helicity gluon along the ladder, also known as the Lipatov 
vertex~\cite{Lipatov:1976zz,Lipatov:1991nf,DelDuca:1995zy},
is given by
\beq
  V_+ (q_1, q_2) = \sqrt{2} \, \frac{q_1^{\perp\ast} 
  q_2^\perp}{k_4^\perp} \, ,
\label{Lipvert}
\eeq
where we use the complex momentum notation $k^\perp = k^1 + 
{\rm i} k^2$. The corresponding vertex for a negative helicity gluon,
$V_- (q_1, q_2)$, is obtained by taking the complex conjugate of 
\eqn{Lipvert}. One may generalize \eqn{3rapord} to $2\to n$ 
scattering, which is usually referred to as multi-Regge kinematics. 
The emission of a single gluon along the ladder iterates in 
an obvious fashion, so as to describe the emission of any number 
gluons (see~\cite{DelDuca:1995hf} for a pedagogical review), and
\eqn{Mggg} generalizes accordingly~\cite{Kuraev:1976ge}. 
The amplitude for $2\to n$ scattering, with emission of many 
gluons along the ladder, is related via unitarity to the BFKL 
equation~\cite{Kuraev:1977fs,Balitsky:1978ic}, and it will be 
discussed in \sect{sec:multiregge} in the context of the dipole 
formula.

Reggeization of the gluon has also been proven to next-to-leading 
logarithmic (NLL) accuracy in QCD~\cite{Fadin:2006bj}, implying that only 
the octet contributes to the $\alpha_s^n\, (\ln(s/|t|))^{n - 1}$ 
term of the $n$-loop amplitude, with $\alpha^{(2)}$ the two-loop 
gluon Regge trajectory~\cite{Fadin:1995xg,Fadin:1996tb,
Fadin:1995km,Blumlein:1998ib,DelDuca:2001gu}, given 
by\footnote{Note that the sign of the double pole term in
\eqn{eq:2loop} is the opposite of the one found, for example, 
in~\cite{DelDuca:2001gu}: this is due to the fact that here we 
are expanding the Regge trajectory in terms of the renormalized
coupling, rather than the bare coupling, as was done 
in~\cite{DelDuca:2001gu}.}
\beq
  \alpha^{(2)} \, = \, C_A \left[ - {b_0 \over \eps^2} +
  \widehat{\gamma}_K^{(2)} \,
  {2 \over \eps} + C_A \left( {404 \over 27} - 2 \zeta_3 \right) 
  + n_f \left(- {56 \over 27} \right) \right] \, ,
\label{eq:2loop}
\eeq
where we introduced the one-loop $\beta$-function coefficient 
$b_0$, and the two-loop cusp anomalous dimension 
$\widehat{\gamma}_K^{(2)}$, given by~\cite{Korchemsky:1988si,Korchemsky:1988hd,Korchemsky:1987wg,Ivanov:1985np,Korchemsky:1985xj}
\beq 
  b_0 = {11C_A - 2 n_f \over 3}, \qquad {\rm and} \qquad 
  \widehat{\gamma}_K^{(2)} = \left({67 \over 18} -
  {\pi^2 \over 6} \right) C_A - {5\over 9} n_f \, .
\label{eq:beta}
\eeq
Together with the coefficient functions for the emission of two 
gluons or two quarks along the ladder~\cite{Fadin:1989kf,
DelDuca:1995ki,Fadin:1996nw,DelDuca:1996me}, and the 
one-loop corrections to the emission of one gluon along the 
ladder~\cite{Fadin:1993wh,Fadin:1994fj,Fadin:1996yv,
DelDuca:1998cx,Bern:1998sc}, the two-loop gluon Regge 
trajectory constitutes a building block of the BFKL equation 
at NLL accuracy~\cite{Fadin:1998py,Camici:1997ij,
Ciafaloni:1998gs}. 

\begin{figure}
\begin{center}
\scalebox{0.8}{\includegraphics{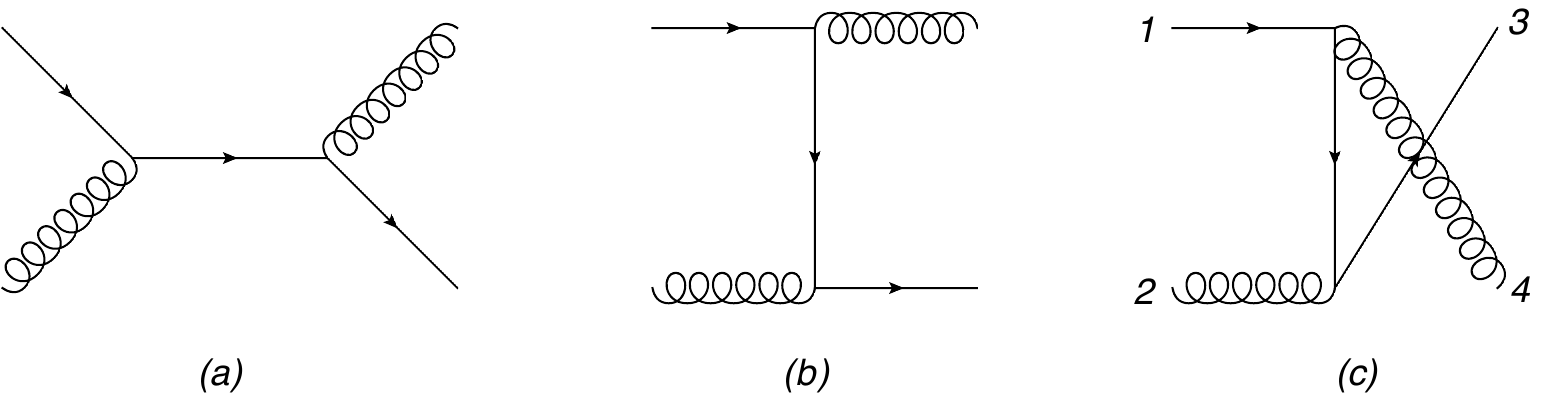}}
\caption{The leading order Feynman diagrams for $q g \rightarrow 
q g$ scattering which are mediated by quark exchange: (a) $s$ 
channel; (b) $t$ channel; (c) $u$ channel. Only diagram (c) 
contributes in the limit $|u/s| \to 0$.}
\label{LOqg}
\end{center}
\end{figure}

With similar methods, one may also examine Reggeization of the 
quark, using the scattering process $q g \rightarrow q g$, as 
shown in ~\fig{LOqg}. 
In the limit $|t/s| \to 0$ this process is dominated by gluon exchange
in the $t$-channel (not displayed in \fig{LOqg}), as was the case for 
gluon-gluon scattering. One may however consider the alternative
high-energy limit $|u/s| \to 0$, $t \simeq - s$. Note that in the 
center-of-mass frame of two-particle scattering one has $u = - s \, 
(1 + \cos \theta)/2$, so the limit $|u/s| \to 0$ corresponds to 
backward scattering, in contrast to the forward scattering associated 
with $t$-channel exchange in the Regge limit $|t/s| \to 0$. 
This alternative high-energy limit is dominated by quark exchange: 
again, there are three diagrams at tree level, displayed 
in \fig{LOqg}, and only the diagram of figure~\ref{LOqg}(c), 
representing scattering via $u$-channel quark exchange, contributes 
at leading power in $u/s$. As in the case of $t$-channel exchange, 
Reggeization amounts to the statement that the propagator for 
the exchanged quark becomes dressed, through virtual corrections, 
by a factor
\beq
  \left( \frac{s}{ - u} \right)^{\alpha_q (u)} \, ,
\label{sufac}
\eeq
where $\alpha_q (u)$ is the quark Regge trajectory. In particular, as 
was the case for gluon exchange, the color structure of the tree level 
interaction is preserved to all orders in the perturbation expansion, at 
least at LL level. The relevant color factor in this case has positive 
parity under the interchange of particles 2 and 3 (with labels as in 
figure~(\ref{LOqg}c)), corresponding to the interchange $s 
\leftrightarrow t$. The amplitude may thus be written in a form
which has manifestly positive signature in the $u$-channel. 
 
By analogy to what was done in \eqn{alphb}, one may expand the 
quark trajectory as
\beq
  \alpha_q (u) \, = \, \frac{\alpha_s (- u, \epsilon)}{4 \pi} 
  \, \, \alpha^{(1)}_q +
  \left( \frac{\alpha_s (- u, \epsilon)}{4 \pi} \right)^2 
  \, \alpha^{(2)}_q + \, \ord \left( \alpha_s^3 \right) \, ,
\label{alphd}
\eeq
where the running coupling, defined as in \eqn{rescal}, is now 
evaluated with reference scale $u$. The result for the one-loop 
quark Regge trajectory is~\cite{Fadin:1977jr}
\beq
  \alpha^{(1)}_q \, = \, C_F \, 
  \frac{\widehat{\gamma}_K^{(1)}}{\epsilon} \, = \,
  C_F \, {2 \over \epsilon} \, ,
\label{alphaq1}
\eeq
with $\CF = (N_c^2 - 1)/(2 N_c)$ the quadratic Casimir invariant of 
the fundamental representation, appropriate to the exchange of a 
quark\footnote{Note that a similar result holds in QED, where the 
electron also Reggeizes~\cite{PhysRevLett.9.275,PhysRev.133.B145,
PhysRev.133.B161, McCoy:1976ff}.}. Truncating \eqn{alphd} to first 
order is equivalent to claiming that only the triplet of positive signature 
contributes to the $\alpha_s^n \, (\ln(s/|u|))^n$ term of the $n$-loop 
amplitude, {\it i.e.} states the Reggeization of the quark to leading 
logarithmic accuracy~\cite{Bogdan:2006af}. The two-loop quark 
Regge trajectory was computed in~\cite{Bogdan:2002sr}, and reads
\beq
  \alpha_q^{(2)} = \CF \left[ - {b_0 \over \eps^2} +
  \widehat{\gamma}_K^{(2)} {2 \over \eps} + \CA 
  \left({404 \over 27} - 2 \zeta_3 \right) + n_f \left(- {56
  \over 27}\right) + (\CF - \CA) \left(16 \zeta_3 \right) \right] \, .
\label{twolooptraja}
\eeq
Note that \eqn{twolooptraja} has the remarkable feature that if 
one replaces everywhere $\CF$ with $\CA$ one obtains the 
two-loop gluon Regge trajectory in \eqn{eq:2loop}. Specifically, 
the pole terms in $\eps$ are the same as in the two-loop gluon 
Regge trajectory, up to the interchange of the overall factor 
$\CF \leftrightarrow \CA$, a fact that will be precisely understood 
in our approach. Finite terms contain a contribution proportional
to the difference $C_F - C_A$: this would vanish for fermions
in the adjoint representation, characteristic of supersymmetric 
gauge theories. Note that Reggeization of the quark to NLL accuracy 
has never been proven: in fact, for both the gluon and the quark 
trajectories, the calculation of the $(k +1)$-loop Regge trajectory, 
which requires a $(k +1)$-loop fixed-order calculation, has so far 
always predated the proof of the corresponding Reggeization to 
${\rm N}^k{\rm LL}$ accuracy, which requires an all-order analysis.

In this section we have reviewed various details regarding 
Reggeization, which are relevant for the remainder of this paper. 
In particular, we have seen that both the quark and the gluon Reggeize 
in QCD, at least at LL level. Furthermore, the singular parts of their 
Regge trajectories are completely determined, at least at two loops, 
by the cusp anomalous dimension and by the beta function, and they 
only differ by the replacement of the overall quadratic Casimir 
invariant corresponding to the exchanged particle (in the $u$ and 
$t$ channels for quark and gluon exchanges, respectively).

\subsection{The dipole formula}
\label{sec:sumodipoles}

In this section, we review the dipole formula of~\cite{Gardi:2009qi,
Becher:2009cu,Becher:2009qa}, which will be used in 
\sect{sec:reggeproof} to investigate Reggeization. The formula 
is a closed form result for the anomalous dimension matrix which 
generates all infrared (soft and collinear) singularities of arbitrary 
fixed-angle scattering processes involving only massless external 
partons, and was first derived in~\cite{Gardi:2009qi,Becher:2009cu,
Becher:2009qa}\footnote{For a pedagogical review, see 
also~\cite{Gardi:2009zv}. The proportionality of the all-order 
anomalous dimension matrix to the one-loop result was conjectured
in~\cite{Bern:2008pv}, after the two-loop calculation 
of~\cite{Aybat:2006mz}.}. Here we briefly summarise the 
derivation of~\cite{Gardi:2009qi}, in order to make clear the 
origin (and possible limitations) of the dipole formula, as well as to 
introduce notation which will be useful in what follows.

\begin{figure}[b]
\begin{center}
\scalebox{0.8}{\includegraphics{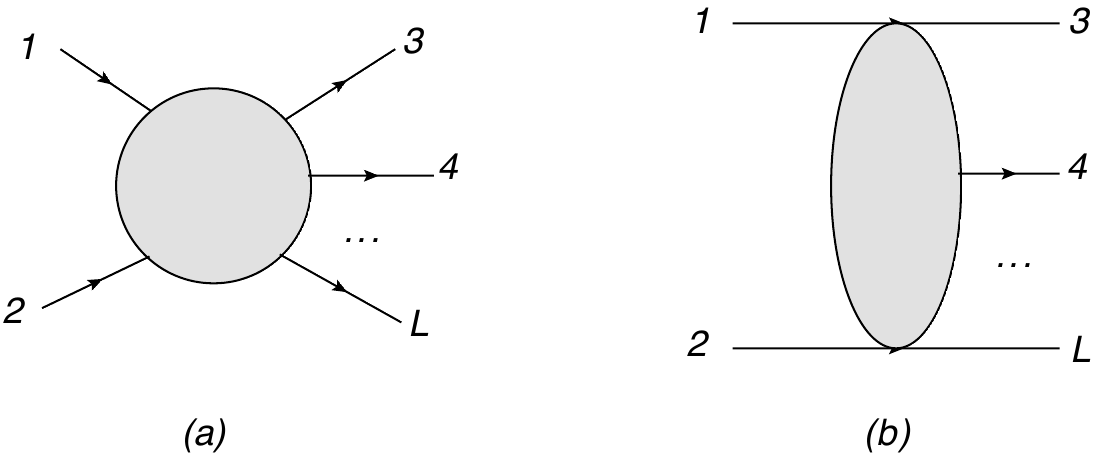}}
\caption{Schematic depiction of: (a) a fixed angle scattering 
amplitude with $L$ massless partons; (b) the same amplitude 
in the forward (Regge) limit $|t/s| \to 0$.}
\label{ampfig}
\end{center}
\end{figure}

Our starting point is a generic fixed-angle scattering amplitude 
for $L$ massless partons, shown schematically in \fig{ampfig}(a). 
Parton momenta $p_l$ satisfy $p_l^2 = 0$, and the invariants 
$p_i \cdot p_j$ are all taken to be large relative to 
$\Lambda^2_{\rm QCD}$, and are assumed to be parametrically 
of the same size. Each parton carries a color index $a_l$, and one 
may write the scattering amplitude as a vector in the space of 
possible color flows,
\beq
  {\cal M}_{a_1\ldots a_L} \left(\frac{p_i}{\mu}, \alpha_s (\mu^2),
  \epsilon \right) \, = \,
  \sum_J  {\cal M}_J \left( \frac{p_i}{\mu}, \alpha_s (\mu^2), 
  \epsilon \right) \, (c^J)_{a_1 \ldots a_L} \, ,
\label{ampcol}
\eeq
where $\{c^J\}$ is a suitable basis of color tensors for the process 
at hand. The amplitude contains long distance singularities, which 
may be traced to soft and collinear regions of integration
in loop momentum space (see {\it e.g.}~\cite{Sterman:1995fz}). 
Many years of studies~\cite{Mueller:1979ih,Collins:1980ih,
Sen:1981sd,Sen:1982bt,Magnea:1990zb,Kidonakis:1998nf,
Sterman:2002qn,Aybat:2006mz,Dixon:2008gr}, have established 
that soft and collinear radiation has universal properties which lead 
to the fact that the associated singularities can be factorized from 
the complete amplitudes. Specifically, one may write the subamplitudes 
${\cal M}_J$ in \eqn{ampcol} in the factorized form\footnote{A 
similar form has recently been explored in the context of perturbative 
quantum gravity~\cite{Naculich:2011ry,White:2011yy,Akhoury:2011kq}.}
\beqa
  {\cal M}_{J} \left(\frac{p_i}{\mu}, \alpha_s (\mu^2), 
  \epsilon \right)  & = & \sum_K {\cal S}_{J K} \left(\beta_i  
  \cdot \beta_j, \alpha_s (\mu^2), 
  \epsilon \right) \,  H_{K} \left( \frac{2 p_i \cdot p_j}{\mu^2},
  \frac{(2 p_i \cdot n_i)^2}{n_i^2 \mu^2}, \alpha_s (\mu^2),
  \epsilon \right) \nonumber \\ & & \times \,
  \prod_{i = 1}^L \, \frac{{\displaystyle J_i 
  \left(\frac{(2 p_i \cdot n_i)^2}{n_i^2 \mu^2},
  \alpha_s (\mu^2), \epsilon \right)}}{{\displaystyle {\cal J}_i 
  \left(\frac{2 (\beta_i \cdot n_i)^2}{n_i^2}, \alpha_s (\mu^2), 
  \epsilon \right)} \,} \,\, ,
\label{fact_massless}
\eeqa
where $\beta_l$ is the 4-velocity of parton $l$, and $n_l$ an 
auxiliary vector associated with each hard parton\footnote{The 
factors of two in the arguments of the various functions in 
\eqn{fact_massless} are conventional and do not play a significant 
role in the present discussion.}, and such that $n_l^2 \neq 0$. 
Here $H_K$ is the {\it hard function}, which is free of infrared 
singularities and thus finite as $\epsilon \rightarrow 0$ after 
renormalization. The {\it soft function} $S_{JK}$ collects all 
soft singularities (including those which are both soft and collinear), 
and acts as a matrix in color flow space, owing to that fact that soft 
gluon emissions transfer color between the external hard parton lines. 
The soft function may be written as a vacuum expectation value of 
a renormalized product of Wilson-line operators acting on the color 
flow basis $\{c^J\}$. One finds
\beq
  \left( c^J \right)_{\{a_k\}} {\cal S}_{J K} \left(\beta_i 
  \cdot \beta_j, \alpha_s (\mu^2), \epsilon \right) 
  \equiv \sum_{\{j_k\}} \, \, \left\langle 0 \left| 
  \, \prod_{i = 1}^L \Big[ \Phi_{\beta_i} (\infty, 0)_{a_k b_k} 
  \Big]  \, \right| 0 \right\rangle_{ \!\! {\rm ren.}}  \, 
  \left( c^K \right)_{\{b_k\}} \, ,
\label{softcorr_massless}
\eeq
where each Wilson-line operator may be written, as usual, as a 
path ordered exponential
\beq
  \Phi^{(l)}_{a_l b_l} \, = \, \left[ {\cal P} \exp \left( {\mathrm i} 
  g_s \int_0^{\infty} d t \, \beta_l \cdot {A} (t \beta_l ) \right) 
  \right]_{a_l b_l} \, .
\label{Wilson_line_def}
\eeq
The {\it jet functions} $J_l$ in \eqn{fact_massless} collect collinear 
singularities associated with parton line~$l$, including those that are 
both collinear and soft. In terms of the auxiliary vector $n_l$, and 
taking as an example the case of an external quark, one has
\beq
  \overline{u}(p_l) \, J_l \left( \frac{(2 p_l 
  \cdot n_l)^2}{n_l^2 \mu^2}, 
  \alpha_s(\mu^2), \e \right) \, = \, \langle p_l \, | \, 
  \overline{\psi} (0) \, \Phi_{n_l} (0, - \infty) \,  | 0 \rangle\, .
\label{Jdef}
\eeq
It is important to note that the jet functions are diagonal in color 
flow space, and they depend only on the quantum numbers of the 
single parton $l$. Note further that singularities which are both soft 
and collinear appear in both the soft function $S_{J K}$ and in the 
jet functions $J_l$. One corrects for this double counting, as shown 
in \eqn{fact_massless}, by dividing by the {\it eikonal jet functions} 
${\cal J}_l$, which are simply defined as the eikonal approximations 
to the partonic jet functions $J_l$. They can thus be expressed in 
terms of Wilson lines as 
\beq
  {\cal J}_l \left( \frac{2 (\beta \cdot n_l)^2}{n_l^2}, 
  \alpha_s(\mu^2), \e \right)  \, = \, \langle 0 | \, 
  \Phi_{\beta_l}(\infty, 0) \, 
  \Phi_{n_l} (0, - \infty) \, | 0 \rangle~.
\label{calJdef}
\eeq
The soft function of \eqn{softcorr_massless} satisfies the evolution
equation 
\beq
  \mu  \frac{d}{d \mu} \, {\cal S}_{J K} \left(\beta_i \cdot 
  \beta_j, \alpha_s(\mu^2), \epsilon \right) \, = \, - \, 
  {\cal S}_{J I} \left(\beta_i \cdot \beta_j, \alpha_s(\mu^2), 
  \epsilon \right) \, \Gamma^{{\cal S}}_{I K} \left(\beta_i 
  \cdot \beta_j, \alpha_s(\mu^2), \epsilon \right) \, ,
\label{renS_massless}
\eeq
a consequence of the fact that Wilson lines renormalize 
multiplicatively~\cite{Polyakov:1980ca,Arefeva:1980zd,
Dotsenko:1979wb,Brandt:1981kf}. The anomalous dimension 
$\Gamma^S_{I K}$, however, is singular as $\epsilon\rightarrow
0$, due to the fact that the soft function still contains collinear 
singularities. Related to this is the fact that the functional
dependence of the soft function involves the scalar products 
$\beta_i \cdot \beta_j$, which are not invariant under rescalings 
of the 4-velocities $\beta_i$, as one would expect from the formal 
definition of the soft function in terms of semi-infinite Wilson lines. 
As analysed in detail in~\cite{Gardi:2009qi}, these facts are both 
consequences of the cusp singularity of massless Wilson lines~\cite{Korchemsky:1988si,Korchemsky:1988hd,Korchemsky:1987wg,Ivanov:1985np,Korchemsky:1985xj}, whose 
properties are dictated by the cusp anomalous dimension 
$\gamma_K(\alpha_s)$ to all orders in perturbation theory. 

One may restore rescaling invariance by considering the {\it reduced 
soft function}~\cite{Dixon:2008gr,Gardi:2009qi}
\beq
  \overline{{\cal S}}_{J K} \left(\rho_{i j}, \alpha_s(\mu^2), \epsilon 
  \right) \, = \,  \frac{{\cal S}_{J K} \left(\beta_i \cdot \beta_j, 
  \alpha_s (\mu^2), \epsilon \right)}{\displaystyle \prod_{i = 1}^L 
  {\cal J}_i \left(\frac{2 (\beta_i \cdot 
  n_i)^2}{n_i^2}, \alpha_s(\mu^2), \epsilon \right)} \, .
\label{reduS}
\eeq
This function is free of collinear poles, which are removed by dividing 
out the eikonal jets. It must then follow that the anomaly in rescaling 
invariance, which was due to collinear singularities, has also been 
cancelled. The reduced soft function must then depend on the 
velocities in a rescaling-invariant manner, and this requirement leads 
to the fact that the kinematic dependence on the left-hand side of 
\eqn{reduS} is through the quantities
\beq
  \rho_{ij} \, \equiv \, 
  \frac{ \, \left| \beta_i \cdot \beta_j \right|^2 \, 
  {\rm e}^{-2 {\rm i} \pi \lambda_{ij}} }
  {\displaystyle \frac{2 \, (\beta_i \cdot n_i)^2}{n_i^2}
  \frac{2 \, (\beta_j \cdot n_j)^2}{n_j^2}} \,  ,
\label{rhoij}
\end{equation}
which are indeed manifestly invariant under the transformation 
$\beta_i \rightarrow \kappa_i \beta_i$. The phases $\lambda_{ij}$ 
are defined by $- \beta_i \cdot \beta_j =  |\beta_i \cdot \beta_j |
{\rm e}^{- {\rm i} \pi \lambda_{ij}}$, where $\lambda_{ij} = 1$ if
$i$ and $j$ are both initial-state partons, or both final-state partons,
and  $\lambda_{ij} = 0$ otherwise.

The reduced soft function in \eqn{reduS} satisfies an evolution 
equation identical in form to \eqn{renS_massless},
\beq
  \mu  \frac{d}{d \mu} \, \overline{\cal S}_{J K} \left(\rho_{ij}, 
  \alpha_s(\mu^2), \epsilon \right) \, = - \, \, \overline{\cal S}_{J I} 
  \left(\rho_{ij}, \alpha_s(\mu^2), \epsilon \right) \, 
  \Gamma^{\overline{\cal S}}_{I K} \left(\rho_{ij}, \alpha_s(\mu^2)
  \right) \, .
\label{renSbar_massless}
\eeq
In this case however the anomalous dimension matrix 
$\Gamma^{\overline{\cal S}}_{I K}$ is finite as $\epsilon 
\rightarrow 0$, since the reduced soft function is free of collinear 
singularities. The restoration of the symmetry under rescaling 
transformations $\beta_i \to \kappa_i \beta_i$ can be further 
exploited, using (\ref{reduS}) and the properties of the jet functions 
${\cal J}_i$, to derive a set of equations~\cite{Gardi:2009qi} that 
tightly constrain the functional dependence of the anomalous 
dimension matrix $\Gamma^{\overline{\cal S}}_{I J}$. They take 
the form
\beq
  \sum_{j \neq i} \, \frac{\partial}{\partial \ln(\rho_{i j})} \,
  \Gamma^{{\overline{\cal S}}}_{IJ} 
  \left( \rho_{i j}, \alpha_s \right) \, = \, \frac{1}{4} \, 
  \gamma_K^{(i)} \left( \alpha_s \right) \, \delta_{IJ} 
  \, ,\qquad \quad \forall i \, .
\label{oureq_reformulated}
\eeq
This is a set of $L$ independent differential equations for the 
matrix-valued soft anomalous dimension, which explicitly couple 
color and kinematic degrees of freedom. In order to write down 
the minimal solution to \eqn{oureq_reformulated}, which leads to 
the announced dipole formula, it is useful to switch to a slightly more 
formal, basis-independent notation for color exchange. This is achieved 
by introducing color-insertion operators ${\bf T}_i$, following the 
notation of Catani and Seymour~\cite{Bassetto:1984ik,
Catani:1996vz}. The color operator ${\bf T}_i$ acts as the identity 
on the color indices of all external partons other than parton $i$, and 
it inserts a color generator in the appropriate representation on
the $i$-th leg. Using this compact notation, color conservation is 
simply expressed (upon choosing a suitable sign convention) by the 
operator identity $\sum_{i = 1}^L {\bf T}_i = 0$, which is understood 
as acting on the hard part of the matrix element. One may furthermore 
define the product ${\bf T}_i \cdot {\bf T}_j \equiv \sum_a 
{\bf T}_i^a \, {\bf T}_j^a$, where $a$ is the adjoint index 
enumerating the color generators. In this language ${\bf T}_i^2
\equiv {\bf T}_i \cdot {\bf T}_i = C_i$, where $C_i$ is the quadratic 
Casimir eigenvalue appropriate for the color representation of parton 
$i$. When employing this notation, one does not need to display 
explicitly the matrix indices of the soft functions and anomalous 
dimensions, since they are understood as operators acting in the 
color flow vector space.

Having introduced the appropriate notation, we can now write 
down the minimal solution to \eqn{oureq_reformulated}. It is 
given by~\cite{Gardi:2009qi}
\beq
  \left. \Gamma^{\overline{S}} \left(\rho_{i j}, \alpha_s \right) 
  \right\vert_{\rm dip} \, = \, - \frac18 \,
  \widehat{\gamma}_K \left(\alpha_s \right) \sum_{i = 1}^L 
  \sum_{j \neq i} \, \ln(\rho_{ij}) \,  {\bf T}_i \cdot  {\bf T}_j 
  \, + \, \frac12 \, \widehat{\delta}_{{\overline{\cal S}}} (\alpha_s)
  \sum_{i = 1}^L  {\bf T}_i \cdot  {\bf T}_i \, ,
\label{ansatz}
\eeq 
where $\widehat{\gamma}_K$, $\widehat{\delta}_{{\overline{\cal 
S}}}$ are anomalous dimensions which have been normalized by 
extracting from the perturbative result the quadratic Casimir 
eigenvalue of the appropriate representation, making $\widehat
{\gamma}_K$ and $\widehat{\delta}_{{\overline{\cal S}}}$ 
representation-independent. We emphasize that \eqn{ansatz}
only provides a solution to \eqn{oureq_reformulated} if the cusp 
anomalous dimension admits Casimir scaling, namely if
$\gamma_K^{(i)}$ corresponding to parton $i$ may be 
written as
\beq
  \gamma_K^{(i)} (\alpha_s) \, = \, C_i \, 
  \widehat{\gamma}_K(\alpha_s)\, = \, {\bf T}_i \cdot {\bf T}_i \,
  \widehat{\gamma}_K(\alpha_s) \, , 
\label{gamKdef}
\eeq
which assumes that there are no quartic (or higher-rank) Casimir 
invariants contributing to $\gamma_K^{(i)}$ at high orders.
Casimir scaling of the cusp anomalous dimension has been checked 
by explicit calculation up to three loops~\cite{Moch:2004pa}. 
Four loops is the first order where quartic Casimirs may appear.
Nevertheless, arguments were given
in~\cite{Becher:2009qa} indicating that quartic Casimirs do not appear in 
$\gamma_K^{(i)}$ at this order. If higher-rank Casimir operators
turn out to contribute to $\gamma_K^{(i)}$ at some order, 
also $\Gamma^{\overline{S}}$ would receive corrections at that order. 
We shall return to this point below. Note that only the first term in 
\eqn{ansatz} has a non-trivial matrix structure in color flow space, 
and furthermore this term is governed solely by the cusp anomalous 
dimension and by the running of the coupling. 

Substituting \eqn{ansatz} into \eqn{renSbar_massless}, one may
solve for the reduced soft function; one may then combine this 
solution with \eqn{reduS} and \eqn{fact_massless} and use the 
known structure of the jet functions (eq.~(2.2) in 
\cite{Dixon:2009ur}). The scattering amplitude may finally 
be written in a simple factorized form, as
\beq
  {\cal M} \left(\frac{p_i}{\mu}, \alpha_s (\mu^2),  \epsilon 
  \right) \, = \, Z \left(\frac{p_i}{\mu_f}, \alpha_s(\mu_f^2),  
  \epsilon \right) \, \, {\cal H} \left(\frac{p_i}{\mu}, 
  \frac{\mu_f}{\mu}, \alpha_s(\mu^2), \epsilon \right) \, ,
\label{Mfac}
\eeq
with the $Z$ matrix given by~\cite{Gardi:2009qi,Gardi:2009zv,
Becher:2009cu,Becher:2009qa}
\beqa
\label{sumodipoles}
  &&Z \left(\frac{p_l}{\mu}, \alpha_s(\mu^2), \epsilon \right) 
   =  \\&&=\exp \Bigg\{ \frac12 \int_0^{\mu^2}  \nonumber
  \frac{d \lambda^2}{\lambda^2}
  \bigg[ \frac{\widehat{\gamma}_K \left(\alpha_s (\lambda^2) 
  \right)  }{4} \sum_{(i,j)} \ln \left(\frac{- s_{ij}}{\lambda^2} 
  \right) {\bf T}_i \cdot {\bf T}_j  
- \, \sum_{i = 1}^L
  \gamma_{J_i} \left(\alpha_s (\lambda^2)
  \right) \bigg] \Bigg\} \, ,
\eeqa
where $(-s_{ij})\equiv 2 \, \left\vert p_i \cdot p_j \right\vert \, 
e^{-{\rm i} \pi \lambda_{i j}}$ and $(i, j)$ is a shorthand notation 
for summing over all pairs of hard partons $i\neq j$, where each pair 
is counted twice (once for $1 \leq j < i \leq L$ and once for $1\leq 
i < j \leq L$). Finally, $\gamma_{J_i}$ in \eqn{sumodipoles}  is the 
anomalous dimension for the partonic jet function $J_i$. Infrared 
singularities are generated in \eqn{sumodipoles}) as poles in 
$\epsilon$ through the integration over the $d$-dimensional 
coupling, which obeys the renormalization group equation
\beq
  \mu \frac{\partial \alpha_s}{\partial \mu} \, = 
  \beta (\epsilon, \alpha_s) \, = \, - \, 2 \epsilon \alpha_s - 
  \frac{\alpha_s^2}{2 \pi} \, \sum_{n = 0}^\infty
  b_n \left( \frac{\alpha_s}{\pi} \right)^n \, .
\label{running_coupling_RGE}
\eeq
In this paper, we will refer to \eqn{sumodipoles} as the {\it dipole 
formula}. The name emphasizes the fact that the non-trivial matrix 
structure of $Z$ is determined solely by pairs of color operators 
on distinct parton lines ({\it i.e.} color dipole operators). This is 
already evident in \eqn{ansatz}, which we also sometimes call 
the dipole formula.

As discussed above, \eqn{sumodipoles} arises as the simplest 
solution of \eqn{oureq_reformulated}. One may then ask what 
corrections to \eqn{sumodipoles}, if any, are compatible with 
\eqn{oureq_reformulated}. As first pointed out in 
Ref.~\cite{Gardi:2009qi}, and then discussed in detail in 
Refs.~\cite{Gardi:2009zv,Becher:2009cu,
Becher:2009qa}, for massless particles there are only two 
possible sources of corrections to the dipole formula. 
\begin{itemize}
\item{}  First of all, recall our assumption that the cusp anomalous 
dimension admits Casimir scaling, as expressed in \eqn{gamKdef}. 
Additional contributions to the soft anomalous dimension going 
beyond the dipole formula will be present if this is not true. 
\item {} Next, one may add to \eqn{ansatz} any solution of the 
homogeneous equations obtained from \eqn{oureq_reformulated}.
Such solutions must be functions of conformally-invariant cross 
ratios of the form $\rho_{ijkl} \equiv \rho_{ij} \rho_{kl}/(\rho_{ik}
\rho_{jl})$, and may therefore exist for amplitudes with four or 
more hard partons. Such corrections may potentially arise starting 
at the three-loop order, which is beyond the state of the art of 
explicit calculations of multiparton amplitudes.  If such corrections
are present, the full soft anomalous dimension matrix can be 
written as 
\beq
  \Gamma^{\overline{S}} \left(\rho_{i j}, \alpha_s \right) \, = \,
  \left. \Gamma^{\overline{S}} \left(\rho_{i j}, \alpha_s \right) 
  \right\vert_{\rm dip} \, +\, \Delta (\rho_{ijkl}, \alpha_s) \, .
\label{Delta}
\eeq
In this case, the matrix $\Delta$ would of course appear under 
the integral in the exponent of the $Z$ function in \eqn{sumodipoles}, 
leading to a single pole in $\epsilon$ at ${\cal O}(\alpha_s^3)$.
\end{itemize}
The form of potential quadrupole corrections $\Delta(\rho_{ijkl}, \alpha_s)$ has been studied in detail in Refs.~\cite{Dixon:2009ur,
Becher:2009qa}.
It was shown there that the set of admissible 
functions at three loops in any multi-leg amplitude is severely constrained by various properties.
\begin{enumerate}
\item{} The correction function $\Delta$ must only depend on the kinematics via conformally-invariant cross ratios~\cite{Gardi:2009qi}.
\item{} Given its origin in soft singularities, $\Delta$ depends only on color and kinematics. The colour structure is ``maximally non-Abelian"~\cite{Gatheral:1983cz,
Frenkel:1984pz}\footnote{The diagrammatic approach to non-Abelian exponentiation has been recently extended to the multi-leg case~\cite{Gardi:2010rn,Mitov:2010rp}.}. Furthermore, owing to the fact that the eikonal lines are effectively scalars, there must be Bose symmetry among all external partons. This correlates parity under color with parity under kinematics for each pair of partons.
\item{} 
The behaviour of an $L$-parton amplitude in the limit where two outgoing partons become collinear is constrained~\cite{Becher:2009qa} by its relation to the corresponding $L-1$ parton amplitude. Given that there are no corrections for the three-parton amplitude~\cite{Gardi:2009qi},  $\Delta$~corresponding to the four-parton amplitude must vanish in all collinear limits~\cite{Becher:2009qa,Dixon:2009ur}.
\item{} Based on the fact that $\Delta$, at three loops, is the same as in ${\cal N}=4$ supersymmetric Yang-Mills theory, it is expected to have the maximal permissible transcendentality, which is $\tau=5$~\cite{Dixon:2009ur}.
\end{enumerate}

As shown in Ref.~\cite{Dixon:2009ur}, these constraints, 
while very restrictive, still do not completely rule out three-loop 
corrections to the anomalous dimension: some specific 
functions consistent with all constraints were presented in 
Ref.~\cite{Dixon:2009ur}. Most of our analysis in the present 
paper relies on the dipole formula alone: indeed, one may observe 
that corrections going beyond the dipole formula may only be 
relevant starting at the next-to-next-to-leading order (NNLO) in 
the exponent, and are therefore entirely irrelevant to LL and NLL 
Reggeization. Moreover, as we explain in \sect{sec:beyond}, it 
can easily be seen that these corrections, if present, cannot 
affect our arguments concerning the breaking of Reggeization 
at NNLL level. We shall nevertheless return to analyse possible
contributions to the function $\Delta$ in \sect{sec:beyond_dipoles}, 
where we show that an additional constraint based on the Regge 
limit allows to rule out all explicit three-loop examples for $\Delta$
constructed in 
Ref.~\cite{Dixon:2009ur}, thus giving further support to the 
validity of the dipole formula beyond two-loop order. 

In this section we have reviewed the features of the dipole
formula, \eqn{sumodipoles}, emphasizing the possible sources 
of corrections. We will now show how this result can be used to 
study the high-energy limit of scattering amplitudes.

\section{The infrared approach to the high-energy limit}
\label{sec:reggeproof}

In the preceding sections, we reviewed existing results on the 
Reggeization of fermions and gauge bosons, and we presented
the dipole formula for the infrared singularity structure of general 
fixed-angle scattering amplitudes involving massless partons. 
The aim of this section is to demonstrate how the latter result can 
be used as a tool to study the high-energy limit for general
massless gauge theory amplitudes. In the present section we will
focus on four-point amplitudes, and derive a general expression
for the high-energy limit of the infrared operator $Z$, valid up
to corrections suppressed by powers of $|t/s|$, and thus to all logarithmic accuracies. In \sect{sec:hard} we shall use this expression to derive Reggeization at LL level for general color exchanges.

Our strategy is as follows~\cite{DelDuca:2011xm}. First, we examine the dipole formula in 
the specific case of $2 \rightarrow 2$ scattering, writing it in terms 
of the Mandelstam invariants $s$, $t$ and $u$. In the Regge limit,
$|t/s| \to 0$, we will see explicitly that the $Z$ matrix becomes 
proportional to a color operator corresponding to definite 
$t$-channel exchanges. We will then be able to interpret $Z$ as 
a ``Reggeization operator'': when acting on hard interactions  
consisting of a given $t$-channel exchange, such an operator 
automatically guarantees Reggeization, allowing the singular parts 
of the Regge trajectory to be simply read off. The divergent 
contributions to the corresponding Regge trajectory will automatically 
be proportional to the quadratic Casimir eigenvalue in the appropriate 
representation of the gauge group, as already observed for quarks 
and gluons in \sect{sec:Reggeization}. 

Before proceeding, let us briefly pause to comment on the applicability
of the dipole formula in the Regge limit. Recall that the dipole formula
was derived under the explicit assumption that all kinematic invariants 
$|p_i \cdot p_j|$ be large compared with $\Lambda^2_{\rm QCD}$, 
and parametrically of similar size. This assumption is no longer valid 
in the Regge limit, where one neglects $t$ with respect to $s$ and 
$u$. We note however that, for any fixed number of external legs, 
the amplitude is an analytic function of the available kinematic 
invariants, as well as a meromorphic function of the dimensional 
regularization parameter $\epsilon$. All infrared poles in $\epsilon$ 
arising in the fixed-angle amplitude are correctly generated by the 
dipole formula. Now, when taking the Regge limit starting from  the 
fixed-angle configuration, it is important to note that no new poles 
in $\epsilon$ are generated: the factorized form of the fixed-angle 
amplitude breaks down only because a new class of large logarithms 
becomes dominant: these are the Regge logarithms of the ratio 
$|t/s|$. The Regge logarithms which appear together with poles in 
$\epsilon$, however, are still correctly generated by the 
dipole formula, which controls all infrared and collinear singularities.
What is lost is just control over those Regge logarithms that are
associated with contributions that are finite as $\epsilon \to 0$. 
As a consequence, the evidence we provide in favor of Reggeization 
is limited to infrared-singular contributions, and we can only expect 
to compute correctly the divergent part of the Regge trajectory. On 
the other hand, the evidence we will provide against Reggeization
at NNLL in \sect{sec:beyond} is solid, since clearly full Reggeization
must in particular imply Reggeization of infrared poles.

\subsection{The Regge limit of the dipole formula}
\label{sec:forward}

We begin by considering a generic $2 \rightarrow 2$ scattering 
process involving massless external partons whose momenta satisfy 
momentum conservation
\beq
  p_1 + p_2 = p_3 + p_4 
\label{mom_con}
\eeq
and color conservation
\beq
  {\bf T}_1 + {\bf T}_2 + {\bf T}_3 + {\bf T}_4 \, = \, 0 \, ,
\label{color_conservation} 
\eeq
where ${\bf T}_1$ and ${\bf T}_2$ act as
insertions of  the color generators of the two incoming particles, while
${\bf T}_3$ and ${\bf T}_4$ act as insertions of \emph{minus} the 
color generators of the outgoing ones. Introducing as usual the 
Mandelstam variables
\beq
  s \, = \, (p_1+ p_2)^2 \, , \qquad  
  t \, = \, (p_1 - p_3)^2 \, , \qquad 
  u \, = \, (p_1 - p_4)^2 \, ,
\label{Mandel}
\eeq
where $s >0$ and $s + t + u = 0$ (with $t, u < 0$), we find that the 
$Z$ operator of \eqn{sumodipoles} takes the form
\beqa
  Z \left( \frac{p_i}{\mu}, \alpha_s (\mu^2), \epsilon \right) & = &
  \exp \left\{\int_0^{\mu^2} \frac{d \lambda^2}{\lambda^2}
  \Bigg[ \frac{1}{4} \, \widehat{\gamma}_K \left( \alpha_s
  (\lambda^2, \epsilon) \right) \left[ \ln \left(\frac{s \, 
  {\rm e}^{ - {\rm i} \pi}}{\lambda^2} \right) 
  \left({\bf T}_1 \cdot {\bf T}_2 + 
  {\bf T}_3 \cdot {\bf T}_4 \right) \right. \right. \nonumber \\
  & & \left. \left. + \, \, \ln \left(\frac{- t}{\lambda^2} \right)
  \left({\bf T}_1 \cdot {\bf T}_3 + {\bf T}_2 \cdot {\bf T}_4 
  \right) \, + \, \ln \left(\frac{- u}{\lambda^2} \right) 
  \left({\bf T}_1\cdot {\bf T}_4 + {\bf T}_2 \cdot {\bf T}_3 
  \right) \right] \right. \nonumber \\
  & & \left. - \, \, \frac{1}{2} \sum_{i = 1}^4 \gamma_{J_i}
  \left(\alpha_s(\lambda^2, \epsilon) \right) \Bigg] \right\} \, .
\label{sumodipoles2}
\eeqa
One may write this in a more suggestive form by introducing
operators associated with the color flow\footnote{Care is needed 
here with minus signs. Recall that we defined ${\bf T}_3$ and 
${\bf T}_4$ to be the negative of the color generators for the 
outgoing partons, so that color conservation was expressed by 
\eqn{color_conservation}.} in each channel~\cite{Dokshitzer:2005ig}.
They are
\beqa
  {\bf T}_s & = & {\bf T}_1 + {\bf T}_2 \, = \, 
  - \left( {\bf T}_3 + {\bf T}_4 \right) \, , \nonumber \\
  {\bf T}_t & = & {\bf T}_1 + {\bf T}_3 \, = \, 
  - \left( {\bf T}_2 + {\bf T}_4 \right) \, , \nonumber \\
  {\bf T}_u & = & {\bf T}_1 + {\bf T}_4 \, = \, 
  - \left( {\bf T}_2 + {\bf T}_3 \right) \, .
\label{Tstudef}
\eeqa
In terms of these operators, color conservation may be written as
\beq
  {\bf T}_s^2 + {\bf T}_t^2 + {\bf T}_u^2 \, = \, 
  \sum_{i = 1}^4 C_i \, ,
\label{colcon2}
\eeq
where the right-hand side contains a sum over the quadratic Casimir
eigenvalues of all four external partons. Armed with this notation, 
we may rewrite \eqn{sumodipoles2} as 
\beqa
  Z \left( \frac{p_i}{\mu}, \alpha_s (\mu^2), \epsilon \right) & = &
  \exp \left\{ \int_0^{\mu^2} \frac{d \lambda^2}{\lambda^2}
  \Bigg[ \frac{1}{4} \, \widehat{\gamma}_K 
  \left(\alpha_s (\lambda^2, \epsilon) \right) 
  \left[ \ln \left(\frac{s \, {\rm e}^{- {\rm i} 
  \pi}}{\lambda^2} \right) \left({\bf T}_s^2 - \frac{1}{2}
  \sum_{i = 1}^4 C_i \right) \right. \right. \nonumber  \\
  & & \left. \left. + \, \, \ln \left( \frac{- t}{\lambda^2} \right) 
  \left({\bf T}_t^2 - \frac{1}{2} \sum_{i = 1}^4 C_i \right)
  + \ln \left(\frac{- u}{\lambda^2} \right) \left({\bf T}_u^2
  - \frac{1}{2} \sum_{i = 1}^4 C_i \right) \right] \right. \nonumber \\
   & & \left. - \, \, \frac{1}{2} \sum_{i = 1}^4 \gamma_{J_i}
  \left(\alpha_s (\lambda^2, \epsilon) \right) \Bigg] \right\} \, .
\label{sumodipoles3}
\eeqa
So far our manipulations are exact. Let us now consider the Regge 
limit, $|t/s| \to 0$, which allows us to replace $u$ with $- s$, up 
to corrections suppressed by powers of $t/s$. Using color 
conservation, as given in \eqn{colcon2}, we find that 
\eqn{sumodipoles3} becomes
\beqa
  Z \left( \frac{p_i}{\mu}, \alpha_s (\mu^2), \epsilon \right) & = &
  \exp \left\{ \int_0^{\mu^2} \frac{d \lambda^2}{\lambda^2}
  \Bigg[ \frac{1}{4} \, \widehat{\gamma}_K 
  \left(\alpha_s(\lambda^2, \epsilon) \right) 
  \left[- \, {\bf T}_t^2 \, \ln \left(\frac{s}{- t}\right) -
  {\rm i} \pi \, {\bf T}_s^2 \right. \right. \nonumber \\
  & & \left. \left. + \, \, \frac{1}{2} \left({\rm i}\pi- \ln 
  \left(\frac{-t}{\lambda^2}
  \right)\right) \sum_{i = 1}^4 C_i \right] 
  - \, \frac{1}{2} \sum_{i = 1}^4 \gamma_{J_i}
  \left(\alpha_s (\lambda^2, \epsilon) 
  \right) \Bigg] \right\} \, .
\label{sod_regge}
\eeqa
Notice that \eqn{sod_regge} is correct to all logarithmic orders, 
and only receives corrections suppressed by powers of $t/s$. Notice 
also that only the first two terms in the exponent have a non-trivial 
color structure, and only the first term depends on $s$. This suggests
writing $Z$ in factorized form, as
\beq
  Z \left( \frac{p_i}{\mu}, \alpha_s (\mu^2), \epsilon \right)
  \, = \, \widetilde{Z} \left( \frac{s}{t}, \alpha_s (\mu^2), \epsilon 
  \right) \, Z_{\bf 1} \left( \frac{t}{\mu^2}, \alpha_s (\mu^2), 
  \epsilon \right) \, ,
\label{Zfac}
\eeq
where 
\beq
  \widetilde{Z} \left( \frac{s}{t}, \alpha_s (\mu^2), \epsilon \right) 
  \, = \, \exp \left\{ K \Big(\alpha_s (\mu^2), \epsilon \Big)
  \Bigg[ \ln \left( \frac{s}{- t} \right) {\bf T}_t^2 + 
  {\rm i} \pi \, {\bf T}_s^2 \Bigg] \right\} \, ,
  \label{Ztildedef}
\eeq
and
\beqa
  & & Z_{\bf 1} \left( \frac{t}{\mu^2}, \alpha_s (\mu^2), 
  \epsilon \right) \, = \,\exp \left\{ \sum_{i = 1}^4 B_i 
  \Big(\alpha_s (\mu^2), \epsilon \Big) \right. \nonumber \\
  & & \qquad \qquad \left. + \,
  \, \frac12 \left[K \Big(\alpha_s (\mu^2), \epsilon 
  \Big) \, \left(\ln \left(\frac{-t}{\mu^2}\right)-{\rm i}\pi\right)  + 
  D \Big(\alpha_s (\mu^2), \epsilon \Big) \right] \, 
  \sum_{i = 1}^4 C_i \right\} \, .
\label{Zddef}
\eeqa
Note that $Z_{\bf 1}$, as suggested by the notation, is proportional
to the unit matrix in color space. In \eqns{Ztildedef}{Zddef} we have 
introduced  the integrals
\begin{subequations}
 \label{Jintdef}
\begin{align}
  K \Big(\alpha_s (\mu^2), \epsilon \Big) & \equiv 
  - \frac14 \int_0^{\mu^2} \frac{d \lambda^2}{\lambda^2} \, 
  \widehat{\gamma}_K \left(\alpha_s(\lambda^2, \epsilon) 
  \right) \, , \label{Kdef} \\
  D \Big(\alpha_s (\mu^2), \epsilon \Big) & \equiv 
  - \frac14 \int_0^{\mu^2} \frac{d \lambda^2}{\lambda^2} \, 
  \widehat{\gamma}_K \left(\alpha_s(\lambda^2, \epsilon) \right)
  \ln \left( \frac{\mu^2}{{\lambda^2}} \right) \, ,
  \label{Idef} \\
  B_i \Big(\alpha_s (\mu^2), \epsilon \Big) & \equiv 
  - \frac12 \int_0^{\mu^2} \frac{d\lambda^2}{\lambda^2} \, 
  \gamma_{J_i} \left(\alpha_s (\lambda^2, \epsilon) \right) \, ; 
\end{align}
\end{subequations}
these integrals\footnote{Note that the $K$ integral defined here 
differs from the one used, for example, in~\cite{Dixon:2008gr} by 
a factor of $2 C_i$.} contain all the infrared singularities, which are 
explicitly generated upon substituting the form of the $d$-dimensional 
running coupling and integrating. 

One sees that in the Regge limit the $Z$ operator factorizes into 
a product of operators, the first of which is both $s$ dependent 
and non-trivial in color flow space, while the second is independent 
of $s$, and proportional to the unit matrix. Furthermore, the $s$ 
dependence has a particularly simple form: as \eqn{Ztildedef} 
shows, this dependence is associated with a quadratic color 
operator whose eigenstates correspond to definite $t$-channel 
exchanges (as we will see in more detail in the following section). 
Beyond leading logarithms,  there is a correction to this simple 
behavior, given by the second term in the exponent of \eqn{Ztildedef}. 
This term in general does not commute with the first, since 
$[{\bf T}_s^2, {\bf T}_t^2] \neq 0$; furthermore, it does not generically 
admit $t$-channel exchanges as eigenstates, so it signals possible 
violations of the Reggeization picture beyond LL (which will be 
discussed in detail in \sect{sec:beyond}). Confining ourselves, for 
the time being, to LL accuracy, we may ignore the ${\rm i} \pi$ 
term. The matrix $\widetilde{Z}$ becomes then a pure 
$t$-channel operator
\beq
  \left.\widetilde{Z} \left( \frac{s}{t}, \alpha_s (\mu^2), 
  \epsilon \right)\right\vert_{\rm LL} \, = \, 
  \exp \left[ K \Big( \alpha_s (\mu^2), 
  \epsilon \Big) \, \ln \left(\frac{s}{- t} \right) \,
  {\bf T}_t^2 \right] \, .
\label{ZLLdef}
\eeq
In the following section we will interpret \eqn{ZLLdef} as a Reggeization
operator. This will lead to an expression for the singular part of the trajectory in terms of an integral over the cusp anomalous dimension, consistent with the Wilson line derivation of  Ref.~\cite{Korchemskaya:1996je} (see Eq.~(33) there).

Before doing this, however, it is important to note that the
reasoning developed so far for the conventional Regge limit $|t/s| 
\to 0$, $u \simeq - s$, can be precisely repeated with similar results
for the alternative Regge limit $|u/s| \to 0$, $t \simeq - s$, as a 
consequence of the fact that the dipole formula treats all dipoles in
an essentially symmetric way. By taking the alternative Regge limit, 
one may easily verify that the $Z$ matrix factorizes as in \eqn{Zfac},
\beq
  Z \left( \frac{p_i}{\mu}, \alpha_s (\mu^2), \epsilon \right)
  \, = \, \widetilde{Z}^{(u)} \left( \frac{s}{u}, \alpha_s (\mu^2), 
  \epsilon \right) \, Z_{\bf 1}^{(u)} \left( \frac{u}{\mu^2}, 
  \alpha_s (\mu^2), \epsilon \right) \, ,
\label{Zufac}
\eeq
where the factors can be obtained from \eqns{Ztildedef}{Zddef} by
simply replacing $t$ by $u$ and ${\bf T}_t^2$ by ${\bf T}_u^2$.

\section{The Reggeization operator at leading logarithmic 
accuracy}
\label{sec:hard}

In the previous section, we saw that the dipole formula has a 
particularly simple form in the high-energy limit. In particular, the 
$s$-dependent poles of the scattering amplitude are generated by a 
$Z$-factor whose color structure coincides with that of a pure 
$t$ or $u$-channel exchange. In this section, we interpret this result 
in terms of Reggeization. For the sake of simplicity, we begin by 
considering $g g\rightarrow g g$ scattering, which was discussed 
in \sect{sec:Reggeization}. We then proceed to generalize our 
considerations to color exchanges in arbitrary representations 
of the gauge group.

\subsection{Reggeization for gluons and quarks}
\label{sec:gluqua}

Based on eqs.~(\ref{Mfac}, \ref{Zfac}, \ref{ZLLdef}), any four-point 
scattering amplitude in the Regge limit $|t/s| \to 0$ may be 
written, to leading logarithmic accuracy, as
\beq
  {\cal M} \left( \frac{p_i}{\mu}, 
  \alpha_s(\mu^2), \epsilon \right) \, = \, 
  \exp \left\{ K \Big( \alpha_s (\mu^2), 
  \epsilon \Big) \, \ln \left(\frac{s}{- t} \right) \,
  {\bf T}_t^2 \right\} \, Z_{\bf 1} \, {\cal H} \left( \frac{p_i}{\mu}, 
  \alpha_s(\mu^2), \epsilon \right) \, ,
\label{Mggdef}
\eeq
where ${\cal H}$ is the appropriate hard interaction, and where we 
chose $\mu_f = \mu$ for simplicity. Note that the hard scattering 
vector ${\cal H}$ is the only process-dependent factor on the 
right-hand side of \eqn{Mggdef}. Consider now for example 
the process $ g g \rightarrow g g $, as discussed in 
\sect{sec:Reggeization}. In that case the hard interaction,
at tree level, contains three different color structures, corresponding 
to $s$, $t$ and $u$-channel exchanges, depicted in \fig{LOgg}. 
As remarked in \sect{sec:Reggeization}, however, only the 
$t$-channel diagram survives in the Regge limit, with the other 
diagrams being kinematically suppressed by powers of $t/s$. 
The $t$-channel exchange color structure is, by construction, an 
eigenstate of the operator ${\bf T}_t^2$, so that
\beq
  {\bf T}_t^2 \, {\cal H}^{gg\rightarrow gg} \,
  \xrightarrow{|t/s| \to 0} \, C_t \,
  {\cal H}^{gg\rightarrow gg}_t \, ,
\label{eigen}
\eeq
where $C_t$ is the quadratic Casimir eigenvalue corresponding 
to the representation of the exchanged particle, and 
${\cal H}^{gg\rightarrow gg}_t$ is the $t$-channel component
of the hard interaction. In this case clearly $C_t = C_A$, given 
that the exchanged particle is a gluon belonging to the adjoint 
representation. 

For gluon-gluon scattering, to leading power in $|s/t|$, and to 
leading logarithmic accuracy, \eqn{Mggdef} then becomes
\beq
  {\cal M}^{gg\rightarrow gg}  \, = \,
  \left(\frac{s}{-t}\right)^{C_A \, K \left(\alpha_s (\mu^2),
  \epsilon \right)} Z_{\bf 1} \, {\cal H}^{gg\rightarrow gg}_t \, .
\label{Mgg3}
\eeq
Comparing this with \eqn{Mgg}, we see that $C_A K(\alpha_s, 
\epsilon)$ must correspond to the singular parts of the LL Regge 
trajectory of the gluon, as this is the only source of $s$-dependent 
$\epsilon$ poles in \eqn{Mgg3}\footnote{The factor $Z_1$ 
generates $s$-independent collinear singularities, which in 
\eqn{Mgg} are contained in the impact factors.}. In other words, 
the dipole formula implies the LL Reggeization of the gluon: 
technically, we have shown this only for the divergent part of  
the Regge trajectory, however at LL this is trivially related to the 
complete result. All that was necessary for the specific process 
at hand was to demonstrate that only the $t$-channel exchange 
graph survives in the Regge limit at tree level; higher-order 
contributions to the hard function may bring in other exchanges, 
and other color representations, however these would contribute 
only to subleading logarithms. Note that eq.~(\ref{Mgg3}) is similar to the result obtained in the Wilson-line approach in Ref.~\cite{Korchemskaya:1996je}.

We may verify the above statements by computing the integral 
$K (\alpha_s, \epsilon)$ defined in \eqn{Kdef}. To this end we 
just need the leading order cusp anomalous dimension
\beq
  \widehat{\gamma}_K (\alpha_s) \, = \, 2 \, \frac{\alpha_s}{\pi}
  \, + \, {\cal O}(\alpha_s^2) \, ,
\label{gamKLO}
\eeq
and the LO $d$-dimensional running coupling
\beq
  \alpha_s (\lambda^2,\epsilon) \, = \, \left(
  \frac{\lambda^2}{\mu^2} \right)^{- \epsilon} 
  \Big[ \alpha_s (\mu^2, \epsilon) + {\cal O}(\alpha_s^2) \Big] \, .
\label{LOalpha}
\eeq
Substituting \eqns{gamKLO}{LOalpha} into \eqn{Kdef}, one finds
\beq
  K (\alpha_s, \epsilon) \, = \, \frac{1}{2 \epsilon} \, 
  \frac{\alpha_s}{\pi} + {\cal O}(\alpha_s^2) \, ,
\label{KLO}
\eeq
so that the singular part of the Regge trajectory at one-loop order
is given by
\beq
  \alpha^{(1)} \, =  \, C_A \, \frac{2}{\epsilon} \, + \,
  {\cal O} \left(\epsilon^0\right) \, ,
\label{alpha1pole}
\eeq
which indeed agrees exactly with \eqn{alpha1}. 

Some comments are in order. First, we note again that our method for 
deriving Reggeization allows us in general to extract only the singular 
parts of the Regge trajectory: finite corrections are not determined 
by the dipole formula. At LL accuracy, however, the only finite 
corrections are those corresponding to the rescaling of the
coupling given in \eqn{rescal}, which is essentially a choice of 
renormalization scale, so the complete answer is easily recovered. At
NLL non-trivial finite contributions to the Regge trajectory do arise.

We note also that, while we considered gluon scattering above, we 
could equally have chosen \emph{any} scattering process such that the hard 
interaction, at leading order and in the Regge limit, would consist of 
a single $t$-channel gluon exchange, as is the case for example
$q q \rightarrow q q$ scattering. For any such scattering process,
the hard function is an eigenstate of the Reggeization operator
in \eqn{ZLLdef}, and this immediately leads to an equation of the 
same form as \eqn{Mgg3}, with the same exponent of $s/t$, as 
expected. This follows immediately from the fact that the Reggeization 
operator in \eqn{ZLLdef} is process-independent. It acts on any hard
interaction dominated by a definite $t$-channel exchange to give 
a corresponding Regge trajectory.

Finally, we note that only the color octet exchange has Reggeized 
in \eqn{Mgg3}. In the usual proofs of the form of the Reggeized 
amplitude in \eqn{Mgg}, much work must be invested in order to 
show that only the octet contributes at each order in the perturbative
expansion. Here we see explicitly why the color octet exchange is 
picked out: it is the only $t$-channel exchange which survives at 
tree level, and thus immediately Reggeizes upon application of the 
Reggeization operator. We will shortly discuss the more general case 
in which several possible representations contribute to $t$-channel 
color exchanges at leading order.

Let us now briefly discuss the issues related to the signature of the 
amplitude under  $s \leftrightarrow u$ exchange. As remarked in
\sect{sec:Reggeization}, it is conventional to rewrite the Reggeized 
amplitude to display explicitly its definite parity under $s 
\leftrightarrow u$ interchange, corresponding to the fact that 
the octet exchange has negative signature. In the present formalism, 
one may carry out this procedure at the level of the Reggeization 
operator. Indeed, it is easy to check that \eqn{Ztildedef} can be 
identically rewritten as
\beq
  \widetilde{Z} \left( \frac{s}{t}, \alpha_s(\mu^2), \epsilon \right)
  \, = \, \exp \left\{ K \Big( \alpha_s (\mu^2), \epsilon \Big) 
  \left[ \ln \left(\frac{- s}{- t} \right) {\bf T}_t^2 + {\rm i} \pi
  {\bf T}_u^2 \right] \right\} \, .
\label{Ztildedef2}
\eeq
For the case at hand (gluon-gluon scattering), in which the octet 
exchange has negative signature, it makes sense to use for the 
Reggeization operator the symmetric form
\beqa
  \widetilde{Z} \left( \frac{s}{t}, \alpha_s(\mu^2), \epsilon \right)
  & = & \frac{1}{2} \left\{ 
  \exp \left\{ K \Big( \alpha_s (\mu^2), \epsilon \Big) 
  \left[ \ln \left(\frac{s}{- t} \right) {\bf T}_t^2 + {\rm i} \pi
  {\bf T}_s^2 \right] \right\} \right. \nonumber \\
  & & \left. \, \, + \, 
  \exp \left\{ K \Big( \alpha_s (\mu^2), \epsilon \Big) 
  \left[ \ln \left(\frac{- s}{- t} \right) {\bf T}_t^2 + {\rm i} \pi
  {\bf T}_u^2 \right] \right\} \right\} \, .
\label{regg2}
\eeqa
At leading logarithmic order one can drop the imaginary parts in 
the exponents: having done that, both the original Reggeization 
operator, \eqn{Ztildedef}, and its signaturized form, \eqn{regg2},
become pure $t$-channel operators. Acting upon the hard interaction, 
they clearly reproduce the kinematic structure of \eqn{Mgg2} for 
the singular parts of the amplitude.

Having described how the singular part of the one-loop gluon 
Regge trajectory can be extracted using the dipole formula, 
we now briefly turn our attention to Reggeization of the 
quark. As discussed in \sect{sec:Reggeization}, this proceeds 
by considering the alternative Regge limit $|u/s| \to 0$, in which 
backward scattering dominates. To this end, one may use the 
appropriate limit of the dipole formula, given by \eqn{Zufac}. 
The argument for Reggeization is exactly analogous to the 
$t$-channel case: one considers tree level $q g \rightarrow  q g$ 
scattering in the (backward) Regge limit, which consists of a single 
$u$-channel exchange graph; this graph, shown in figure~\ref{LOqg} 
(c), has a color factor which is an eigenstate of the $u$-channel 
Reggeization operator; the analogue of \eqn{eigen} is then
\beq
  {\bf T}_u^2 \, {\cal H}^{q g \rightarrow q g} \,
  \xrightarrow{|u/s| \to 0} \, C_u \,
  {\cal H}^{q g \rightarrow q g}_u \, ,
\label{eigen2}
\eeq
where ${\cal H}^{q g \rightarrow q g}_u$ is the $u$-channel 
contribution to the hard interaction, and the eigenvalue $C_u$ 
is the quadratic Casimir invariant associated with the representation 
of the $u$-channel exchange, which in this case is $C_u = C_F$, for 
a fermion in the fundamental representation. One then finds, by
analogy with \eqn{Mgg3},
\beq
  \left.{\cal M}^{q g \rightarrow q g} \right\vert_{\rm LL} \, = \,
  \left(\frac{s}{- u}\right)^{C_F \, K \left( \alpha_s (\mu^2),
  \epsilon \right)} Z_{\bf 1} \, {\cal H}^{q g \rightarrow q g}_u \, .
\label{Mqg}
\eeq
One reads off the one-loop Regge trajectory for the quark,
\beq
  \alpha_q^{(1)} \, =  \, C_F \, \frac{2}{\epsilon} \, + \, 
  {\cal O} \left(\epsilon^0\right) \, ,
\label{alpha1qpole}
\eeq
in direct agreement with \eqn{alphaq1}. As was the case for gluon 
scattering, one may choose to rewrite the Reggeization operator as 
a sum of two terms related by crossing symmetry, for those cases in 
which the color factor of the tree-level interaction has a definite 
signature. 

In this section, we have seen how Reggeization of the quark and 
gluon at one loop follows from the dipole formula, reproducing the 
results of \sect{sec:Reggeization}. In the following section, we 
generalise this result to particle exchanges in arbitrary color 
representations.

\subsection{Reggeization of arbitrary particle exchanges}
\label{sec:general}

In the previous section, we saw how Reggeization at leading logarithmic
accuracy follows from the dipole formula. The crucial steps in the 
argument were the following.
\begin{itemize}
\item{} For $2 \rightarrow 2$ scattering, in the Regge limit 
$|t/s| \to 0$, and at LL order, the dipole formula becomes a 
pure $t$-channel operator . Alternatively, it becomes a pure 
$u$-channel operator in the limit $|u/s| \to 0$. The exponent 
is proportional to the quadratic Casimir operator corresponding 
to this channel (${\bf T}_t^2$ or ${\bf T}_u^2$).
\item{} In the chosen limit, tree level scattering of quarks or gluons
becomes dominated by a single color structure, which is a $t$-channel 
or $u$-channel color factor for gluon or quark exchange respectively. 
\item{} The tree level hard interaction then becomes an eigenstate 
of the dipole operator $Z$, so that the latter plays the role of a 
Reggeization operator. The allowed color structure at tree level 
selects the particle which Reggeizes, and the Regge trajectory 
necessarily contains the quadratic Casimir invariant of the 
appropriate representation, multiplied by the universal factor 
$K \left(\alpha_s (\mu^2), \epsilon \right)$.
\end{itemize}
As perhaps is already clear, this argument is not restricted to gluon 
and quark Reggeization, but easily generalizes to arbitrary color 
structures being exchanged in the $t$ or $u$ channel. In what follows 
we will consider, without loss of generality, $t$-channel exchanges.

\begin{figure}
\begin{center}
\scalebox{0.8}{\includegraphics{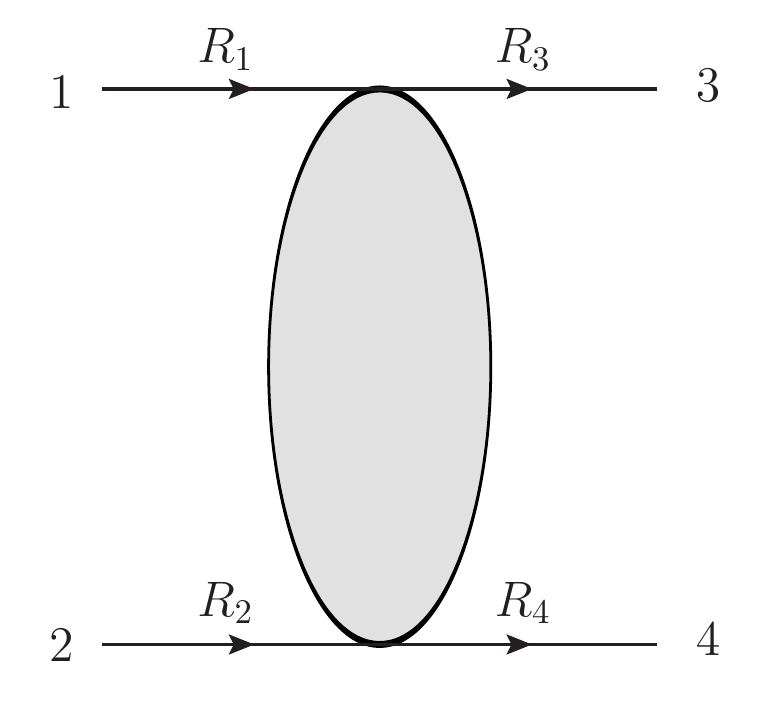}}
\caption{A general hard interaction dominated by $t$-channel 
scattering in the Regge limit, where $R_l$ is the representation 
of particle $l$. }
\label{tchannelfig}
\end{center}
\end{figure}

Consider a general scattering process involving particles belonging 
to different irreducible representations of the gauge group. Such a 
process is depicted in \fig{tchannelfig}, where $R_l$ denotes 
the irreducible representation of particle $l$. 
It is possible to enumerate the color representations that can be 
exchanged in the $t$ channel in full generality, and indeed one can 
explicitly construct projection operators that extract from the full 
amplitude the contribution of each representation. A detailed 
discussion is given in~\cite{Beneke:2009rj}\footnote{Note that
Ref.~\cite{Beneke:2009rj} works in an $s$-channel basis rather 
than in a $t$-channel basis. The arguments are however the same 
in both cases.}, while the case of gluon-gluon scattering at one loop 
was studied in~\cite{Kidonakis:1998nf,Dokshitzer:2005ig,
Sjodahl:2008fz}. An explicit analysis in terms of Clebsch-Gordan 
coefficients for the purposes of the present paper in given in 
App.~\ref{app:clebsch}. For $t$-channel exchanges, the result 
of the analysis can be summarized as follows. With representation 
labels as in \fig{tchannelfig}, one must first construct the tensor 
products $R_{13} \equiv R_1 \otimes R_3$ and $R_{24} \equiv 
R_2 \otimes R_4$. One then decomposes each of the two product 
spaces into a sum of irreducible representations according to
\beq
  R_{13} = \bigoplus_\alpha \, m_\alpha \, R^{(13)}_\alpha 
  \, , \qquad
  R_{24} = \bigoplus_\beta \, n_\beta \, R^{(24)}_\beta \, ,
\label{scompirrep}
\eeq
where $m_\alpha$ and $n_\beta$ are the multiplicities with which
each representation recurs in the given tensor product. The list of 
possible $t$-channel exchanges for the scattering process at hand 
is then the intersection of the sets $\{R^{(13)}_\alpha\}$, 
$\{\overline{R}^{(24)}_\beta\}$ (where the bar denotes complex 
conjugation\footnote{The need to consider the conjugate 
representation follows from our choice of momentum flow.}), 
counting multiple occurrences of equivalent representations 
as distinct. We denote the resulting set by  $\{R^{(t)}_\alpha\}$:
this is the set of permissible representations which can flow in the $t$
channel.

In this general case, it is to be expected that several representations
$\{R^{(t)}_\alpha\}$ will contribute to the high-energy limit of the 
tree-level amplitude. On the basis of the arguments given for gluon 
and quark scattering, we can anticipate that each such representation 
will Reggeize independently. In order to see that this is indeed the case,
it is convenient to choose a color flow basis where each element 
consists of a definite  irreducible representation being exchanged 
in the $t$-channel. Each color tensor in this basis represents an 
abstract vector in color space,
\beq
  \left( c^\alpha \right)_{a_1\cdots a_4} \longrightarrow  
  | \alpha \rangle \, ,
\label{cvec}
\eeq
and each such vector is an eigenvector of the color operator 
${\bf T}^2_t$, according to
\beq
  {\bf T}_t^2 \, | \alpha\rangle \, = \, C_{R^{(t)}_\alpha} 
  \, |\alpha \rangle \, ,
\label{alphadef}
\eeq
where $C_{R^{(t)}_\alpha}$ is the quadratic Casimir invariant in 
the representation $R^{(t)}_\alpha$, corresponding to the basis 
element $|\alpha\rangle$. We may then write the hard interaction 
in the Regge limit as
\beq
  | {\cal H} \, \rangle \, = \, \sum_\alpha{\cal H}_\alpha \, 
  |\alpha\rangle \, ,
\label{Halpha}
\eeq
In this basis, the factorization formula in \eqn{Mfac} can be written
in components as
\beq
  {\cal M}_\beta \left(\frac{p_i}{\mu}, \alpha_s (\mu^2),  
  \epsilon \right)  \, = \, 
  Z_\beta^{\phantom{\beta} \alpha} \left(\frac{p_i}{\mu_f}, 
  \alpha_s(\mu_f^2),  
  \epsilon \right) \,\, {\cal H}_\alpha \left(\frac{p_i}{\mu}, 
  \frac{\mu_f}{\mu}, \alpha_s(\mu^2), \epsilon \right) \, ,
\label{Malpha}
\eeq
Substituting the Regge limit of the dipole operator $Z$, given by 
\eqns{Zfac}{ZLLdef}, gives
\beqa
  {\cal M}_\beta \left( \frac{p_i}{\mu}, 
  \alpha_s(\mu^2), \epsilon \right) & = & 
  \left\{ \exp \left[ K \Big( \alpha_s (\mu^2), 
  \epsilon \Big) \, \ln \left(\frac{s}{- t} \right) \,
  {\bf T}_t^2 \right] \right\}_\beta^{\phantom{\beta} \, \alpha} 
  \, Z_{\bf 1} \, {\cal H}_\alpha \left( \frac{p_i}{\mu}, 
  \alpha_s(\mu^2), \epsilon \right) \nonumber \\
  & = & \, \exp \left[ K \Big( \alpha_s (\mu^2), 
  \epsilon \Big) \, \ln \left(\frac{s}{- t} \right) \,
  C_{R^{(t)}_\beta} \right] \,Z_{\bf 1} \, {\cal H}_\beta \, .
\label{Mbeta2}
\eeqa
The interpretation of the final result is straightforward: if the hard 
interaction consists of a number of possible $t$-channel exchanges, 
each exchange independently Reggeizes, with a trajectory containing 
the relevant quadratic Casimir. This is a consequence of the fact that 
the Reggeization operator is process-independent, and that different 
color exchanges combine additively in the hard interaction. The 
argument was formulated here for the $t$-channel, but it clearly
also applies in the limit $|u/s| \to 0$ for $u$-channel exchanges. 
Given the somewhat abstract nature of the above discussion, 
it is perhaps useful to see explicitly how the color algebra 
operates in terms of partonic indices, and specifically how the 
representations occurring in $t$-channel exchange can be explicitly
identified. The interested reader is referred to App.~\ref{app:clebsch}.

One may further clarify the above result using a few examples. First,
let us return to the familiar example of gluon-gluon scattering. At
LO in the hard interaction, in QCD, only color octet exchange is
present. At higher orders, all the representations on the right hand
side of \eqn{8*8} may appear. In quark-quark scattering, the
representations on the upper and lower quark lines are $R_1 =
{\bf 3}$ and $R_3 = \overline{\bf 3}$ (recall that color generators
are reversed in sign for outgoing particles), so the allowable
$t$-channel exchanges are given by
\beq
 {\bf 3} \otimes \overline{\bf 3} \, = \, {\bf 1} \oplus {\bf 8}_a \, .
\label{3*3bar}
\eeq
Both of these occur in the hard interaction in QCD, with the octet
appearing at tree level, and the singlet appearing at NLO.

So far, we considered examples in which the upper and lower lines 
are in the same representations (that is, $R_1 = R_2$ and $R_3 =
R_4$). A simple example where this is not the case is gluon-mediated
$q g \rightarrow q g$ scattering, where $R_1 = {\bf 3}$ and $R_3 = 
\overline{\bf 3}$, while $R_2 = R_4 = {\bf 8}_a$. Decomposing 
the product of representations on upper and lower lines gives 
\eqn{3*3bar} and \eqn{8*8}, respectively. Therefore, the only 
allowable $t$-channel exchanges are again singlet and 
octet\footnote{In this case taking the conjugate representations 
on the lower line has no effect, since ${\bf 1}$ and ${\bf 8}_a$ 
are both self-conjugate.}. As before, the hard interaction picks 
out which exchanges actually occur: one finds again tree-level 
octet exchange and higher-order singlet exchange. This essentially
completes the discussion of quark-gluon scattering in QCD.

\begin{figure}
\begin{center}
\scalebox{0.9}{\includegraphics{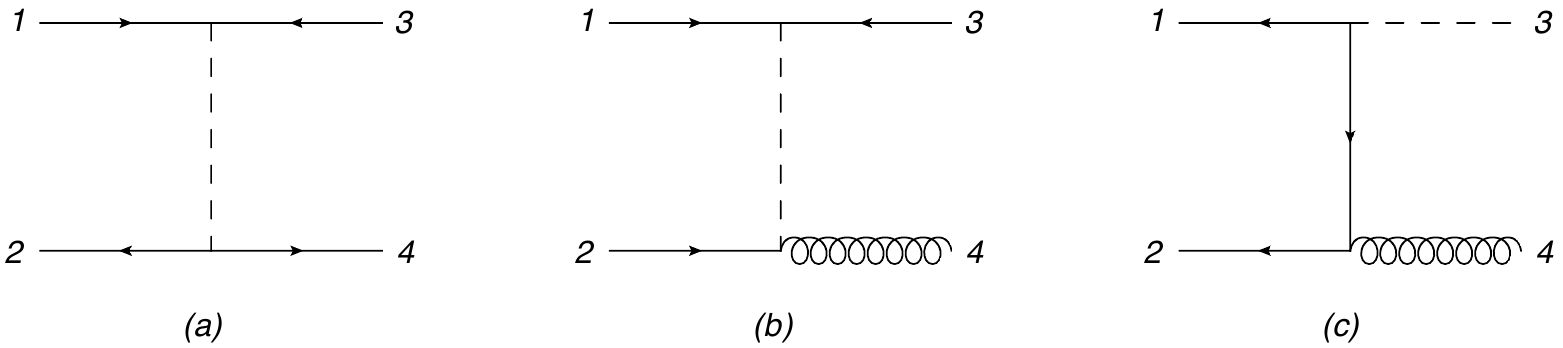}}
\caption{Examples of tree-level interactions involving quarks and 
gluons in a new physics model with extended particle content. 
The dashed line represents an arbitrary color representation, and 
explicit examples are considered in the text.}
\label{RPVfig}
\end{center}
\end{figure}

Scattering processes of partons in exotic color representations 
may be of interest for several reasons. First of all, from a theoretical 
perspective, this is an obvious generalization of the processes 
considered above, and therefore interesting to study. We will
indeed see that Reggeization is a very general phenomenon, 
which applies to arbitrary representations. Second, within QCD, 
one may consider scatterings of unconventional hadron
constituents such as diquarks, which play a role in models of
hadronic phenomenology~\cite{Anselmino:1992vg}. Finally, 
more exotic scattering processes are possible in theories other 
than QCD, including some viable new physics models. For example, 
one may envisage flavour-violating interaction vertices, which allow 
for scattering processes such as the one shown in \fig{RPVfig}(a), 
in which four different (anti)quark species scatter, in potentially 
different color representations; a concrete example is the 
$R$-parity violating supersymmetric model considered 
in~\cite{Han:2009ya}. 
In the case where the solid lines in figure~\ref{RPVfig}(a) represent 
ordinary quarks (in the fundamental representation), the upper and
lower lines give a $t$-channel color decomposition
\beq
  {\bf 3} \otimes {\bf 3} \, = \, \overline{\bf 3} \oplus {\bf 6} \, ,
\label{3*3}
\eeq
and
\beq
  \overline{\bf 3} \otimes \overline{\bf 3}\, = \, {\bf 3} \oplus
  \overline{\bf 6} \, .
\label{3bar*3bar}
\eeq
Possible $t$-channel exchanges are thus color triplet and sextet, 
which match up since \eqn{3bar*3bar} is the conjugate of \eqn{3*3}. 
As a consequence, sextet and triplet automatically Reggeize at 
LL accuracy, if they are present in the tree-level hard interaction
dictated by the chosen new physics model (for example, this is indeed 
the case in~\cite{Han:2009ya}, where $t$-channel exchange 
represents a scalar diquark). 

Consider now \fig{RPVfig}(b), representing $q q \to 
\tilde{q}^\dagger g$ scattering, where $\tilde{q}^\dagger$ 
denotes, for example, an antisquark. In this case the color 
decompositions on the upper and lower lines are given by 
\eqn{3*3}, and  by
\beq
  {\bf 3} \otimes {\bf 8}_a \, = \, {\bf 3} \oplus
  \overline{\bf 6} \oplus{\bf 15} \, ,
\label{3*8}
\eeq
respectively. One may again conclude that the triplet and sextet 
Reggeize, as in the case of \fig{RPVfig}(a).

Finally, in \fig{RPVfig}(c) we consider a case in which an exotic 
particle occurs as an external leg. Taking this to be, for example,
in the  $\overline{\bf 6}$ representation (for the outgoing particle),
the color decompositions on the upper and lower lines are
\beq
  \overline{\bf 3} \otimes {\bf 6}\, = \, {\bf 3} \oplus{\bf 15} \, ,
\label{3*6bar}
\eeq
and
\beq
  \overline{\bf 3} \otimes {\bf 8}_a \, = \, \overline{\bf 3}
  \oplus{\bf 6} \oplus \overline{\bf 15} \, ,
\label{3bar*8}
\eeq
respectively. One sees that in this case the ${\bf 3}$ and ${\bf 15}$ 
exchanges in \eqn{3*6bar} both Reggeize, as they match up with their 
conjugates in \eqn{3bar*8}. 

We have now seen a number of examples of how Reggeization of 
$t$-channel exchanges follows quite generally from the dipole 
formula. Let us however stress again that color information alone 
is not sufficient to guarantee Reggeization: a given representation 
which is permissible in the $t$-channel ({\it i.e.} it belongs to the 
set $\{R_\alpha^{(t)}\}$) must be shown to arise in the hard 
interaction. If this is the case, then it automatically Reggeizes. 
Note also that different exchanges may show up at different orders 
in the perturbation expansion. In such cases, the representations 
which arise at higher orders are logarithmically suppressed. 

The general picture of LL Reggeization which emerges from 
\eqn{Mbeta2} is that the hard interaction may be decomposed
in the Regge limit into a series of $t$-channel exchanges, each 
corresponding to a distinct irreducible representation of the gauge 
group. All such exchanges Reggeize separately, and the one-loop 
Regge trajectory in each case is given by
\beq
  \alpha_R^{(1)} \, = \, C_R \, \frac{2}{\epsilon} \, + \, 
  {\cal O} \left( \epsilon^0 \right) \, .
\label{genalpha}
\eeq

In this section we have outlined how the Reggeization operator 
stemming from the dipole approach automatically Reggeizes any 
allowable $t$-channel exchange at one-loop order. We now turn to
the study of what happens at higher logarithmic accuracy.

\section{The high-energy limit beyond leading logarithms}
\label{sec:beyond}

In the previous sections, we have used the dipole formula to 
provide a novel derivation of Reggeization for $t$-channel 
(or $u$-channel) exchanges, which allows the singular parts 
of the Regge trajectory to be easily read off, in terms of the 
quadratic Casimir eigenvalues of the exchanged particles. Our
explicit discussion, however, has so far been limited to leading 
logarithmic accuracy, since we considered the Reggeization 
operator in \eqn{ZLLdef}, neglecting the additional term in the 
exponent of \eqn{Ztildedef} (and likewise for the corresponding 
$u$-channel operator in \eqn{Zufac}). The dipole formula, however, 
is an all-order ansatz, which furthermore is known to be exact up to 
two loops in the exponent. We can therefore explore the consequences 
of employing the complete result, \eqn{Ztildedef}, which is accurate 
up to corrections suppressed by powers of $t/s$.

The first obvious thing to note is that \eqn{Ztildedef}, unlike 
\eqn{ZLLdef}, is not a pure $t$-channel operator. The exponent
involves both ${\bf T}_t^2$ and ${\bf T}_s^2$, which are not 
mutually commuting in general. The coefficient of the ${\bf T}_s^2$ 
term, however, is independent of $s/t$, and imaginary. We expect
then that this term will influence the result starting at NLL, and
it will affect the real and imaginary parts of the scattering amplitude 
in a different way. The main conclusion, however, is that eigenstates 
of the dipole operator are, in general, no longer eigenstates of 
${\bf T}_t^2$, a fact that was already established in the case of 
quark-quark scattering in Ref.~\cite{Korchemskaya:1994qp}.
This means that the eigenstates can no longer be interpreted 
as definite $t$-channel exchanges, and this implies that Reggeization 
generically breaks down beyond leading logarithmic order. 

In order to verify our expectations, we may start by expressing
the full Reggeization operator, \eqn{Ztildedef}, in terms of
a product of exponentials involving nested commutators of the
color operators ${\bf T}_t^2$ and ${\bf T}_s^2$, using an 
appropriate version of the Baker-Campbell-Hausdorff formula, 
sometimes referred to as the Zassenhaus formula (see {\it e.g.}
\cite{Magnus}). The formula states that given two non-commuting 
objects $X$ and $Y$, and a $c$-number function $K$, and 
having defined exponentials in terms of their Taylor expansion, 
one finds
\beqa
  \exp \Big[ K \, (X + Y) \Big] & = & \exp \big( K X \big) 
  \, \times \,
  \exp \big( K Y \big) 
  \, \times \, 
  \exp \left( - \frac{K^2}{2} \left[ X, Y \right] \right) 
  \label{Zassen} \\ 
  & & \times \, \exp \Bigg( \frac{K^3}{3!} \bigg( 2 \, 
  \Big[Y, [X, Y] \Big] + 
  \Big[X, [X, Y] \Big] \bigg) \Bigg) \times \, 
  \exp \Big( {\cal O} \left( K^4 \right) \Big) \, . \nonumber
\eeqa
In the present case we may define
\beq
  X \, = \, \ln \left(\frac{s}{- t} \right) \, {\bf T}_t^2 \, , 
  \qquad Y \, = \, {\rm i} \, \pi \, {\bf T}_s^2 \, ,
  \qquad K \, = \, K \left(\alpha_s(\mu^2), \epsilon \right) \, ,
\label{XY}
\eeq
and exploit the fact that the function $K (\alpha_s, \epsilon)$ 
begins at order $\alpha_s$. As a consequence, the commutator 
terms in \eqn{Zassen} will start contributing at NLL, as expected, 
and can be organized in order of decreasing logarithmic relevance. 
Applying \eqn{Zassen} to \eqn{Ztildedef}, with the definitions in 
\eqn{XY}, we find
\beqa
  &&\widetilde{Z} \left( \frac{s}{t}, \alpha_s (\mu^2) , 
  \epsilon \right)  =  \left( \frac{s}{- t} \right)^{K 
  \left( \alpha_s, \epsilon \right) \, \, {\bf T}_t^2} \, \, 
  \exp \left\{ {\rm i} \, \pi \, K \big( \alpha_s, 
  \epsilon \big) \, {\bf T}_s^2 \right\}
  \nonumber \\  & & \,\, \times \,
  \exp \left\{ - \, {\rm i} \, \frac{\pi}{2}
  \Big[ K \big( \alpha_s, \epsilon \big) \Big]^2 \, 
  \ln \left(\frac{s}{- t} \right) \,  
  \Big[{\bf T}_t^2, {\bf T}_s^2 \Big] \right\} 
  \nonumber \\ & & \,\, \times \, 
  \exp \left\{ \frac{1}{6} \, \Big[ K \big( \alpha_s, 
  \epsilon \big) \Big]^3 \, \left(- 2 \pi^2 \ln 
  \left(\frac{s}{- t}\right) 
  \Big[ {\bf T}_s^2, \big[{\bf T}_t^2, {\bf T}_s^2 \big] \Big] 
  + \,  {\rm i} \pi \, \ln^2 \left(\frac{s}{- t} \right) 
  \Big[{\bf T}_t^2, \big[{\bf T}_t^2, {\bf T}_s^2 
  \big] \Big] \right) \right\}
  \nonumber \\ & &  \,\, \times \, 
  \exp \left\{ {\cal O} \left( \Big[ K \big( 
  \alpha_s, \epsilon \big) \Big]^4 \right) \right\} \, . 
\label{Zcomm}
\eeqa
This generalises \eqn{ZLLdef} to arbitrary logarithmic accuracy 
in $\ln(s/t)$. By factoring the color-non-diagonal operator in 
\eqn{Ztildedef} into separate exponentials, we have generated an 
infinite product of factors, having increasing powers of $K(\alpha_s, 
\epsilon)$, alongside increasingly nested commutator terms.
Working with a fixed logarithmic accuracy in the high-energy limit 
requires expanding $K(\alpha_s, \epsilon)$ in powers of $\alpha_s$,
and then collecting all terms in the various exponentials in \eqn{Zcomm}
that behave as $\alpha_s^k \left(\alpha_s \ln(s/t) \right)^p$ for 
fixed $k$. At leading logarithmic accuracy ($k = 0$), one therefore 
returns to the high-energy asymptotic behaviour of \eqn{ZLLdef}.  

In order to achieve next-to-leading logarithmic accuracy (NLL), one 
must expand the function $K(\alpha_s, \epsilon)$ to two loops in the
LL operator, given by the first factor in \eqn{Zcomm}, and further one
must include all terms in \eqn{Zcomm} with precisely one power of 
$K(\alpha_s, \epsilon)$ not accompanied by $\ln(s/(-t))$. Clearly 
these are all terms in which the nested commutators contain the 
operator ${\bf T}^2_s$ only once. An infinite sequence of exponentials 
becomes relevant then, but in each one of them only one commutator 
contributes. Furthermore, in all such terms one may retain only the 
leading-order contributions in $K(\alpha_s, \epsilon)$. The relevant 
operator can be written as
\beqa
  \left. \widetilde{Z} \left( \frac{s}{t}, \alpha_s, 
  \epsilon \right) \right\vert_{\rm NLL} 
  & = & \left(\frac{s}{- t} \right)^{K \left( \alpha_s, 
  \epsilon \right) \, \, {\bf T}_t^2}
  \Bigg\{ 1 + {\rm i} \, \pi  K \left( \alpha_s, 
  \epsilon \right) \, \bigg[ {\bf T}_s^2 - 
  \frac{K \left( \alpha_s, \epsilon \right)}{2!}
  \ln \left(\frac{s}{- t} \right) \,  
  \big[{\bf T}_t^2, {\bf T}_s^2 \big] \,
  \nonumber \\
  & &  \quad + \, \, \frac{K^2 \left( \alpha_s, \epsilon \right)}{3!} 
  \, \ln^2 \left(\frac{s}{- t}\right) \Big[ {\bf T}_t^2, 
  \left[{\bf T}_t^2, {\bf T}_s^2 \right] \Big]  \nonumber \\ 
  & & \quad - \, \, \frac{K^3 \left( \alpha_s, \epsilon \right)}{4!} \,
  \ln^3 \left(\frac{s}{- t} \right) \Big[{\bf T}_t^2, 
  \big[{\bf T}_t^2, \left[{\bf T}_t^2, {\bf T}_s^2 \right]
  \big] \Big] \, + \, \ldots \bigg] \Bigg\} \, .
\label{ZNLL}
\eeqa
It is evident that at this logarithmic order only the imaginary part 
of $Z$ contains non-diagonal color matrices, when working in the 
$t$-channel-exchange basis. We conclude that for the real part of 
the amplitude we still have Reggeization at NLL, and the trajectory for a 
given $t$-channel exchange is still given by the function $K(\alpha_s, 
\epsilon)$ times the quadratic Casimir eigenvalue of the appropriate 
color representation. It is straightforward to test the result by 
evaluating the function $K(\alpha_s, \epsilon)$ at NLO, generalizing
\eqn{KLO}. Using the NLO expression for the $d$-dimensional 
running coupling $\alpha_s (\mu^2, \epsilon)$, solution of
\eqn{running_coupling_RGE}, one readily finds
\beq
  K (\alpha_s, \epsilon) \, = \, \frac{\alpha_s}{\pi} \, 
  \frac{1}{2 \epsilon} \, + \left(\frac{\alpha_s}{\pi}\right)^2 \,
  \left( \frac{\widehat{\gamma}_K^{(2)}}{8 \epsilon} -
  \frac{b_0}{16 \epsilon^2} \right) + {\cal O}(\alpha_s^3) \, ,
\label{KNLO}
\eeq
where $b_0$ and $\widehat{\gamma}_K^{(2)}$ are given in eq.~(\ref{eq:beta}).
Using \eqn{KNLO}, one then recovers the (universal) result 
for the divergent parts of the two-loop Regge trajectory given in
\eqns{eq:2loop}{twolooptraja}. 

Interestingly, it is possible to write a closed form expression summing 
the series of commutators in \eqn{ZNLL}. To this end we take the 
Taylor expansion of \eqn{Zcomm} for small $K = K(\alpha_s, 
\epsilon)$ and fixed $\widetilde{X} = K(\alpha_s, \epsilon) \ln 
\left(s/(-t)\right) \, {\bf T}_t^2$, using the general result
\beq
  {\rm e}^{\widetilde{X} + K Y} \, = \, {\rm e}^{\widetilde{X}} \, 
  \bigg[ 1 + K \left(\int_0^1 d a \, {\rm e}^{- a\widetilde{X}} \,
  Y \, {\rm e}^{a \widetilde{X}} \right) + {\cal O}(K^2) \bigg] \, . 
\label{genres}
\eeq
One finds then
\beqa
  \left.  \widetilde{Z} \left( \frac{s}{t}, \alpha_s, 
  \epsilon \right) \right\vert_{\rm NLL} 
  & = & \left( {\frac{s}{- t}} \right)^{K (\alpha_s, \epsilon) 
  \, {\bf T}_t^2} \Bigg[1 \, + \, {\rm i} \, \pi \, 
  K (\alpha_s, \epsilon) 
  \label{Zres} \\ 
  & & \times
  \left(\int_0^1 d a \, \left({\frac{s}{- t}}\right)^{- a \, 
  K (\alpha_s, \epsilon) \, {\bf T}_t^2}  {\bf T}_s^2 \, 
  \left({\frac{s}{- t}}\right)^{a \, K (\alpha_s, \epsilon) 
  \, {\bf T}_t^2} \right) + {\cal O}(K^2) \Bigg] \, . 
\nonumber 
\eeqa
Using the Hadamard lemma
\beq
  {\rm e}^{ - a \widetilde{X}} \, Y \, {\rm e}^{ a \widetilde{X}} 
  \, = \, Y - a \, \left[\widetilde{X}, Y \right] \, + \, 
  \frac{a^2}{2!} \, \left[ \widetilde{X}, \big[\widetilde{X}, Y 
  \right] \Big] \, - \, \frac{a^3}{3!}  \left[\widetilde{X},
  \left[\widetilde{X}, \left[ \widetilde{X}, Y \right] \right]
  \right] +{\cal O} \left( a^4 \right)
\label{hada}
\eeq
it is straightforward to see that \eqn{Zres} is indeed equivalent to 
\eqn{ZNLL}. We conclude that \eqn{Zres} provides a compact 
expression for the singularities of the amplitude to NLL accuracy, 
in the high energy limit, including both real and imaginary parts. 
Working in the $t$-channel exchange basis, where ${\bf T}_t^2$ 
is diagonal and ${\bf T}_s^2$ is not, it is evident that the NLL 
${\cal O}(K)$ term in the square brackets mixes between different 
components of the hard interaction. Thus, while Reggeization extends 
to NLL for the real part of the amplitude, it does not for the 
imaginary part.

Considering now next-to-next-to-leading logarithmic accuracy (NNLL), 
where two powers of $K$ not accompanied by $\ln(s/t)$ must be 
included, \eqn{Zcomm} tells us that also the real part of the amplitude 
becomes non-diagonal in the $t$-channel-exchange basis. 
In particular,  already at ${\cal O}(\alpha_s^2)$ we encounter
a NNLL correction to the real part of the amplitude which is non-diagonal in the $t$-channel exchange basis: this contribution originates in the expansion of the exponential $\exp \left\{ {\rm i} \, \pi \, K \big( \alpha_s,   \epsilon \big) \, {\bf T}_s^2 \right\}$ in the first line of \eqn{Zcomm} to second order, giving, 
\begin{equation}
\label{first_violation}
-\,\frac{1}{2} \pi^2 \, K^2 \big( \alpha_s,   \epsilon \big) \, \left({\bf T}_s^2\right)^2\,.
\end{equation}
Furthermore, at ${\cal O}(\alpha_s^3)$, and at the same logarithmic order (NNLL) one encounters in the exponent of \eqn{Zcomm} the operator  
\beq
  {\cal E} \left( \frac{s}{t}, \alpha_s, 
  \epsilon \right) \equiv - \, \frac{\pi^2}{3} \,
  {K^3 (\alpha_s, \epsilon)} \, \ln \left(\frac{s}{- t} \right) 
  \Big[{\bf T}_s^2, \big[{\bf T}_t^2, {\bf T}_s^2 \big] \Big] \, ,
\label{NNLL}
\eeq
which is also real.  
Since it mixes between components of the hard interaction 
corresponding to different $t$-channel exchanges, it leads generically
to a breakdown of the Reggeization picture at NNLL, also for the real 
part of the scattering amplitude. 

Several remarks are in order.  First we note that evidence for a 
possible breakdown of the Reggeization picture, for specific 
amplitudes and beyond NLL, has already been presented  in the 
literature. In Refs.~\cite{Korchemskaya:1994qp} and 
\cite{Korchemskaya:1996je}, the problem of Reggeization was 
studied, for the case of quark scattering (albeit in a somewhat 
different kinematic limit, where the eikonal lines are massive)
by diagonalizing the soft anomalous dimension matrix defined in
\eqn{renS_massless}. In the case studied there, it was found that 
the eigenvectors of the matrix  are not given by pure $t$-channel 
exchanges. Furthermore, in Ref.~\cite{DelDuca:2001gu}, two-loop
scattering amplitudes for gluon-gluon, quark-gluon and quark-quark
scattering were exploited to determine the two-loop gluon Regge 
trajectory and the two-loop gluon and quark impact factors. The 
knowledge of these data allows to set up a consistency test for
Regge factorization at the two-loop level. This test was found to
fail at the level of constant (non-logarithmic) terms, in that impact factors
became process-dependent, with a 
discrepancy from the predictions of Regge factorization 
proportional to $\alpha_s^2 \pi^2/\epsilon^2$. We note 
that this failure is consistent with our predictions, as given 
in \eqn{Zcomm}. Indeed, while the operator ${\cal E}$ in 
\eqn{NNLL} acts non-trivially starting at order $\alpha_s^3 
\ln(s/t)$, a discrepancy of precisely the form suggested 
in~\cite{DelDuca:2001gu} can be generated within our 
approach by expanding the exponential in the first line of 
\eqn{Zcomm} to ${\cal O} (\alpha_s^2)$, as shown in (\ref{first_violation}) above. 
Our results are thus consistent with existing evidence
for a breakdown of the Reggeization picture, but place it in a 
completely general context, hopefully allowing in the future for 
definite tests in specific cases at the three-loop level. 

On the face of it, a potential loophole in our argument could be 
the fact that the Reggeization breaking operator ${\cal E}$ arises 
at the same order (${\cal O} (\alpha_s^3)$) where the first 
possible corrections to the dipole formula, $\Delta (\rho_{ijkl},
\alpha_s)$, might arise, as explained in \sect{sec:sumodipoles}.
It is, however, easy to see on very general grounds that such 
corrections to the anomalous dimension cannot cancel (or indeed 
modify) the Reggeization breaking operator. To this end, it is 
sufficient to recall that any three-loop correction to the anomalous 
dimension only generates a single pole, ${\cal O}(1/\epsilon)$, 
at the three-loop order, whereas the Reggeization breaking 
operator ${\cal E} = {\cal O}(1/\epsilon^3)$, as follows from 
its proportionality to the third power of $K$ in \eqn{KNLO}. 
The conclusion is that corrections to the anomalous dimension, 
which may indeed arise at ${\cal O} (\alpha_s^3)$, are entirely 
irrelevant to Reggeization breaking. The Reggeization breaking 
argument is robust. 

Finally, it is important to emphasize that Reggeization breaking 
is \emph{always} suppressed in the large-$N_c$ limit\footnote{Reggeization 
breaking has also been discussed in~\cite{Bartels:1999xt}. Also 
in that case the effect arises from non-planar diagrams, thus is 
again subleading in the large $N_c$ limit.}. To see this recall that the operators ${\bf T}_s^2$ and ${\bf T}_t^2$ we use to express the $\widetilde{Z}$ factor in (\ref{Ztildedef}) and in (\ref{Zcomm}) are related by colour conservation, (\ref{colcon2}), a relation which involves the third operator, ${\bf T}_u^2$. For general $N_c$ diagonalizing ${\bf T}_t^2$ leaves ${\bf T}_s^2$ and ${\bf T}_u^2$ non-diagonal. However at large $N_c$ one of the three must be proportional to the unit matrix up to $1/N_c^2$ corrections\footnote{This is equivalent to the statement that in a colour-ordered amplitude, the corresponding dipoles generate non-planar graphs.}, and thus the other two are diagonalised simultaneously. As a consequence the $\widetilde{Z}$ factor in (\ref{Ztildedef}) and in (\ref{Zcomm}) is always colour diagonal at large $N_c$.

It should also be pointed out that, while 
Reggeization breaking emerges as a generic feature, there may be 
special cases, for example particular gauge theories, or specific
scattering processes in certain representations, where the 
Reggeization breaking operator ${\cal E}$ might have vanishing 
eigenvalues. It would be interesting to investigate the circumstances 
under which this situation may arise, so that Reggeization might generalize to higher logarithmic accuracy, or indeed might even 
be exact to leading power. In such a circumstance one might, for 
example, viably attempt to extend the techniques of Regge 
factorization and study the resummation of high-energy effects 
at NNLL order and beyond\footnote{An approximate NNLL BFKL 
kernel was presented in~\cite{Marzani:2007gk}, and discussed 
further in~\cite{Marzani:2007wx}, which also briefly comments 
on some aspects of Reggeization breaking.}.

\section{Possible corrections to the dipole formula at three loops}
\label{sec:beyond_dipoles}

The purpose of this section is to briefly analyse the form of 
potential corrections to the dipole formula at three-loop order, 
following Refs.~\cite{Gardi:2009qi,Becher:2009qa,Dixon:2009ur},
in order to verify whether they can be constrained by studying
the high-energy limit. Any potential correction to the dipole formula, 
at three-loops and beyond (excluding corrections due to the presence of higher
order Casimir invariants), must depend on the kinematics exclusively through conformally-invariant cross ratios of the form
\beq
  \rho_{ijkl} \equiv \frac{(- s_{i j}) (- s_{k l})}{(- s_{i k})(- s_{j l})} \, ,
\label{cicr}
\eeq
where, as usual, $-s_{ij}\equiv 2 \,\left\vert p_i \cdot p_j
\right\vert \, e^{-{\rm i} \pi \lambda_{ij}}$. For four partons there 
are three cross ratios: $\rho_{1234}$, $\rho_{1423}$ and 
$\rho_{1342}$, and they are related through the identity
$
\rho_{1234} \, \rho_{1423} \, \rho_{1342} \, = 1 \, .
$
Specializing to forward kinematics, where $s_{12} \gg - t > 0$, we 
find
\beqa
  - s_{12} & = & - s_{34} = s \, {\rm e}^{- {\rm i} \pi} < 0 \, , 
  \nonumber \\
  - s_{13} & = & - s_{24} = - t > 0 \, , \\
  - s_{14} & = & - s_{23} = - u = s + t > 0 \, , \nonumber
\label{mandelhigh}
\eeqa
so we obtain the following conformally-invariant cross ratios:
\begin{subequations}
\label{Lijkl_forward}
\begin{align}
  \rho_{1234} & \equiv \frac{(- s_{12})(- s_{34})}{(- s_{13})(-
  s_{24})} = \left(\frac{s}{- t}\right)^2 \, {\rm e}^{- 2 {\rm i} \pi}
  \, ; & \qquad L_{1234} & = 2( L - {\rm i} \pi) \, ;
\label{L1234_forward}
  \\
  \rho_{1342} & \equiv \frac{(- s_{13})(- s_{24})}{(- s_{14})(-
  s_{23})} = \left(\frac{- t}{s + t}\right)^2 
  \,; & \qquad L_{1342} & \simeq - 2 L \, ;
\label{L1342_forward}
  \\
  \rho_{1423} & \equiv \frac{(- s_{14})(- s_{23})}{(- s_{12})(-
  s_{34})} = \left(\frac{s + t}{s}\right)^2 \, {\rm e}^{2{\rm i} \pi}
  \,; & \qquad L_{1423} & \simeq 2 {\rm i} \pi \, ,
\label{L423_forward}
\end{align}
\end{subequations}
where we carefully extracted the phase factors. Perturbative 
corrections are expected to depend on the kinematic variables 
via logarithms (or polylogarithms, see below); thus we wrote
explicitly also the logarithm of each cross ratio, $L_{ijkl} \equiv
\ln(\rho_{ijkl})$, defining $L = \ln(s/(- t))$ and neglecting power 
corrections in the ratio $|t/s|$. We see that while the logarithms 
of the first two cross ratios are large in the high-energy limit, the 
logarithm of the third ratio is not, and in fact it would have 
vanished if not for the phase. This may be compared to the 
collinear limit (where the two collinear particles belong to the 
final state) analysed in \cite{Dixon:2009ur}, where indeed the 
logarithm of the third cross ratio vanishes up to power corrections.
The non-vanishing phase of $L_{1423}$ in the high-energy limit 
will be crucial in what follows.

Next we recall that owing to Bose symmetry, collinear limits and 
transcendentality constraints, only a small set of functions could 
potentially appear as a three-loop correction to the soft anomalous 
dimension for massless partons. In particular, according to the 
analysis of Ref.~\cite{Dixon:2009ur} there is only one candidate 
function composed of products of logarithms $L_{ijkl}$. It is 
given by
\beqa
\label{Delta_case212}
  \hspace{-5mm}
  \Delta^{(212)} (\rho_{ijkl}, \alpha_s) \, = \, 
  \left( \frac{\alpha_s}{\pi} \right)^3 \,
  {\bf T}_1^{a} {\bf T}_2^{b} {\bf T}_3^{c} {\bf T}_4^{d} \,
  \bigg[ && \! \! f^{ade} f^{cbe} \,
  L_{1234}^2 \, \Big(L_{1423} \, L_{1342}^{2} \, + \, 
  L_{1423}^{2} \, L_{1342} \Big) \\
  + && \! \! f^{cae} f^{dbe} \, 
  L_{1423}^2 \, \Big(L_{1234} \, L_{1342}^{2} \, + \, 
  L_{1234}^{2} \, L_{1342} \Big) \nonumber \\
  + && \! \! f^{bae} f^{cde} \,
  L_{1342}^{2} \, \Big(L_{1423} \, L_{1234}^{2} \, + \,
  L_{1423}^{2} \, L_{1234} \Big) \bigg] \, .
  \nonumber
\eeqa
Substituting \eqn{Lijkl_forward} into \eqn{Delta_case212}
we obtain
\beqa
  \Delta^{(212)} (\rho_{ijkl}, \alpha_s)) & = & 
  \left(\frac{\alpha_s}{\pi}\right)^3 \, {\bf T}_1^{a} {\bf T}_2^{b} 
  {\bf T}_3^{c} {\bf T}_4^{d} \, \,
  32 \, {\rm i} \, \pi \Big[ \Big( - L^4 - {\rm i} \pi L^3
  - \pi^2 L^2 - {\rm i} \pi^3 L \Big) f^{ade}f^{cbe}
  \nonumber \\ && \qquad + \, \,
  \Big(2 {\rm i} \pi L^3 - 3 \pi^2 L^2 - {\rm i} \pi^3 L \Big) 
  f^{cae}f^{dbe} \Big] \, + {\cal O} \left( |t/s| \right)\, ,
\label{Delta_4_large_s}
\eeqa
where we used the Jacobi identity to eliminate the third color 
factor. This result is interesting because it cannot be consistent 
with known properties of the high-energy limit: indeed, Reggeization
at LL level implies that the highest power of $\ln(s/(-t))$ at $n$ 
loops must be $n$. A contribution of the form $\alpha_s^3 \, L^4$, 
as in \eqn{Delta_4_large_s}, is then super-leading and must be 
discarded. We conclude that, based on the high-energy limit, 
a function $\Delta$ of the form of \eqn{Delta_case212}, which 
was proposed in Ref.~\cite{Dixon:2009ur}, may be excluded. 

As emphasized in Ref.~\cite{Dixon:2009ur}, other functions of 
conformally-invariant cross ratios are possible as well. Functions 
involving polylogarithms were examined there, and two viable 
candidates, consistent with all available constraints, were 
proposed\footnote{The proposed polylogarithmic functions
are given in eqs.~(5.40) and (5.41) of Ref.~\cite{Dixon:2009ur}.}.
Interestingly, we find that those two examples also have 
super-leading logarithms of the form of \eqn{Delta_4_large_s}
in the high-energy limit. Therefore, neither of those functions can, 
by themselves, be consistent with the Regge limit. The only way in which 
the latter two examples, or \eqn{Delta_case212}, could still be 
relevant for a three-loop correction to the dipole formula is if 
they appear in a specific linear combination, chosen so as to 
eliminate all super-leading $L^4$ terms. We emphasize that
also $L^3$ terms (and possibly $L^2$ terms) would need to be 
eliminated so as to be consistent with leading (or next-to-leading) 
logarithmic Reggeization, making it all the more challenging to find 
a candidate function.

A further comment is due concerning the origin of the high 
power of the logarithms (or polylogarithms) in the examples of 
Ref.~\cite{Dixon:2009ur}. Beyond mere transcendentality arguments, 
which could be satisfied by numerical factors such as $\zeta(n)$, 
the high multiplicity of logarithms is a consequence of the need to satisfy the collinear constraint for any pair of partons, which
requires that $L_{1234}$, $L_{1423}$ and $L_{1342}$ appear 
to power $p \geq 1$ in every term. This requirement leads to a direct 
conflict with the Regge-limit constraint, and this incompatibility
appears to be a generic feature.

To conclude, we have provided an additional constraint on potential corrections to the dipole formula, based on the high-energy limit. 
We find that all explicit candidate functions proposed in 
Ref.~\cite{Dixon:2009ur}, which were found to be consistent with all other 
constraints, fail to adhere to the known structure of high-energy 
logarithms, and may therefore be excluded (with the only possible 
exception of finding a linear combination of these functions which is 
consistent with what is known about the Regge limit). This, of course, 
gives additional support to the validity of the dipole formula beyond 
two loops. Nevertheless, it does not preclude three-loop corrections 
altogether, and it is still possible that proper counter examples might
be found that satisfy the high-energy constraints.

\section{The dipole formula in multi-Regge kinematics}
\label{sec:multiregge}

In the preceding sections, we concentrated on the simplest 
and most commonly studied case in which Reggeization governs 
the high-energy limit: that of $2 \rightarrow 2$ scattering. We 
were able to confirm and extend known results regarding the 
Reggeization of $t$ and $u$-channel particle exchanges, and we provided a general expression for singular contributions to 
high-energy amplitudes, valid at leading power in $|t/s|$ and 
going beyond the limitations of Regge factorization. As 
outlined in \sect{sec:Reggeization}, however, Reggeization in 
the four-point amplitude is tightly connected, via unitarity, to 
the behavior of $2 \rightarrow n$ scattering amplitudes in 
multi-Regge (MR) kinematics, in which the $n$ final-state particles 
are emitted with a strong rapidity ordering, implying a hierarchy 
in the corresponding two-particle invariant masses.

As for the case of $2 \rightarrow 2$ scattering, we can investigate 
$(n + 2)$-parton scattering using the infrared approach and the 
dipole formula, which is valid for any number of massless external 
partons. Once again, the dipole formula will provide a compact and 
general derivation of the asymptotic properties of the scattering 
amplitude in the MR limit, at least for divergent contributions to 
the corresponding Regge trajectories and impact factors. 
Furthermore, we will be able to investigate what happens at 
subleading logarithmic order, and we will find evidence for a 
breakdown of Reggeization starting at NNLL, as was already 
discussed in the context of $2 \rightarrow 2$ scattering. 

\begin{figure}[htb]
\begin{center}
\scalebox{1.0}{\includegraphics{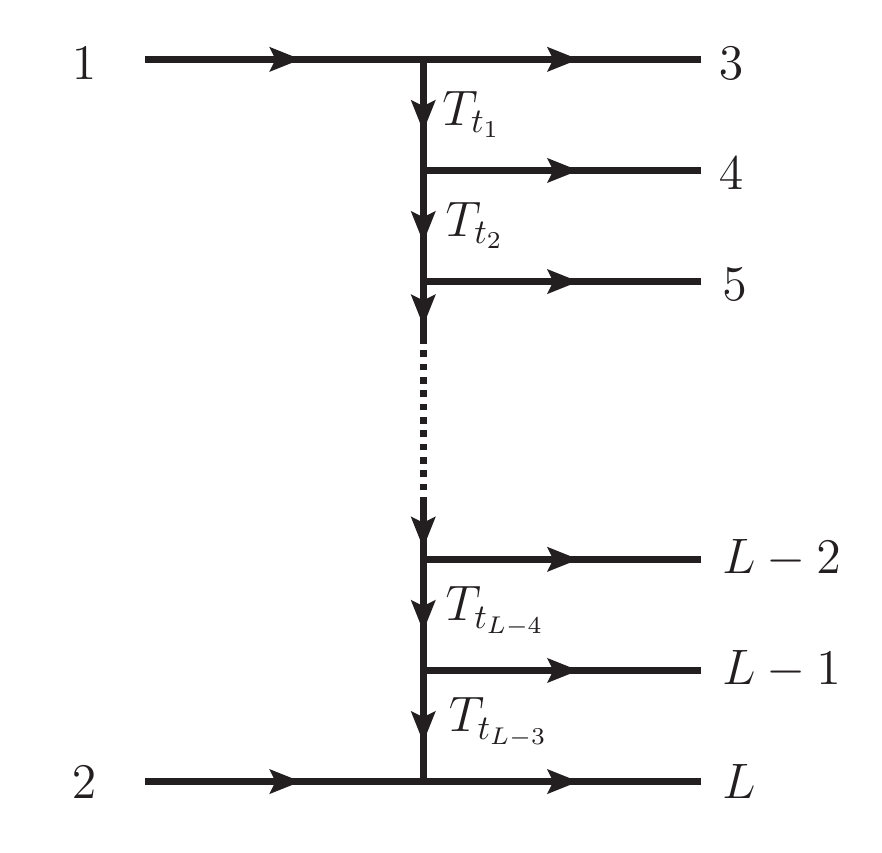}}
\caption{A general $L$-parton scattering process in the MR limit, 
consisting of strongly ordered rapidities in the final state. The 
$t$-channel Casimir operator ${\bf T}_{t_k}$ is defined in \eqn{Tkdef}.}
\label{MRfig}
\end{center}
\end{figure}

Let us begin by considering the general $L$-parton scattering 
amplitude, depicted in figure~\ref{MRfig}. More precisely, let 
$y_i$ ($3 \leq i \leq L$) be the rapidity of final state parton $i$, 
and consider the MR limit of strongly ordered rapidities, with 
comparable transverse momenta,
\beq
  y_3 \gg y_4 \gg \ldots \gg y_L \,, \qquad\qquad  
  |k_i^\perp| \simeq |k_j^\perp| \,,\,\, \quad \forall i,j  \, ,
\label{rapord}
\eeq
where we use the complex momentum notation $k_i^\perp = k_i^1 
+ i k_i^2$, and where, without loss of generality, we may take 
rapidities decreasing down the final state parton ladder, as shown 
in figure~\ref{MRfig}.
Also in the figure, we have defined the $t$-channel color operator
\beq
  {\bf T}_{t_k} \, = \, {\bf T}_1 + \sum_{p = 1}^k 
  {\bf T}_{p + 2} \, ,
\label{Tkdef}
\eeq
whose eigenstates are definite $t$-channel exchanges occurring 
between partons $k + 2$ and $k + 3$. In the MR limit the invariants 
$- s_{i j}\equiv  2 \, \left\vert p_i \cdot p_j \right\vert \, e^{- {\rm i} \pi \lambda_{ij}}$ to be inserted into the dipole operators in 
\eqn{sumodipoles}  are approximated (see {\it e.g.} eq.~(66) 
of~\cite{DelDuca:1995hf}) by their leading rapidity dependence,
\beqa
  - \, s & \equiv & - \, s_{12} \, \simeq \, \, |k_3^\perp| 
  \, |k_L^\perp| \, e^{- {\rm i} \pi } \,
  {\rm e}^{y_3 - y_L} \, , \nonumber \\
  - \, s_{1 i} & \simeq &  \, |k_3^\perp| \, |k_i^\perp| 
  \, {\rm e}^{y_3 - y_i} \, , \nonumber \\
  - \, s_{2 i} & \simeq &  \, |k_L^\perp| \, |k_i^\perp| 
  \, {\rm e}^{y_i - y_L} \, , \nonumber \\ 
  - \, s_{i j} & \simeq &  \, |k_i^\perp| \, |k_j^\perp| 
  \, {\rm e}^{y_i - y_j} \, e^{- {\rm i} \pi } \,,
  \quad  \quad \quad  3 \leq i < j \leq L \, ,
\label{invars}
\eeqa
where we keep track of the phases, and the notation is chosen 
so as to explicitly distinguish initial and final state particles. 

With this parametrization of momentum invariants, the dipole 
operator in \eqn{sumodipoles} becomes, in the MR limit, 
\begin{align}
\label{sumodipolesMR}
\begin{split}
  &  Z \left(\frac{p_l}{\mu}, \alpha_s(\mu^2), \epsilon \right)
  \, = \, \exp \left[ \int_0^{\mu^2} \frac{d \lambda^2}{\lambda^2}
  \left\{ \frac{1}{4} \, \widehat{\gamma}_K \Big( \alpha_s 
  \left(\lambda^2, \epsilon \right) \Big) \left[ \sum_{i = 1}^{L - 1}
  \sum_{j > i} \left|y_i - y_j \right| {\bf T}_i \cdot {\bf T}_j 
  \right. \right. \right.  \\
  &  \qquad \quad \left. \left. \left.  - \, {\rm i} \pi
  \left({\bf T}_1 \cdot {\bf T}_2 + \sum_{i = 3}^{L - 1}
  \sum_{j > i}{\bf T}_i \cdot{\bf T}_j \right) \,  + \, 
  \ln \left(\frac{|k_3^\perp|}{\lambda}
  \right) \left({\bf T}_1 \cdot {\bf T}_2 + \sum_{i = 3}^L 
  {\bf T}_1 \cdot {\bf T}_i \right) 
  \right. \right. \right.  \\
  &  \qquad \left. \left. \left. \quad
  + \, \ln \left(\frac{|k_L^\perp|}{\lambda} \right) 
  \left({\bf T}_1 \cdot {\bf T}_2 + \sum_{i = 3}^L{\bf T}_2
  \cdot{\bf T}_i \right) 
  \, + \, \sum_{i = 3}^L 
  \ln \left(\frac{\, |k_i^\perp|}{\lambda} \right)
  \bigg({\bf T}_1 \cdot {\bf T}_i + {\bf T}_2 \cdot {\bf T}_i
  \right. \right. \right. \\
  &  \qquad \left. \left. \left. \qquad \quad
  + \sum_{j = 3, j\neq i}^L{\bf T}_i \cdot{\bf T}_j \bigg)
  \right] - \frac{1}{2} \sum_{i = 1}^L \gamma_{J_i} 
  \Big( \alpha_s \left(\lambda^2, \epsilon \right) \Big)\right\} 
  \right] \, .
\end{split}
\end{align}
Note that we have defined here the (unphysical) rapidities $y_1 = 
y_3$ and  $y_2 = y_L$ for the initial state particles, which 
simplifies notation in the first sum on the right-hand side. For later 
convenience, we also introduce the (unphysical) transverse momenta 
$|k_1^\perp| = |k_3^\perp|$ and $|k_2^\perp| = |k_L^\perp|$.
Concentrating on the contents of the square brackets in the exponent, the first (and only rapidity-dependent) term may be rewritten 
according to
\beq
  \sum_{i = 1}^{L - 1} \, \sum_{j > i}  \left|y_i - y_j \right| \,
  {\bf T}_i \cdot {\bf T}_j \, = \, - \, \sum_{k = 3}^{L - 1} \,
  {\bf T}_{t_{k - 2}}^2 \, \Delta y_k \, ,
\label{rapres}
\eeq
where we have introduced the difference of two consecutive rapidities
\beq
  \Delta y_k \, \equiv \, y_k - y_{k+1} \, ,
\label{ykkdef}
\eeq
while the $t$-channel color operators are defined in \eqn{Tkdef}.
A proof of \eqn{rapres} may be found in App.~\ref{app:rapproof}. 
One may simplify the coefficients of the other terms by using the color 
conservation equations
\beq
  \sum_{i = 1}^L {\bf T}_i \, = \, 0 \, , \qquad
  \left( \sum_{i = 1}^L {\bf T}_i \right)^2 = \, 
  \sum_{i = 1}^L  C_i \, + \, 2 \sum_{j > i} {\bf T}_i \cdot 
  {\bf T}_j \, = \, 0 \, .
\label{colcongen}
\eeq
\Eqn{sumodipolesMR} now becomes
\beqa
  & & \hspace{-1cm}
  Z \left(\frac{p_l}{\mu}, \alpha_s(\mu^2), \epsilon \right)
  \, = \, \exp \left[ \int_0^{\mu^2} \frac{d \lambda^2}{\lambda^2}
  \left\{\frac{1}{4} \, \widehat{\gamma}_K \Big( \alpha_s 
  \left(\lambda^2, \epsilon \right) \Big) \left[ - 
  \sum_{k = 3}^{L-1} \, {\bf T}^2_{t_{k - 2}} \, \Delta y_k
  - {\rm i} \pi \, {\bf T}_s^2 
  \right. \right. \right. \nonumber \\
  & & \hspace{0cm} \left. \left. \left.
  - \,\sum_{i=1}^L C_i \left(\ln \left(\frac{|k_i^\perp|}{\lambda}
   \right)-{i\pi\over 2}
  \right)  \right] - \frac{1}{2} 
  \sum_{i = 1}^L\gamma_{J_i} \Big( \alpha_s 
  \left(\lambda^2, \epsilon \right) \Big)\right\} \right] \, ;
\label{sumodipolesMR2}
\eeqa
as a consequence, the multiparticle dipole operator, in the MR limit, 
may be written in a factorized form similar to \eqn{Zfac}, as
\beq
  Z \left( \frac{p_l}{\mu}, \alpha_s (\mu^2), \epsilon \right)
  \, = \, \widetilde{Z}^{\rm MR} \Big( \Delta y_k, 
  \alpha_s (\mu^2), \epsilon \Big) \, \, Z_{\bf 1}^{\rm MR} 
  \left( \frac{|k_i^\perp|}{\mu}, 
  \alpha_s (\mu^2), \epsilon \right) \, ,
\label{ZMR} 
\eeq
where
\beqa
  \widetilde{Z}^{\rm MR} \Big( \Delta y_k, \alpha_s (\mu^2), 
  \epsilon \Big) & = & \exp \left\{ K \Big( \alpha_s (\mu^2), 
  \epsilon \Big) \left[\sum_{k = 3}^{L-1} {\bf T}^2_{t_{k - 2}}
  \, \Delta y_k \, + \, {\rm i} \pi {\bf T}_s^2 \right] \right\} \, ,
  \label{ZtildeMR} \\
  Z_{\bf 1}^{\rm MR} \left( \frac{|k_i^\perp|}{\mu}, 
  \alpha_s (\mu^2), \epsilon \right) & = & 
  \exp \left\{ \sum_{i = 1}^L B_i \Big( \alpha_s (\mu^2), 
  \epsilon \Big) \, + \, \frac{1}{2} \, D \Big( \alpha_s (\mu^2), 
  \epsilon \Big) \sum_{i = 1}^L C_i \right. \nonumber \\ 
  & & \left.  \hspace{-4cm} + \, K \Big( \alpha_s (\mu^2), 
  \epsilon \Big) \, \sum_{i = 1}^L C_i \,\left[ 
  \ln \left( \frac{ \, |k_i^\perp|}{\mu} \right)  -{i\pi\over2}
 \right] \right\} \, .
\label{Z1MR}
\eeqa
Eqs.~(\ref{ZtildeMR}, \ref{Z1MR}) are generalisations 
to multiparticle amplitudes of the results obtained in the four-point 
case, expressed in \eqns{Ztildedef}{Zddef}. As in that example,
the non-trivial color matrix $\widetilde{Z}^{\rm MR}$ contains 
the dominant contribution in the high-energy limit, consisting of an 
operator whose eigenstates are definite $t_k$-channel exchanges. 
Furthermore, there is a correction which affects the imaginary part 
of the amplitude starting at NLL, and involving the $s$-channel 
operator ${\bf T}_s^2$. The factor $Z_{\bf 1}^{\rm MR}$, 
proportional to the unit matrix in color space, collects collinear 
singularities in the form of jet functions for each external particle, 
as well as terms which only depend upon the quadratic Casimir 
invariants of individual particles.  It is straightforward to check 
that \eqns{ZtildeMR}{Z1MR} reduce to \eqns{Ztildedef}{Zddef} 
in the special case $L = 4$. Indeed, the non-diagonal factor has 
only a single $t$-channel exchange with ${\bf T}_{t_1} \equiv 
{\bf T}_t$, so that \eqn{ZtildeMR} obviously reduces to
\eqn{Ztildedef}, upon using the fact that the rapidity difference 
may be expressed as $y_3 - y_4 = \ln(|s/t|)$. In \eqn{Z1MR}, 
one may use the fact that if only two final state particles are present, 
one has $|k_3^\perp| = |k_L^\perp| \equiv |k^\perp|$. One then 
readily recovers \eqn{Zddef}.

Having derived the form of the dipole operator in MR kinematics, we
see that implications regarding Reggeization are directly analogous
to the four-point case, and may be stated as follows. If a given 
$L$-parton hard interaction is dominated by a ladder exchanged in the 
$t$-channel at leading order, in the MR limit, these automatically 
Reggeize. By this, we mean that each $t_k$-channel propagator 
factor (connecting particles $k$ and $k+1$) is dressed according 
to
\beq
  \frac{1}{t_k} \longrightarrow \frac{1}{t_k} \,
  {\rm e}^{\alpha_k (t) \, (y_k - y_{k+1})}
  \equiv \frac{1}{t_k} \left(- \frac{s_{k, k+1}}{t_k} 
  \right)^{\alpha_k (t)} \, ,
\label{propdress}
\eeq
where $\alpha_k (t)$ is the Regge trajectory of the $t_k$-channel 
exchange, involving the appropriate Casimir $C_k$, and, as before, 
$s_{k, k+1} = 2 p_k \cdot p_{k+1}$, using the known form of 
kinematic invariants in the MR limit~\cite{DelDuca:1995hf}. It must 
of course be stressed again that, as in the $2 \rightarrow 2$ case, our 
approach can only guarantee Reggeization for the singular part of 
the Regge trajectory. 

A further comment is in order, regarding the fact that there are 
different color operators ${\bf T}_{t_k}^2$ for each $t$-channel 
propagator. One could imagine a situation in which 
these operators might not commute with 
each other, requiring the use of the Baker-Campbell-Hausdorff
formula to compute the action of the Reggeization operator 
$\widetilde{Z}^{\rm MR}$ on the hard interaction. This possibility
however does not occur. Indeed, the reasoning of 
App.~\ref{app:clebsch}, where the possibility of choosing a 
basis of $t$-channel eigenstates is proven explicitly, shows that 
the operators always commute,
\beq
  [{\bf T}_{t_k}^2, {\bf T}_{t_{k'}}^2] = 0\,,\qquad \forall k,k'\,.
\label{Tkcomm}
\eeq
A simple way to see this here is as follows: one may use the definition 
of \eqn{Tkdef} to write
\beq
  {\bf T}_{t_{k'}} \, = \, {\bf T}_{t_k} + 
  \sum_{i = k + 1}^{k'}{\bf T}_{i+2} \, ,
\label{Tk'}
\eeq
where, without loss of generality, we have taken $k' > k$. One 
finds then
\beq
  \left[{\bf T}_{t_k}, {\bf T}_{t_{k'}}\right] \, = \,
  \sum_{i = k + 1}^{k'} \left[{\bf T}_{i+2}, {\bf T}_{t_k}\right] \, .
\label{Tkcomm2}
\eeq
The right-hand side vanishes, owing to the fact that ${\bf 
T}_{t_k}$ does not contain ${\bf T}_{i+2}$ for $i > k$, and one 
recovers \eqn{Tkcomm}. As a consequence, at LL accuracy, where 
we can neglect the term in the exponent proportional to 
${\bf T}^2_s$, we can write the Reggeization operator
in a factorized form
\beq
  \left. \widetilde{Z}^{\rm MR} \Big( \Delta y_k, \alpha_s (\mu^2),
  \epsilon \Big) \right\vert_{\rm LL} \, = \, \prod_{k=3}^{L - 1} 
  \exp \left[ K \Big( \alpha_s (\mu^2), \epsilon \Big) \,
  {\bf T}^2_{t_{k - 2}} \, \Delta y_k \, \right] \, .
\label{ZtildeMR2}
\eeq
In App.~\ref{app:clebsch} we show that it is always possible to 
decompose the hard interaction into a colour basis corresponding 
to an exchange of definite states in the $t$ channel. Each $t$ channel
exchange between, say, the emissions of partons $k + 2$ and 
$k + 3$, is an eigenstate of the corresponding color operator
${\bf T}_{t_k}^2$, and its rapidity dependence enters the 
exponent of the amplitude with the corresponding eigenvalue,
so that Reggeization follows. The simplest case is when a single 
particle species is exchanged in the $t$-channel. One then recovers 
the well-known Reggeization of leading logarithms in the form 
of \eqn{propdress}. Reggeization, however, is more general than 
this, as is clear from the structure of \eqn{ZtildeMR2}: in principle, 
different $t$-channel exchanges may occur, so that different 
rapidity intervals exponentiate with different eigenvalues.

Computing the four-point amplitude, we found evidence for a 
breakdown of Reggeization beyond LL order in the imaginary part 
of the amplitude, and beyond NLL order for the real part. This was 
due to the ${\rm i} \pi$ term in \eqn{Ztildedef}, whose coefficient 
does not commute in general with the $t$-channel operator. Exactly 
the same situation occurs in \eqn{ZtildeMR}, and indeed the same 
${\rm i} \pi {\bf T}_s^2$ term occurs, which is independent of 
the number of external partons, as perhaps might be expected for
an $s$-channel color structure. We thus observe a corresponding 
breakdown of Reggeization in the general multiparton case, which 
is entirely consistent with what happens for $L = 4$. As in that case, 
this conclusion is robust with respect to possible corrections to 
the dipole formula arising at three loops.

\section{Discussion}
\label{sec:discuss}

In this paper, we have developed an infrared-based approach to 
the high-energy limit of gauge theory amplitudes, making use of the dipole 
formula, \eqn{sumodipoles}, an explicit ansatz for the all-order infrared 
singularity structure of fixed-angle scattering amplitudes involving 
massless partons. We have seen that in the Regge limit the infrared 
operator $Z$, responsible for all soft and collinear singularities 
in the dipole formula, factors into the product of a color-trivial part, 
multiplied by the universal high-energy operator $\widetilde{Z}$ of
\eqn{Ztildedef}, acting on the appropriate hard interaction. If the 
latter is dominated, as $|s/t| \rightarrow \infty$, by the exchange 
of distinct color states in the $t$-channel, then each such state 
automatically Reggeizes at leading logarithmic accuracy, at least 
for the singular part of the amplitude. The infrared-singular part 
of the renormalized Regge trajectory is given by the function 
$K (\alpha_s, \epsilon)$, which is completely determined by 
the cusp anomalous dimension (and by the $d$-dimensional
beta function), and indeed is a well-known function arising in different
contexts in perturbative QCD (we note for example that this 
function assumes a particularly simple form in conformal gauge 
theories, such as ${\cal N} = 4$ super Yang-Mills theory, as 
discussed in~\cite{Bern:2005iz,Dixon:2008gr}). These results
confirm the calculations of Refs.~\cite{Korchemskaya:1994qp,
Korchemskaya:1996je,Korchemsky:1993hr}, and they imply that
the infrared-singular Regge trajectory is proportional to the 
quadratic Casimir invariant of the appropriate representation, 
but is otherwise universal, as was observed in the past in 
concrete examples. If a number of $t$-channel exchanges are 
possible -- which may occur at different perturbative orders in 
the hard interaction -- then each exchanged state Reggeizes independently. 

Approaching the problem of Reggeization with the dipole formula
gives insights both on the generality of the phenomenon (as we 
discussed, every color state giving leading contributions in the
$t$-channel Reggeizes at LL accuracy), and on its inherent limitations.
We observe that the high-energy operator $\widetilde{Z}$ is diagonal 
in a $t$-channel basis only at LL level, and is corrected by a phase 
which depends on the identity of $s$-channel exchanges at NLL. The 
existence of this phase does not affect Reggeization for the real part 
of the amplitude at NLL, so that all known results are correctly recovered 
and extended. The dipole formula, however, is an all-order ansatz, and 
allows us to explore what happens beyond NLL, and indeed beyond the 
realm of Regge factorization. We find that Reggeization generically 
breaks down at NNLL, also for the real part of the amplitude, and we 
are able to write a completely general form, \eqn{NNLL}, for the 
leading Reggeization-breaking operator, which arises at three loops 
in the exponent.  Note, however, that Reggeization for the infrared-singular terms in the amplitude, is preserved in the large-$N_c$ limit, a fact which deserves further investigation. 

The fact that the simple form of Regge factorization does not fully describe the high-energy limit of amplitudes is perhaps expected based on the argument that in addition to Regge poles, higher-loop corrections may give rise to Regge cuts~\cite{Mandelstam:1965zz}. The possible connection between these analytic structures and the violation of the simple Regge pole picture we observed, requires a dedicated study.

Our conclusion concerning Reggeization breaking
remains valid even if the dipole formula receives quadrupole corrections
at three loops: indeed, possible new contributions to the soft anomalous 
dimension at the three loops would affect only the single-pole term 
at NNLL level, not the triple-pole term generated by the
Reggeization-breaking operator.

We have also demonstrated, in \sect{sec:beyond_dipoles}, that the connection between soft singularities in amplitudes and Reggeization may be exploited to further constrain the soft anomalous dimension
matrix. Three-loop corrections to the soft anomalous dimension going 
beyond the dipole formula have already been shown to be highly 
constrained by factorization and rescaling symmetry, collinear limits, 
Bose symmetry and transcendentality~\cite{Gardi:2009qi,
Becher:2009qa,Dixon:2009ur}. Nevertheless, Ref.~\cite{Dixon:2009ur}
provided explicit examples of functions that are consistent with all 
these constraints. Here, upon considering these functions in the 
high-energy limit, we have found that they all give rise to super-leading 
logarithms which conflict with the known behaviour in the Regge limit. 
Thus, these examples can no longer be considered viable (except if they 
occur in particular combinations in which the super-leading as well as 
the leading logarithms cancel out). This gives further support to the 
validity of the dipole formula beyond two loops. Clearly, however, a 
proof is still missing. 

Finally, we have used the dipole formula to study the high-energy
limit of multiparton scattering amplitudes in multi-Regge kinematics.
Once again, the dipole formalism proves to be an efficient and
appealing way  to study the problem: we recover the known form
of the amplitude in Regge factorization at LL accuracy, we can 
readily read off the (divergent part of) the corresponding Regge 
trajectories, and we can immediately identify the form of
Reggeization-breaking operators starting at NNLL level.

In summary, our infrared-based approach offers new insights into 
the Regge limit, and a particularly clear way of understanding how 
Reggeization arises, and eventually breaks down. Reggeization, 
indeed, appears to be an infrared-dominated phenomenon, and 
one may wonder to what extent the resummation of finite 
contributions, which start arising at two loops in the Regge 
trajectory, might be understood from an infrared point of view:
in fact, in several cases in the past~\cite{Parisi:1979xd,
Sterman:1986aj,Magnea:1990zb,Eynck:2003fn,Bern:2005iz,
Ahrens:2008qu} it was observed that certain classes of 
infrared-finite contributions are carried along with singularities 
when these exponentiate. The study of these issues is left for 
future work.

We believe that our results pave the way for further 
progress in several directions: corrections to the dipole formula may be further constrained, and perhaps shown to be absent; 
Reggeization of finite contributions to the amplitude may be studied from an infrared viewpoint; our results may be used to test the breakdown 
of Reggeization at NNLL and gauge its impact on phenomenology; finally, the infrared singularity structure of amplitudes may be used to study the high-energy limit beyond the realm of Reggeization. 


\section*{Acknowledgements}
We are grateful to Viktor Fadin and Gregory Korchemsky for useful discussions. This work 
was partly supported by the Research Executive Agency (REA) of 
the European Union, through the Initial Training Network LHCPhenoNet, 
under contract PITN-GA-2010-264564, and by MIUR (Italy) under 
contract 2006020509$\_$004. CDW is supported by the STFC 
Postdoctoral Fellowship ``Collider Physics at the LHC''. CDW, CD  and VDD are very 
grateful to the School of Physics and Astronomy at the University 
of Edinburgh for hospitality on a number of occasions. LM gratefully 
acknowledges the award of a SUPA Distinguished Visitor Fellowship. VDD, CD and EG thank the KITP, Santa Barbara for hospitality. Finally, VDD and EG thank the GGI, Firenze, for hospitality at the last stages of writing this paper.

\vspace{1cm}

\appendix

\section{Reggeization using Clebsch-Gordan coefficients}
\label{app:clebsch}

In \sect{sec:general} we have demonstrated in general how 
Reggeization arises after decomposing the hard interaction in the 
Regge limit in a color flow basis consisting of distinct irreducible 
representations in the $t$ channel. These are eigenstates of the 
Reggeization operator, which contains the quadratic Casimir operator 
associated with $t$-channel exchanges. It is instructive to see how 
this works in detail, carrying out the color algebra in full. This is the 
subject of this appendix.

\subsection{Tensor product representations and Clebsch-Gordan coefficients}

Scattering amplitudes transform in general as tensors under gauge 
transformations. In order to lay the ground for our analysis, we start 
by recalling general facts about tensor product representations and 
Clebsch-Gordan decompositions.

Consider two vector spaces $V_1$ and $V_2$ transforming in 
some irreducible representations $R_1$ and $R_2$ of some group 
$G$. More precisely, let $\{|a \, k \rangle\} \equiv\{|a\rangle
\otimes|k\rangle\}$ be a basis of $V_1\otimes V_2$. Then the 
group $G$ acts on the basis vectors via
\beq
  |a \, k \rangle \, \to \, |a' \, k' \rangle \, U^{(R_1)}_{a' a} 
  \, U^{(R_2)}_{k' k} \, ,
\label{tensprod}
\eeq
where $U^{(R)}_{a' a}$ denotes the representation matrices of $G$ 
in the irreducible representation $R$. The tensor product representation 
$R_1\otimes R_2$ will in general be reducible, and we can decompose 
$V_1\otimes V_2$ into a direct sum
\beq
  V_1\otimes V_2 \, = \, \bigoplus_{r} V_r \, ,
\label{decomp}
\eeq
where the sum runs over all irreducible representations of $G$ 
that appear in the decomposition\footnote{If a given irreducible 
representation appears with multiplicity $m > 1$, we consider each 
replica separately.} of $R_1\otimes R_2$, and $V_r$ is the subspace 
of $V_1\otimes V_2$ that transforms in the irreducible representation 
$r$. Let $\{| r, \alpha \rangle\}$ be a basis of $V_r$, which under 
the group action transforms as
\beq
  | r, \alpha \rangle \, \to \, | r, \alpha' \rangle \, 
  U^{(r)}_{\alpha' \alpha} \, .
\label{rot}
\eeq
The change of basis is expressed by the unitary transformation 
whose matrix elements are the Clebsch-Gordan coefficients,
\beq
  | a \, k \rangle \, = \, \sum_r  | r, \alpha \rangle \, \langle
  r, \alpha \, | \, a \, k \rangle \, \equiv \, \sum_r  | r, \alpha 
  \rangle \, C(R_1, R_2; r)_{\alpha a k} \, \, .
\label{eq:basis_change}
\eeq
As the change of basis is unitary, the Clebsch-Gordan coefficients 
must satisfy the relations
\beqa
  \sum_r C(R_1, R_2; r)^\ast_{\alpha a k} \, 
  C(R_1, R_2; r)_{\alpha a' k'} & = &
  \delta_{a a'} \, \delta_{k k'} \, , \nonumber \\
  C(R_1, R_2; r)^\ast_{\alpha a k} 
  \, C(R_1, R_2; r')_{\alpha' a k} & = & \delta_{r r'} \,
  \delta_{\alpha \alpha'} \, ,
\label{eq:unitarity_relations}
\eeqa
as well as
\beq
  C (\overline{R}_1, \overline{R}_2; \overline{r})_{\alpha a k} 
  \, = \, C(R_1, R_2; r)^\ast_{\alpha a k} \, ,
\label{eq:CG_complex_conj}
\eeq
where $\overline{R}$ denotes the irreducible representation complex conjugate to $R$.

Let now $| X \rangle$ be an arbitrary vector in $V_1\otimes V_2$. 
We can write
\beq
  | X \rangle \, = \, | a \, k \rangle \, X_{a k} \, = \, \sum_r 
  | r, \alpha \rangle \, X(r)_\alpha \, .
\label{eq:X}
\eeq
Inserting \eqn{eq:basis_change} into \eqn{eq:X}, we immediately 
see that
\beq
  X(r)_\alpha \, = \, C(R_1,R_2;r)_{\alpha a k} \, X_{a k} \, ,
\label{eq:X_decomp}
\eeq
and, using the unitarity relations~\eqn{eq:unitarity_relations} we can invert this relation to obtain
\beq
  X_{a k} \, = \, \sum_r X(r)_\alpha \, 
  C(R_1, R_2; r)^\ast_{\alpha a k} \, .
\label{eq:contract}
\eeq
The components of the vector $| X \rangle$ transform under the 
group action as
\beq
  X_{a k} \, \to \, U^{(R_1)}_{a a'} \, U^{(R_2)}_{k k'} \, X_{a' k'} 
  \, \, {\rm~~and~~} \, \,
  X(r)_\alpha \, \to \, U^{(r)}_{\alpha \alpha'} \, X(r)_{\alpha'} \, .
\label{and}
\eeq
Compatibility of these transformations with \eqn{eq:X_decomp} 
then implies the relation
\beq
  C(R_1, R_2; r)_{\alpha a' k'} \, U^{(R_1)}_{a' a} \,
  U^{(R_2)}_{k' k} \, = \, U^{(r)}_{\alpha \alpha'} \,
  C(R_1, R_2; r)_{\alpha' a k} \, ,
\label{Ctransf}
\eeq
or, equivalently, in infinitesimal form,
\beq
  C(R_1, R_2; r)_{\alpha a' k} \, \left(T_{R_1}^c \right)_{a'a} + 
  C(R_1, R_2; r)_{\alpha a k'} \, \left(T_{R_2}^c \right)_{k' k}
  \, = \,  \left(T_{r}^c \right)_{\alpha \alpha'} \, 
  C(R_1, R_2; r)_{\alpha' a k} \, ,
\label{Ctransfinf}
\eeq
where $T_R^c$ denote the generators of the irreducible 
representation $R$. In the following we will need the equivalent of 
this relation for the complex conjugated Clebsch-Gordan coefficients 
$C(R_1, R_2; r)_{\alpha a k}^\ast$. Taking the complex conjugate 
of the previous equations, and using the fact that the generators are 
hermitian, we arrive at
\beq
  \hspace{-3mm}
  \left(T_{R_1}^c \right)_{a a'} \, 
  C(R_1, R_2; r)_{\alpha a' k}^\ast + \left(T_{R_2}^c 
  \right)_{k k'} \, C(R_1, R_2; r)_{\alpha a k'}^\ast \, = \, 
  C(R_1, R_2; r)_{\alpha'a k}^\ast \, \left(T_{r}^c 
  \right)_{\alpha' \alpha} .
\label{eq:CB_transform_star}
\eeq
Note that this relation is consistent with \eqn{eq:CG_complex_conj}.

\subsection{Decomposition of a scattering amplitude in a $t$ 
channel basis}

\begin{figure}
\begin{center}
\scalebox{0.8}{\includegraphics{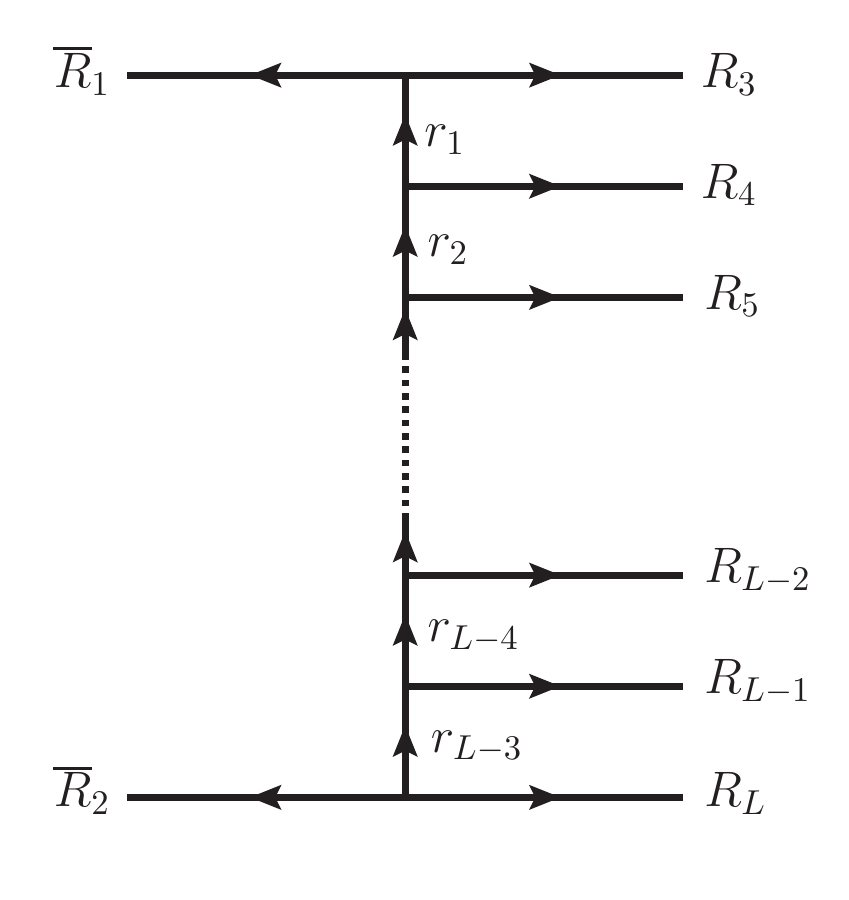}}
\caption{Diagrammatical representation in colour flow space of the 
the subamplitude $\M_J \, (c^J)_{a_1 \ldots a_L}$ for $J = (r_1, 
\ldots,r_{L-3})$. Each three-point vertex is proportional to a 
Clebsch-Gordan coefficient.}
\label{fig:t_channel_decomp}
\end{center}
\end{figure}

In this section we prove that every scattering amplitude can be decomposed into a $t$ channel basis, and that the subamplitudes 
that appear in this decomposition are eigenstates of the $t$ 
channel colour operators defined in Sec.~\ref{sec:multiregge},
\beq
  {\bf T}_{t_k}^2 \, = \, \left({\bf T}_1 + \sum_{p = 1}^k 
  {\bf T}_{p + 2} \right)^2 \, .
\label{eq:app:t_channel_operators}
\eeq
Let us consider a scattering of $L$ particles, $1, 2 \to 3, \ldots,
L$, transforming in the representations $R_i$, $1 \le i \le L$.
The scattering amplitude for this process  can be seen as a vector 
in colour space,
\beq
  | \M \rangle \, = \, | a_1 \, \ldots \, a_L \rangle \, 
  \M_{a_1 \ldots a_L} \, .
\label{Mvec}
\eeq
The basis vectors transform in the representation $\overline{R}_1
\otimes \overline{R}_2 \otimes R_3 \otimes \ldots \otimes R_L$ of 
the gauge group\footnote{We consider initial state particles as 
outgoing antiparticles.},
\beq
  | a_1\, \ldots \, a_L \rangle \, \to \, | b_1 \, \ldots \, b_L \rangle
  \, U^{(\overline{R}_1)}_{b_1 a_1} 
  \, U^{(\overline{R}_2)}_{b_2 a_2}
  \, U^{(R_3)}_{b_3 a_3} \ldots U^{(R_L)}_{b_L a_L} \, ,
\label{multivectransf}
\eeq
so that the amplitude $\M_{a_1\ldots a_L}$ transforms as
\beq
  \M_{a_1 \ldots a_L} \, \to 
  \, U^{(\overline{R}_1)}_{a_1 b_1}
  \, U^{(\overline{R}_2)}_{a_2 b_2}
  \, U^{(R_3)}_{a_3 b_3}
  \ldots U^{(R_L)}_{a_L b_L} \, \M_{b_1 \ldots b_L} \, .
\label{multiMtransf}
\eeq
The colour operators ${\bf T}_i$ are defined by their action 
on colour space, according to~\cite{Bassetto:1984ik,Catani:1996vz}\footnote{In this section we follow strictly the conventions of~\cite{Catani:1996vz}, which differ slightly from the convention used in the main text.}
\beq
  {\bf T}_i^c \, | a_1 \, \ldots \,a_L \rangle \, \equiv \, \left\{ 
  \begin{array}{ll}
  | a_1 \, \ldots \, b_i \, \ldots \, a_L \rangle \, 
  \left( T^c_{R_i} \right)_{b_i a_i} \, , & \quad 3 \le  i \le L \, , \\
  | a_1 \, \ldots \, b_i \, \ldots \, a_L \rangle \,
  \left( T^c_{\overline{R}_i} \right)_{b_i a_i}  , 
  & \quad i = 1, 2 \, .
  \end{array} \right.
\label{eq:T_operator}
\eeq
or equivalently,
\beq
  {\bf T}_i^c \, \M_{a_1 \ldots a_L} \, = \, \left\{
  \begin{array}{ll}
  \left( T^c_{R_i} \right)_{a_i b_i} \, \M_{a_1 \ldots b_i 
  \ldots a_L} \, , & \quad 3 \le i \le L \, , \\
  \left( T^c_{\overline{R}_i} \right)_{a_i b_i} \M_{a_1 \ldots 
  b_i \ldots a_L} , & \quad i = 1, 2 \, ,
  \end{array} \right.
\label{altTop}
\eeq
Note that in this notation the generators of the representation 
$\overline{R}_1 \otimes \overline{R}_2 \otimes R_3 \otimes
\ldots \otimes R_L$ are just given by $\sum_{i = 1}^L{\bf T}^c_i$.
Colour conservation implies that $|\M\rangle$ must be a colour 
singlet, or in other words, $\M_{a_1 \ldots a_L}$ is an invariant 
tensor transforming in the representation $\overline{R}_1 \otimes
\overline{R}_2 \otimes R_3 \otimes \ldots \otimes R_L$. As a 
consequence, $\M_{a_1\ldots a_L}$ must be annihilated by the 
generators of the gauge group, leaving us with the usual constraint 
expressing colour conservation,
\beq
  \left(\sum_{i = 1}^L{\bf T}^c_i \right) \, 
  \M_{a_1 \ldots a_L} \, = \, 0 \, .
\label{eq:app:colour_conserve}
\eeq
Let us now turn to the proof that we can always decompose 
$\M_{a_1 \ldots a_L}$ into a colour basis corresponding to 
definite $t$ channel exchanges. Using \eqn{eq:contract}, one 
may sequentially multiply representations starting at the top 
of the ladder in \fig{fig:t_channel_decomp}, and ending at the 
bottom. At each step, one  eliminates one of the outgoing parton 
indices in favour of an index associated with the corresponding 
vertical strut of the ladder, yielding
\beqa
  \M_{a_1\ldots a_L} & = & \sum_{r_1} 
  \M(r_1)_{\alpha_1 a_2 a_4 \ldots a_L} \, 
  C(\overline{R}_1, R_3; r_1)^\ast_{\alpha_1a_1a_3} 
\label{M_ai} \\
  & = & \sum_{r_1, r_2} \M(r_1, r_2)_{\alpha_2 a_2 a_5 \ldots a_L}
  \, C(r_1, R_4; r_2)^\ast_{\alpha_2 \alpha_1 a_4}
  \, C(\overline{R}_1, R_3; r_1)^\ast_{\alpha_1a_1 a_3}
  \,\, = \, \ldots \nonumber \\ 
  & = & \sum_{r_1, \ldots, r_{L - 2}}
  \M(r_1, r_2, \ldots, r_{L - 2})_{\alpha_{L - 2}a_2}
  \,C(r_{L - 3}, R_L; r_{L - 2})^\ast_{\alpha_{L - 2}
  \alpha_{L - 3} a_L} \, \ldots \nonumber \\
  && \hspace{6cm} \ldots \, \, C(\overline{R}_1, R_3; 
  r_1)^\ast_{\alpha_1 a_1 a_3} \, .
  \nonumber
\eeqa
So far all the manipulations were generic and could have been 
applied to an arbitrary tensor transforming in the representation 
$\overline{R}_1\otimes \overline{R}_2 \otimes R_3 \otimes
\ldots \otimes R_L$. At this stage however we can impose colour 
conservation in the form of \eqn{eq:app:colour_conserve} to 
further constrain the form of the subamplitudes $\M(r_1,\ldots,
r_{L - 2})_{\alpha_{L - 2} a_2}$. Acting with the colour operator
${\bf T}^c_2 \, + \, \sum_{i = 1, \, i \neq 2}^L{\bf T}^c_i = 0$
on the Clebsch-Gordan coefficients, and making repeated use of 
\eqn{eq:CB_transform_star}, we arrive at the identity,
\beqa
  && \hspace{-3mm} \sum_{r_1, \ldots, r_{L - 2}} \left[ 
  (T_{\overline{R}_2}^c)_{a_2 b_2} \, 
  \M(r_1, r_2, \ldots, r_{L - 2})_{\beta_{L - 2}b_2} +
  (T_{r_{L - 2}}^c)_{\beta_{L - 2} \alpha_{L - 2}} \,
  \M(r_1, r_2, \ldots, r_{L - 2})_{\alpha_{L - 2}a_2}
  \right]  \nonumber \\
  && \qquad \times \, \, C(r_{L - 3}, R_L; 
  r_{L - 2})^\ast_{\beta_{L - 2}
  \beta_{L - 3} a_L} \ldots C(\overline{R}_1, R_3; 
  r_1)^\ast_{\beta_1 a_1 a_3} \, = \, 0 \, .
\label{eq:temp_eq}
\eeqa
The unitarity relations for the Clebsch-Gordan coefficients now imply 
that this identity can only be fulfilled if the expression inside the 
square brackets vanishes,
\beq
  (T_{\overline{R}_2}^c)_{a_2 b_2} \, 
  \M(r_1, r_2, \ldots, r_{L - 2})_{\beta_{L - 2} b_2} \, + \,
  (T_{r_{L - 2}}^c)_{\beta_{L - 2} \alpha_{L - 2}} \, 
  \M(r_1, r_2, \ldots, r_{L - 2})_{\alpha_{L - 2} a_2}
  \, = \, 0 \, .
\label{eq:inf_Schur}
\eeq
In order to proceed, we have to recall that Schur's lemma implies 
that if ${\cal T}_{\alpha \beta}$ is an invariant tensor transforming 
in the representation $R \otimes \overline{R}'$, {\it i.e.}
\beq
  {\cal T}_{\alpha \beta} \, = \, U^{(R)}_{\alpha \alpha'} \,
  \, U^{(\overline{R}')}_{\beta \beta'} \, {\cal T}_{\alpha'
  \beta'} \, ,
\label{eq:Schur}
\eeq
then ${\cal T}_{\alpha \beta}$ must be zero, unless $R$ is 
equivalent to $R'$, in which case ${\cal T}_{\alpha \beta}$ must 
be proportional to the identity matrix. \Eqn{eq:inf_Schur} is just 
the infinitesimal form of \eqn{eq:Schur}: thus Schur's lemma 
implies that 
\beq
  \M(r_1, r_2, \ldots, r_{L - 2})_{\alpha_{L - 2} a_2} \, = \, 
  \M(r_1, r_2, \ldots, r_{L - 3}) \, \delta_{r_{L - 2}R_2} \,
  \delta_{\alpha_{L - 2} a_2} \, .
\label{quasidone}
\eeq
One thus concludes that \eqn{M_ai} may be written as
\beq
  \M_{a_1 \ldots a_L} \, = \, \sum_J \M_J \, 
  (c^J)_{a_1\ldots a_L} \, ,
\label{eq:t_decomp}
\eeq
where we defined $J \equiv(r_1, \ldots, r_{L - 3})$, $\M_J 
\equiv \M(r_1, \ldots, r_{L - 3})$, and
\beq
  (c^J)_{a_1 \ldots a_L} \, = \, C(r_{L - 3}, R_L; 
  R_{2})^\ast_{a_2 \alpha_{L - 3} a_L} \, \ldots \, 
  C(\overline{R}_1, R_3; r_1)^\ast_{\alpha_1 a_1 a_3} \, .
\label{appresult}  
\eeq
Eq.~\eqref{eq:t_decomp} is the desired result: we have written 
the scattering amplitude $\M_{a_1\ldots a_L}$ as a sum of
terms, each characterized by a sequence of irreducible 
representations $(r_1, \ldots, r_{L - 3})$, which correspond to 
the irreducible representations of the states propagating in the 
$t$ channel. Figure~\ref{fig:t_channel_decomp} gives a 
diagrammatical representation of a single term $\M_J \, 
(c^J)_{a_1\ldots a_L}$ in colour flow space.

We now want to prove that the colour coefficients $(c^J)_{a_1
\ldots a_L}$, for fixed $J$, are eigenvectors of the $t$ channel 
colour operators defined in \eqn{eq:app:t_channel_operators}, 
and that the eigenvalues are given by the Casimir operators of the 
representation exchanged in the $t$ channel. More precisely,
\beq
  {\bf T}^2_{t_k} \, (c^J)_{a_1 \ldots a_L} \, = \, C_{r_k} 
  \, (c^J)_{a_1\ldots a_L} \, .
\label{eigencas}
\eeq
To prove this, we first act with ${\bf T}^c_{t_k}$ on the colour 
coefficient $(c^J)_{a_1 \ldots a_L}$, and we apply the same reasoning 
as in the derivation of \eqn{eq:temp_eq}, {\it i.e.} we make repeated use of \eqn{eq:CB_transform_star} to obtain
\beqa
  {\bf T}_{t_k}^c \, (c^J)_{a_1 \ldots a_L}  & = & 
  C (r_{L - 3}, R_L; R_{2})^\ast_{a_2 \alpha_{L - 3} a_L}
  \ldots C(r_{k}, R_{k + 3}; r_{k + 1})^\ast_{\alpha_{k + 1} 
  \alpha_k a_{k + 3}}
\label{actone} \\
  && \times \, \, (T_{r_k}^c)_{\beta_k \alpha_k}
  \, C(r_{k - 1}, R_{k + 2}; r_k)^\ast_{\beta_k \beta_{k - 1}
  a_{k + 2}} \ldots
  C(\overline{R}_1, R_3; r_1)^\ast_{\beta_1 a_1 a_3} \, .
  \nonumber 
\eeqa
If we now act a second time with the same operator, we obtain
\beqa
  {\bf T}_{t_k}^2 \, (c^J)_{a_1\ldots a_L} & = & 
  C(r_{L - 3}, R_L; R_2)^\ast_{a_2 \alpha_{L - 3}a_L}
  \ldots C(r_{k}, R_{k + 3}; r_{k + 1})^\ast_{\alpha_{k + 1}
  \alpha_k a_{k +3 }} 
\label{acttwo} \\
  && \times \, \, (T_{r_k}^c)_{\beta_k \alpha_k} \, 
  (T_{r_k}^c)_{\gamma_k \beta_k}
  \, C(r_{k - 1}, R_{k + 2}; r_k)^\ast_{\gamma_k 
  \gamma_{k - 1}a_{k + 2}} \ldots
  C(\overline{R}_1, R_3; r_1)^\ast_{\gamma_1 a_1 a_3}
  \nonumber \\
  & = & C_{r_k} \, (c^J)_{a_1 \ldots a_L} \, . \nonumber
\eeqa
We conclude that, as desired, $(c^J)_{a_1 \ldots a_k}$ is an 
eigenvector of ${\bf T}^2_{t_k}$ with eigenvalue $C_{r_k}$.
Note that, as this argument is independent of $k$, we have at 
the same time shown that the operators ${\bf T}^2_{t_k}$ 
are simultaneously diagonalizable, and hence always commute
\beq
  [{\bf T}^2_{t_k}, {\bf T}^2_{t_{k'}}] \, = \, 0 \, , 
  \quad \forall k, k' \, .
\label{commute}
\eeq

\subsection{The Reggeization operator in the $t$-channel basis}

Let us now return to the actual problem, and consider the action 
of the operator $\widetilde{Z}^{\rm MR}|_{LL}$ of \eqn{ZtildeMR2}
in the high-energy limit,  and let us see how this operator acts on the 
hard amplitude. For simplicity, in this section we concentrate on the 
case of four-point scattering: the generalisation to multiparticle 
production is straightforward, since we have shown that all the 
$t$-channel colour operators commute. The Reggeization operator 
for a four-point amplitude was given in \eqn{ZLLdef},
\beq
  \widetilde{Z}_{LL}  \, = \, \exp \left\{ K (\alpha_s, \epsilon) 
  \, \ln \left( \frac{s}{- t} \right)
  \, {{\bf T}_t}^2 \, \right\} \, .
\label{repetita}
\eeq
We start by decomposing the hard amplitude into a $t$-channel 
colour basis, as given in \eqn{eq:t_decomp}, writing
\beq
  {\cal H}_{a_1 a_2 a_3 a_4} \, = \, \sum_J {\cal H}_J
  \, (c^J)_{a_1 a_2 a_3 a_4} \, .
\label{decomphard}
\eeq
Note that in this case $J$ simply labels the representation 
$r$ of the state exchanged in the $t$ channel. We now act 
with the Reggeization operator $\widetilde{Z}_{LL}$ on 
the hard interaction,
\beqa
  \widetilde{Z}_{LL} \, {\cal H}_{a_1 a_2 a_3 a_4} 
  & = & \sum_{n = 0}^\infty \sum_J {1 \over n!} \, 
  K^n \, \ln^n \left({s \over - t} \right) \, {\cal H}_J \,
  \left( {\bf T}^2_t \right)^n \, (c^J)_{a_1 a_2 a_3 a_4}
  \nonumber \\
  & = & \sum_{n = 0}^\infty \sum_J {1\over n!} \, 
  K^n \, \ln^n \left({s \over - t} \right) \, {\cal H}_J \,
  C_t^n \, (c^J)_{a_1 a_2 a_3 a_4} \nonumber \\
  & = & \sum_J \left[ \sum_{n = 0}^\infty{1 \over n!}
  \, K^n \, \ln^n \left({s \over - t} \right) \,
  C_t^n \right] \, {\cal H}_J \, (c^J)_{a_1 a_2 a_3 a_4}
  \nonumber \\
  & = & \sum_J \left({s \over - t} \right)^{K C_t} \,
  {\cal H}_J \, (c^J)_{a_1 a_2 a_3 a_4} \, .
\label{done}
\eeqa
This formula presents explicitly the Reggeization of the singular 
part of the amplitude. As we already saw each $t$-channel 
exchange Reggeizes separately, with the exponent being 
controlled by the corresponding quadratic Casimir.

\section{Rapidity-dependent contribution in $L$ parton scattering}
\label{app:rapproof}

In this appendix, we prove the result stated in eq.~(\ref{rapres}), namely
that the rapidity-dependent terms in the MR limit of the sum over dipoles
formula decompose into a sum over Casimir operators corresponding to
definite $t$-channel exchanges, where each is associated with a consecutive
rapidity difference.

We start from the left-hand side of eq.~(\ref{rapres}) and rewrite it as
\begin{align}
\sum_{i=1}^{L-1}\sum_{j>i}|y_i-y_j|{\bf T}_i\cdot{\bf T}_j
&={\bf T}_1\cdot{\bf T}_2(y_3-y_L)+{\bf T}_1\cdot\sum_{j=4}^L
{\bf T}_j(y_3-y_j)+{\bf T}_2\cdot\sum_{j=3}^{L-1}{\bf T}_j
(y_j-y_L)\notag\\
&+\sum_{j=3}^{L-1}\sum_{i>j}^L{\bf T}_i\cdot{\bf T}_j
(y_j-y_i).
\label{rapdiff1}
\end{align}
Next, one may eliminate ${\bf T}_2$ by using color conservation, 
eq.~(\ref{colcongen}), to give
\begin{align}
&\sum_{i=1}^{L-1}\sum_{j>i}|y_i-y_j|{\bf T}_i\cdot{\bf T}_j
=-C_1(y_3-y_L)-\sum_{j=3}^L{\bf T}_1\cdot{\bf T}_j(y_3-y_L)
+\sum_{j=4}^L{\bf T}_1\cdot{\bf T}_j(y_3-y_j)\notag\\
&\quad-\sum_{j=3}^{L-1}
{\bf T}_1\cdot{\bf T}_j(y_j-y_L)-\sum_{i=3}^L\sum_{j=3}^{L-1}
{\bf T}_i\cdot{\bf T}_j(y_j-y_L)+\sum_{j=3}^{L-1}\sum_{i>j}^L
{\bf T}_i\cdot{\bf T}_j(y_j-y_i)\notag\\
&=-C_1(y_3-y_L)-2\sum_{j=3}^{L-1}{\bf T}_1\cdot{\bf T}_j(y_j-y_L)
-\sum_{i=3}^L\sum_{j=3}^{L-1}{\bf T}_i\cdot{\bf T}_j(y_j-y_L)
\notag\\
&\quad+\sum_{j=3}^{L-1}\sum_{i>j}^L{\bf T}_i\cdot{\bf T}_j(y_j-y_i),
\label{rapdiff2}
\end{align}
where in the second line we have combined the second, third and fourth terms
from the previous line. Decomposing the third term in the last line of eq.~\eqref{rapdiff2} into three
contributions with $i=j$, $i<j$ and $i>j$ respectively and combining the result with the
fourth term gives
\begin{align}
&-\sum_{j=3}^{L-1}C_j(y_j-y_L)-\sum_{i=4}^{L}\sum_{3\le j<i}{\bf T}_i
\cdot{\bf T}_j(y_j-y_L)-\sum_{i=3}^{L-2}\sum_{j>i}^{L-1}{\bf T}_i
\cdot{\bf T}_j(y_j-y_L)\notag\\
&\quad+\sum_{j=3}^{L-1}\sum_{i>j}^L(y_j-y_i)
{\bf T}_i\cdot{\bf T}_j\notag\\
&=-\sum_{j=3}^{L-1}C_j(y_j-y_L)-\sum_{i=4}^{L}\sum_{3\le j<i}{\bf T}_i
\cdot{\bf T}_j(y_j-y_L)-\sum_{i=3}^{L-2}\sum_{j>i}^{L-1}{\bf T}_i
\cdot{\bf T}_j(y_j-y_L)\notag\\
&\quad+\sum_{i=4}^{L}\sum_{3\le j<i}(y_j-y_i)
{\bf T}_i\cdot{\bf T}_j,
\label{rapdiff3}
\end{align}
where we have interchanged the orders of the summations over $i$ and $j$ in the
final term. Combining this expression with the second term in eq.~\eqref{rapdiff2} gives
\begin{align}
-\sum_{j=3}^{L-1}C_i(y_j-y_L)-\sum_{i=4}^L\sum_{3\le j<i}^{L-1}{\bf T}_i\cdot
{\bf T}_j(y_i-y_L)-\sum_{i=3}^{L-2}\sum_{j>i}^{L-1}{\bf T}_i\cdot
{\bf T}_j(y_j-y_L).
\label{rapdiff4}
\end{align}
Note that the contribution from the $i=L$ term in the second term is zero,
which allows one to replace the upper limit of the sum over $i$ by
$L-1$. One may also relabel $i$ and $j$ in this
term, so that the expression~(\ref{rapdiff4}) becomes 
\begin{align}
&-\sum_{j=3}^{L-1}C_j(y_j-y_L)-\sum_{i=3}^{L-2}\sum_{j>i}^{L-1}{\bf T}_i
\cdot{\bf T}_j(y_j-y_L)-\sum_{i=3}^{L-2}\sum_{j>i}^{L-1}{\bf T}_i
\cdot{\bf T}_j(y_j-y_L)\notag\\
&=-\sum_{j=3}^{L-1}C_j(y_j-y_L)-2\sum_{i=3}^{L-2}\sum_{j>i}^{L-1}{\bf T}_i
\cdot{\bf T}_j(y_j-y_L).
\label{rapdiff5}
\end{align}
Combining with the remaining contributions from 
eq.~(\ref{rapdiff2}), we find
\begin{align}
\sum_{i=1}^{L-1}\sum_{j>i}|y_i-y_j|{\bf T}_i\cdot{\bf T}_j
&=-C_1(y_3-y_L)-\sum_{j=3}^{L-1}C_j(y_j-y_L)-2\sum_{j=3}^{L-1}
{\bf T}_1\cdot{\bf T}_j(y_j-y_L)\notag\\
&\quad-2\sum_{i=3}^{L-2}\sum_{j>i}^{L-1}
{\bf T}_i\cdot{\bf T}_j(y_j-y_L).
\label{rapdiff6}
\end{align}
After interchanging the order of the sums over $i$ and $j$ in the final term of this expression,
we may rewrite eq.~\eqref{rapdiff6}
\begin{align}
\sum_{i=1}^{L-1}\sum_{j>i}|y_i-y_j|{\bf T}_i\cdot{\bf T}_j
&=-C_1(y_3-y_L)-\sum_{j=3}^{L-1}(y_j-y_L)\left[C_j+2\sum_{i<j, i\neq 2}
{\bf T}_i\cdot{\bf T}_j\right],
\label{rapdiff7}
\end{align}
where in the final term one has $1\leq i\leq L-2$. 

We may now rewrite each rapidity difference in terms of the consecutive
differences $\Delta y_k=y_k-y_{k+1}$. That is,
\begin{equation}
y_j-y_L=\sum_{k=j}^{L-1}\Delta y_k.
\label{rapdiffs}
\end{equation}
Substituting this in eq.~(\ref{rapdiff7}) gives
\begin{align}
\sum_{i=1}^{L-1}\sum_{j>i}|y_i-y_j|{\bf T}_i\cdot{\bf T}_j
&=-C_1\sum_{k=3}^{L-1}\Delta y_k-\sum_{j=3}^{L-1}\sum_{k=j}^{L-1}\Delta y_k
\left[C_j+2\sum_{i<j,i\neq2}{\bf T}_i\cdot{\bf T}_j\right]\notag\\
&=-C_1\sum_{k=3}^{L-1}\Delta y_k-\sum_{k=3}^{L-1}\sum_{j=3}^{k}\Delta y_k\left[
C_j+2\sum_{i<j,i\neq2}{\bf T}_i\cdot{\bf T}_j\right],
\label{rapdiff8}
\end{align}
where in the second line we have interchanged the order of summation over $k$
and $j$. The coefficient of $\Delta y_k$ is
\begin{align}
&-C_1-\sum_{j=3}^k\left[C_j+2\sum_{i<j,i\neq2}{\bf T}_i\cdot{\bf T}_j
\right]=-{\bf T}^2_{t_{k-2}},
\label{ycoeff}
\end{align}
where the right-hand side contains the $t$-channel quadratic Casimir operator
defined in eq.~(\ref{Tkdef}). This completes the derivation of 
eq.~(\ref{rapres}).

\bibliographystyle{JHEP}
\bibliography{refs2.bib}
\end{document}